\documentclass[twocolumn,pre,floatfix]{revtex4}
\usepackage{psfrag,epsfig,amsfonts,amssymb,amsmath,wasysym,bm}
\usepackage{dcolumn}
\usepackage{bbold}
\usepackage[normalem]{ulem}
\usepackage{color}
\usepackage{hyperref}
\usepackage{enumitem} % enumerated lists

\newcommand{\ket}[1]{\lvert #1 \rangle} 	% ket
\newcommand{\bra}[1]{\langle #1 \rvert}	% bra
\newcommand{\ketN}[1]{\lvert #1 \rangle_{\! 0}} 	% ket
\newcommand{\braN}[1]{{_0\!}\langle #1 \rvert}	% bra
\newcommand{\mc}{\mathcal}	% calligraphic
\newcommand{\sUp}{\uparrow}		% spin up
\newcommand{\sDn}{\downarrow}	% spin down

\newcommand{\etal}{\emph{et al.}}

\newcommand{\<}{\left\langle}	% angular brackets
\renewcommand{\>}{\right\rangle}	% "

\newcommand{\e}{\mathrm{e}}		% Euler number
\newcommand{\I}{\mathrm{i}}		% imaginary unit
\newcommand{\lvsp}{\varepsilon}

\newcommand{\Ath}{\langle A\rangle_{\!\!\;\rm{mc}}}

\newcommand{\RsvG}{\mathcal{G}}

\newcommand{\ptst}{\lambda}
\newcommand{\rateExp}{\Gamma}

\newcommand{\ldos}{u}
\newcommand{\ldosTilde}{\tilde{u}}
\newcommand{\bos}{\mathrm{B}}
\newcommand{\fer}{\mathrm{F}}
\newcommand{\Ei}{\mathrm{Ei}}

\newcommand{\KK}{\mathcal{K}}
\newcommand{\LL}{\mathcal{L}}

\newcommand{\id}{\mathbb 1}
\newcommand{\RR}{{\mathbb R}}

\newcommand{\NN}{{\mathbb N}}
\newcommand{\ZZ}{{\mathbb Z}}

\newcommand{\tr}{\mbox{Tr}}

\newcommand{\ord}{{\cal O}}

\providecommand{\norm}[1]{\|#1\|}

\providecommand{\av}[1]{[#1]_V}
\providecommand{\avv}[1]{\left[#1\right]_{\!V}}
\providecommand{\avx}[1]{E[#1]}
\providecommand{\avvx}[1]{E\left[#1\right]}

\renewcommand{\d}{\mathrm{d}} % differential
\newcommand{\trans}{\mathsf{T}}     % transpose
\newcommand{\lmat}{\left( \begin{matrix}}	% matrix begin
\newcommand{\rmat}{\end{matrix} \right)}	% matrix end
\newcommand{\str}{\mathrm{str}} % super trace
\newcommand{\mref}[1]{m\ref{#1}}

\begin{document}

\title{Relaxation theory for perturbed many-body 
quantum systems \\ versus numerics and experiment}
\author{Lennart Dabelow}
%\email{ldabelow@physik.uni-bielefeld.de}
\author{Peter Reimann}
%\email{reimann@physik.uni-bielefeld.de}
\affiliation{Fakult\"at f\"ur Physik, 
Universit\"at Bielefeld, 
33615 Bielefeld, Germany}
\date{\today}

\begin{abstract}
An analytical prediction is established of 
how an isolated many-body quantum system 
relaxes towards 
its thermal long-time limit
under the action of a time-independent
perturbation, 
but still remaining sufficiently close to a 
reference case whose temporal relaxation is known.
This is achieved within the conceptual framework of a
typicality approach by showing and exploiting
that the time-dependent expectation values 
behave very similarly for most members 
of a suitably chosen ensemble of 
perturbations.
The predictions are validated by comparison
with various numerical and experimental 
results from the literature.
\end{abstract}

\maketitle

%%%%%%%%%%%%%%%%%%%%%%%%%%%%%%%%%%%%%%%%%%%%%%%%%%%%
The long-standing task to 
adequately 
explain 
the time-dependent relaxation and ultimate equilibration of
isolated many-body quantum
systems has recently witnessed a veritable renaissance,
driven, among others, by very impressive new 
experimental and numerical 
capabilities \cite{gog16,dal16,mor18,gem04}.
On the other hand, quantitative analytical 
interpretations
of the so-acquired data
are still rather scarce.
For instance, a paradigmatic setup for probing the 
temporal relaxation behavior
of 
strongly correlated cold atoms 
was originally proposed in the numerical 
works \cite{cra08,fle08} and 
then 
experimentally explored in \cite{tro12}, concluding 
that ``the exact origin of this enhanced relaxation
in the presence of strong correlations constitutes one of
the major open problems posed by the results 
presented here''
\cite{f1}.
The main objective of our Letter
is to better understand several such 
experimental and numerical findings
by considering
model Hamiltonians of the form
\begin{equation}
	H = H_0 + \lambda V
\label{1}
\end{equation}
and asking how the temporal relaxation of the reference system 
$H_0$ is altered by 
a small perturbation $\lambda V$.

%%%%%%%%%%%%%%%%%%%%%%%%%%%%%%%%%%%%%%%%%%%%%%%%%%%%%%%%
{\em Setting --}
Given a (pure or mixed) initial state $\rho(0)$, 
the system (\ref{1}) evolves in time as 
$\rho(t) = \e^{-\I H t} \rho (0) \, \e^{\I H t}$ 
($\hbar = 1$), resulting in expectation 
values $\langle A \rangle_{\!\rho(t)} := \tr\{ \rho(t) A \}$ 
of an observable (self-adjoint operator) $A$.
Of foremost interest to us are the deviations 
of those $\langle A \rangle_{\!\rho(t)}$ from 
the corresponding expectation values 
$\langle A \rangle_{\!\rho_0(t)}$ when the 
same initial state $\rho(0)$ evolves 
according to the unperturbed 
Hamiltonian $H_0$ in (\ref{1}).

Denoting by $E_n$ and $|n\rangle$ the 
eigenvalues and eigenvectors of $H$, 
we assume as usual \cite{gog16,dal16,mor18,gem04}
that only energies $E_{n}$ within 
a macroscopically small but microscopically 
large interval $I := [E, E + \Delta]$ 
entail non-negligible level populations
$\langle n|\rho(0)|n\rangle$.
Put differently, the system must exhibit a 
macroscopically well-defined 
energy, while the number of $E_n\in I$ 
is still exponentially large in the 
degrees of freedom~\cite{lan70,gol10a},
and the mean level spacing $\lvsp$
is approximately constant throughout $I$
\cite{gog16,dal16,mor18,gem04}.
Very low system energies (close to the 
ground state) are incompatible with these 
requirements and are thus tacitly excluded.
Finally, $\lvsp$ is 
assumed to be (practically) independent of 
the perturbation strengths $\lambda$ considered
in (\ref{1}).
Since the level density $1/\lvsp$ is -- 
according to the textbook microcanonical formalism --
directly connected to the system's 
thermodynamics, we thus
concentrate on sufficiently weak perturbations 
in the sense that phase transitions or 
other significant changes of the 
thermal equilibrium properties 
are ruled out.

Next we adopt the common idea of statistical mechanics
that the system's many-body character  
inevitably implies some uncertainties 
about the microscopic details.
Hence, we temporarily consider
an entire 
class
of similar perturbations instead of one 
particular $V$ in (\ref{1}).
More precisely, denoting by $V^0_{\mu\nu} := \braN{\mu} V \ketN{\nu}$
the matrix elements of $V$ in the eigenbasis $\ketN{\nu}$
of $H_0$,
we choose an ensemble of matrices
whose statistics still reflects the essential 
properties of the ``true'' perturbation $V$
in (\ref{1}) as closely as possible.
For instance, if $H_0$ models noninteracting particles 
and $V$ some few-body interactions, the true matrix 
$V^0_{\mu\nu}$ is known to be sparse 
(most entries are zero) 
\cite{bro81, bor16, fyo96, fla97}.
Similarly, the true perturbation may give rise to a 
so-called banded matrix 
\cite{deu91, fyo96, fei89, wig55, gen12}.
Accordingly, 
it is appropriate to work with 
a possibly (but not necessarily)
sparse and/or banded random matrix ensemble.

Denoting 
ensemble averages 
by $\av{\,\cdots}$, the $V^0_{\mu\nu}$'s are
considered as unbiased and --
apart from $V^0_{\nu\mu}=(V^0_{\mu\nu})^\ast$ --
independent random variables,
with second moments
\begin{equation}
	\avv{ \lvert V^0_{\mu\nu} \rvert^2 }
		= \sigma_v^2 \, F(\lvert \mu - \nu\rvert) \,,
\label{7}
\end{equation}
where the ``band profile'' $F(n)$ approaches unity 
for small $n$, and $\sigma_v^2$ 
sets the overall scale.
Moreover, $F(n)$ is usually a slowly varying
function of $n$ and upper bounded by a constant of 
order unity.
In particular, the matrices $V^0_{\mu\nu}$
may still be sparse, while
$B := \sum_{n=1}^\infty F(n)$ 
characterizes the bandwidth 
if the matrix is banded,
and $B=\infty$ otherwise.
A more detailed discussion is 
provided in \cite{suppl}.

%%%%%%%%%%%%%%%%%%%%%%%%%%%%%%%%%%%%%%%%%%%%%%%%%%%%%%%%
{\em Results --}
Our main result is the following 
analytical prediction of how the unperturbed 
behavior $\< A \>_{\!\rho_0(t)}$ is 
modified by a typical perturbation,
i.e., for the vast majority of $V$'s 
from a given ensemble,
\begin{equation}
	\< A \>_{\!\rho(t)}
		= \Ath
		+ \lvert g(t) \rvert^2 \left\{ \< A \>_{\!\rho_0(t)} - \Ath \right\}
			\ .
\label{22}
\end{equation}
Here $\Ath$ is the microcanonical
expectation value corresponding to the energy 
window $I$,
and $g(t)$ depends on the mean level spacing 
$\lvsp$, the coupling $\lambda$, and the 
properties of the random matrix ensemble 
from (\ref{7}).
In particular, for sufficiently weak 
perturbations we find that
\begin{eqnarray}
g(t)=\exp\{-\Gamma t/2\}\ , \ \ 
\Gamma:= 2 \pi \lambda^2 \sigma_v^2/\lvsp
\ ,
\label{s1}
\end{eqnarray}
where ``sufficiently weak'' means
 -- besides the requirements in ``Setting'' --
that  $\Gamma \ll \lvsp B$.
Likewise, we find that
\begin{eqnarray}
g(t)=
2 J_1(\gamma t) / \gamma t\ , \ \ 
\gamma := \sqrt{8 B} \, \lambda \sigma_v
\ ,
\label{s2}
\end{eqnarray}
for $\gamma \gg \lvsp B$ 
(``sufficiently strong'' perturbations),
where $J_\nu(x)$ are Bessel functions 
of the first kind.
In between, $g(t)$ exhibit as transition from (\ref{s1}) to
(\ref{s2}) as detailed in \cite{suppl}.

%%%%%%%%%%%%%%%%%%%%%%%%%%%%%%%%%%%%
\begin{figure}
\includegraphics[scale=1]{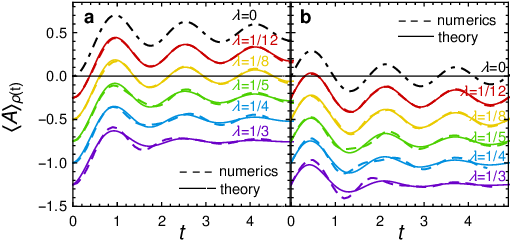}
\caption{
{Bosonic Hubbard chain.}
Dashed: Numerical 
results 
from
\cite{fle08} for the Hubbard model (\ref{23}) with 
$L=32$, $J=1$, and
various $\lambda:=1/U$,
vertically shifted 
in steps of $-0.25$
for better visibility.
The initial state consists of 
singly occupied even and empty odd sites.
The observables are $A:= \hat n_1$ 
in (a) and 
$A:= (\hat b^\dagger_1\hat b_2 - \hat b^\dagger_2\hat b_1)/2\I$
in (b),
implying $\Ath=1/2$ in (a) 
and $\Ath=0$ in (b).
Dash-dotted: Unperturbed analytical solutions 
$\< A \>_{\!\rho_0\!(t)}=[1-J_0(4t)]/2$ in (a) and
$\< A \>_{\!\rho_0\!(t)}=J_1(4t)/2$ in (b) 
\cite{fle08}.
Solid:
Eqs. (\ref{22}), (\ref{s1}) with 
$\Gamma=4.98\,\lambda^2$.
}
\label{fig:ExFlesch}
\end{figure}

Before turning to their derivation, 
we further discuss and exemplify 
those results (\ref{22})--(\ref{s2}):
So far, these are predictions regarding 
the vast majority of a given 
$V$-ensemble.
As usual in 
random matrix theory 
\cite{gol10a,bro81,bor16,fyo96,wig55,deu91},
one thus expects that the ``true''
(non-random) perturbation $V$ in~(\ref{1}) 
belongs to that vast majority, provided its
main properties are well captured by 
the considered ensemble.
Note that the microscopic dynamics
of a many-body system is commonly
expected to be extremely sensitive against 
perturbations (chaotic \cite{haa10}),
so that it is virtually impossible to 
theoretically predict its response 
exactly or in terms of well-controlled 
approximations \cite{kam71}.
Hence, some kind of ``uncontrolled''
approximation is practically unavoidable.
One of them is random matrix theory,
which in fact has been originally 
devised by Wigner 
for the very purpose of exploring  
chaotic quantum many-body systems,
and is 
by now
widely recognized as a remarkably 
effective tool in this context
\cite{haa10}.
Another common justification is
by comparison with particular 
examples, to which we now turn.
These will be state-of-the-art
numerical and experimental results
from previous publications, 
which, however, do not contain the
corresponding (numerical) data for the
variances required in (\ref{7}).
We will thus concentrate 
on the weak perturbations regime, 
where (\ref{s1}) applies.
Again, the quantitative values of
$\sigma_v^2 / \lvsp$ required in~\eqref{s1} 
can in principle be computed, 
but have not been provided 
in those publications, and will 
therefore be treated as fit parameters.

%%%%%%%%%%%%%%%%%%%%%%%%%%%%%%%%%%%%
\begin{figure}
\includegraphics[scale=1]{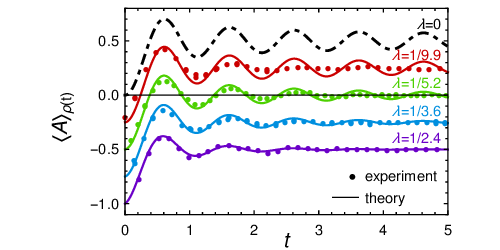}
\caption{
{Cold atom experiments.}
Dots: Experimental data for 
repulsively interacting 
Rb atoms 
in a 
1D
optical superlattice, 
adopted from Fig.~2 in
\cite{tro12}.
Initial condition, observable, and dynamics
experimentally emulate the theoretical 
ones from Fig.~\ref{fig:ExFlesch}(a). 
All further details (dash-dotted and solid 
lines, vertical shifts) are as 
in Fig.~\ref{fig:ExFlesch}.
}
\label{fig:ExTrotzky}
\end{figure}
%%%%%%%%%%%%%%%%%%%%%%%%%%%%%%%%%%%%%

{\em Examples --}
Our first example is the numerical exploration
by Flesch \etal\ \cite{fle08} (see also \cite{cra08})
of the bosonic Hubbard chain
\begin{equation}
	H := -J \sum_{i=1}^L (\hat b^\dagger_{i+1} \hat b_i 
	+ \hat b^\dagger_i \hat b_{i+1}) + \frac{U}{2} \sum_{i=1}^L \hat n_i (\hat n_i - 1)
\label{23}
\end{equation}
with periodic boundary conditions,
creation (annihilation) operators 
$\hat b^\dagger_i$ ($\hat b_i$), 
and $\hat n_i:=\hat b^\dagger_i\hat b_i$.
For the initial state $\rho(0)$
considered in Ref.~\cite{fle08} 
and
Fig.~\ref{fig:ExFlesch},
the model (\ref{23}) can be recast as an effective 
spin-1/2 chain 
by means of a mapping
which becomes asymptotically exact 
for large interaction parameters 
$U$ \cite{giu13}.
In the limit $U\to\infty$, the so-obtained
``unperturbed'' effective Hamiltonian $H_0$ 
amounts to an XX model.
The leading finite-$U$ correction
takes the form $U^{-1}H_1$, where $H_1$
contains nearest neighbor, next-nearest
neighbor, as well as three-spin terms \cite{giu13}.
In other words, $H_1$ plays the role
of the perturbation $V$ in (\ref{1}), 
and $1/U$ that of $\lambda$.
Since $\< A \>_{\!\rho_0\!(t)}$ and $\Ath$ 
are known (see \cite{fle08} and 
Fig.~\ref{fig:ExFlesch}),
the only missing quantity 
is $\sigma_v^2 / \lvsp$ in (\ref{s1}), 
which is, as mentioned above,
not provided by~\cite{fle08}, 
and hence treated as a fit 
parameter, yielding 
$\Gamma=4.98\,\lambda^2$.
The resulting agreement with the numerics 
in Fig.~\ref{fig:ExFlesch} speaks for itself.

An experimental realization of~(\ref{23}) 
by a
strongly correlated Bose gas has been explored
by Trotzky \etal\ in Ref.~\cite{tro12} and is 
compared in Fig.~\ref{fig:ExTrotzky} with 
our theory (\ref{22}), (\ref{s1}).
Adopting $\Gamma = 4.98\,\lambda^2$
from before, this amounts to an entirely 
analytic prediction without any fit 
parameter.
Incidentally, the agreement in Fig.~\ref{fig:ExTrotzky}
improves as $\lambda$ increases.
The same tendency is recovered when 
comparing the numerical results from 
Fig.~\ref{fig:ExFlesch}(a) with the 
experimental data, 
suggesting that the model~(\ref{23})
itself may not capture all experimentally 
relevant details for 
large $U$ (small $\lambda$).

%%%%%%%%%%%%%%%%%%%%%%%%%%%%%%%%%%%%
\begin{figure}
\includegraphics[scale=1]{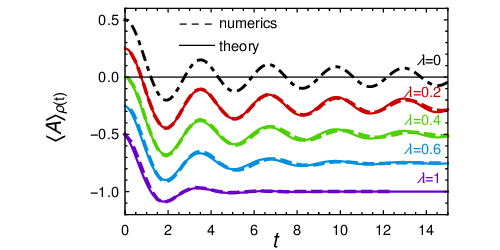}
\caption
{
Spin-1/2 XXZ 
model of the form (\ref{1}) with 
$H_0:= \sum_{i=1}^{L-1} 
(S^x_i S^x_{i+1} + S^y_i S^y_{i+1})$,
$V:=\sum_{i=1}^{L-1} S^z_i S^z_{i+1}$,
observable  $A:= \sum_{i=1}^L (-1)^i S^z_i / L$,
and N\'{e}el initial state
(see  \cite{bar09} for more details).
Dashed: Numerical 
results 
from \cite{bar09}
for different $\lambda$,
vertically shifted in steps of $-0.25$.
Dash-dotted: Unperturbed analytical solution 
$\< A \>_{\!\rho_0\!(t)}=J_0(2t)/2$
\cite{bar09}.
Solid: 
Eqs. (\ref{22}), (\ref{s1}) with $\Ath=0$, 
and $\Gamma=0.46\,\lambda^2$.
}
\label{fig:ExBarmettler}
\end{figure}
%%%%%%%%%%%%%%%%%%%%%%%%%%%%%%%%%%%%

Next we turn to the spin-1/2 
XXZ chain with anisotropy parameter 
$\lambda$ as specified in Fig.~\ref{fig:ExBarmettler},
exhibiting a gapless ``Luttinger liquid''  and a
gapped, Ising-ordered antiferromagnetic 
phase for $\lambda \leq 1$ and
$\lambda > 1$, respectively \cite{bar09}.
Similarly as before,
$\sigma_{\!v}^2/\lvsp$ in~\eqref{s1} 
is unknown and hence treated as a fit 
parameter.
The analytics in Fig.~\ref{fig:ExBarmettler} 
explains the numerics by Barmettler \etal\ 
from Ref. \cite{bar09} remarkably well
all the way up to the critical point at
$\lambda=1$.

\begin{figure}
\includegraphics[scale=1]{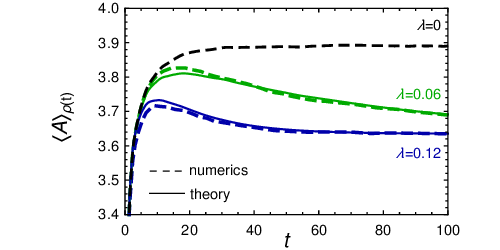}
\caption
{
Hard-core boson model of the form~\eqref{1} with
$H_0 := \sum_i 
%[ -(\hat b_i^\dagger \hat b_{i+1}  
%+ 0.7 \hat b_i^\dagger \hat b_{i+2} + \mathrm{H.c.} 
%+ (\hat n_i - \frac{1}{2}) (\hat n_{i+1} 
%- \frac{1}{2}) + 0.7 (\hat n_i 
%- \frac{1}{2}) (\hat n_{i+2} 
%- \frac{1}{2}) ]$ 
[ -\hat b_i^\dagger(\hat b_{i+1}  + 0.7\hat b_{i+2}) + \mathrm{H.c.} 
+ (\hat n_i - \frac{1}{2}) 
(\hat n_{i+1} + 0.7 \hat n_{i+2} - 0.85)]$ 
and 
$V := \sum_i [\hat b_i 
+ \frac{1}{2}(\hat b_i - b_i^\dagger)\hat b_{i+1}
+ \mathrm{H.c.}]$
(see also (\ref{23}) and \cite{mal19}).
%where $\hat b_i^\dagger$ ($\hat b_i$) creates 
%(annihilates) a hard-core boson at site $i$ 
%in a chain of formally infinite length $L$, 
%and $\hat n_i = \hat b_i^\dagger \hat b_i$.
The initial state is thermal with respect to 
$H_I:=H_0+\frac{1}{2}\sum_i [ \hat b_i^\dagger\hat b_{i+1} + \mathrm{H.c.} 
+ (\hat n_i - \frac{1}{2}) (\hat n_{i+1} - \frac{1}{2})]$,
%with modified coefficients as detailed in \cite{mal19}, 
and the observable is
%$A := \frac{1}{L} \sum_i 
$A := \frac{1}{L} \sum_i 
[\hat b^\dagger_i \hat b_{i+1} 
+ \mathrm{H.c.} ]$.
Dashed: Numerical 
results 
from \cite{mal19} for different $\lambda$.
Solid: Eqs. (\ref{22}), (\ref{s1}) using 
the dashed line for $\lambda=0$
as $\< A \>_{\!\rho_0\!(t)}$,
the numerical values for $\Ath$ 
quoted in \cite{mal19},
and
$\Gamma = 4.2\, \lambda^2$.
}
\label{fig:ExMallayya}
\end{figure}

Our last example in Fig.~\ref{fig:ExMallayya}
is a model of hard-core bosons, 
numerically explored by Mallayya \etal\ 
in Ref.~\cite{mal19},
and exhibiting so-called prethermalization 
\cite{ber04, moe08, mor18}
for small $\lambda$.
Treating $\sigma_v^2 / \lvsp$ again as a fit 
parameter, our theory also explains very well 
such a prethermalization 
scenario.

For lack of space, further examples 
have been moved to the Supplemental Material 
\cite{suppl}, namely the fermionic systems
from Refs. \cite{bal15,ber16} and a case
in which $g(t)$ exhibits a crossover from~(\ref{s1}) to~(\ref{s2})
without any fit parameters.
Moreover,
a scaling behavior $\Gamma\propto\lambda^2$  
as in~\eqref{s1} has also been reported, 
among others, in Refs.\ 
\cite{mor18,igl11,bie17,tan18, mal18}.

%%%%%%%%%%%%%%%%%%%%%%%%%%%%%%%%%%%%
%%%%%%%%%%%%%%%%%%%%%%%%%%%%%%%%%%%%

%%%%%%%%%%%%%%%%%%%%%%%%%%%%%%%%%%%%%%%%%%%%%%%%%%%%%
{\em Derivation --}
Adopting the notation as introduced below (\ref{1}), 
one readily confirms that
\begin{equation}
	\langle A \rangle_{\!\rho(t)}
		= \sum_{m, n} \e^{\I (E_n - E_m) t} \rho_{mn}(0) A_{nm} \,,
\label{2}
\end{equation}
where $\rho_{mn}(t) := \bra{m} \rho(t) \ket{n}$ and $A_{nm} := \bra{n} A \ket{m}$.
Given that in (\ref{2}) only levels 
$E_n\in I$ 
actually matter (see ``Setting''), and that 
their spacings $E_{n+1}-E_n$ generically 
exhibit a Poisson- or Wigner-Dyson-like 
statistics \cite{haa10} of mean value $\lvsp$, 
the random fluctuations of those $E_{n+1}-E_n$
will also be of order $\lvsp$
and (practically) uncorrelated.
Hence, the $E_n-E_m$ exhibit typical 
deviations from their mean value 
$(n-m) \lvsp$ of order $|n-m|^{1/2}\lvsp$.
Recalling that $\lvsp$ is exponentially small \cite{lan70,gol10a}
and $|n-m|\lvsp\simeq |E_n-E_m| \leq\Delta$, 
it is reasonable to expect -- and justified in 
more detail in \cite{suppl} -- that the deviations between 
$(E_n-E_m)t$ and  $(n-m) \lvsp t$ are 
negligible in~\eqref{2} 
for all $t$ of later relevance.
Employing this approximation and the
unitary basis transformation
\begin{equation}
	U_{m\nu} := \langle m | \nu \rangle_{\!0}
\label{3}
\end{equation}
we rewrite~\eqref{2} in terms of the unperturbed 
basis $\ketN\nu$ as
\begin{eqnarray}
	\< A \>_{\!\rho(t)} 
        & = & 
	\sum_{\mu_1, \mu_2, \nu_1, \nu_2} \rho^0_{\mu_1 \nu_2}(0) \, A^0_{\mu_2 \nu_1} 
	W^{\mu_1 \mu_2}_{\nu_1 \nu_2}(t)
	\,,
\label{4}
\\
	W^{\mu_1 \mu_2}_{\nu_1 \nu_2}(t) 
        & := & \sum_{m, n} \e^{\I (n-m) \lvsp t} 
          U_{m \mu_1} U_{n \mu_2} 
          U^*_{m \nu_1} U^*_{n \nu_2} \ ,
\label{6}
\end{eqnarray}
where $\rho^0_{\mu\nu}(0) := \braN{\mu} \rho(0) \ketN{\nu}$, 
and $A^0_{\nu\mu} := \braN{\nu} A \ketN{\mu}$.

The randomness of $V$ (see above (\ref{7}))
is inherited via (\ref{1}) by the eigenvector 
overlaps $U_{m\nu}$ in (\ref{3}).
A particularly important role is played by 
their second moments, $\avv{\lvert U_{m\nu} \vert^2}$,
which due to (\ref{7}) only depend on $m-\nu$,
\begin{equation}
	\avv{\lvert U_{m\nu} \vert^2} =: u(m-\nu) \, .
\label{8}
\end{equation}
Specifically, the function $g(t)$ from (\ref{22})
is defined as their Fourier transform, 
\begin{equation}
	g(t) := \mbox{$\sum_n$} \e^{\I n \lvsp t} \, u(n) \, .
\label{9}
\end{equation}
As a first basic result of our present 
work, we show in \cite{suppl} that
the function $u(n)$ itself follows as 
$\frac{\lvsp}{\pi} \lim_{\eta\to 0+} 
\operatorname{Im} G(n \lvsp - \I \eta)$
from the ensemble-averaged (scalar)
resolvent or Green's function 
$G(z)$ of 
$H$,
which in turn can be obtained as the solution of
\begin{equation}
	G(z) \left[ z - \lambda^2 \sigma_v^2\lvsp^{-1}
\mbox{$\int$}G(z - E) \, F(\lvert E \rvert / \lvsp)\, \d E \right] = 1 \,.
\label{10}
\end{equation}

Along these lines one readily recovers \cite{suppl} 
for weak perturbations as specified below (\ref{s1})
the previously known Breit-Wigner distribution 
\cite{wig55,deu91,fyo96}
\begin{equation}
	u(n) 
        = \frac{1}{2\pi} \frac{\Gamma 
        / \lvsp}{(\Gamma / 2\lvsp)^2 + n^2} \, ,
%\quad\Gamma := \frac{2 \pi \lambda^2 \sigma_v^2}{\lvsp} \ ,
\label{11}
\end{equation}
with $\Gamma$ from (\ref{s1}), and
provided that $\delta := \lvsp / \Gamma\ll 1$. 
Our next goal is to evaluate the ensemble-averaged time evolution~\eqref{4},
hence we need the average of four 
$U_{m\nu}$'s in~\eqref{6}.
The solution of this technically quite 
challenging problem by means of supersymmetry 
methods \cite{efe83,ber10,haa10} 
is provided in the Supplemental 
Material~\cite{suppl}.
There, we also establish that $\delta$ 
is indeed exponentially small in the system's 
degrees of freedom $f$,
\begin{equation}
\label{15}
	\delta \approx \exp\{-\mathcal{O}(f)\} 
\ll 1
\,.
\end{equation}
Introducing those averages of four $U_{m\nu}$'s 
into~\eqref{4} and~\eqref{6} finally yields
\begin{equation}
\label{16}
	\av{ \< A \>_{\!\rho(t)} }
		= \< A \>_{\!\tilde\rho}
			+ \lvert g(t) \rvert^2 \left\{ \< A \>_{\!\rho_0(t)} - \< A \>_{\!\tilde\rho} \right\}
			+ R(t) \ .
\end{equation}
Here, $\tilde\rho$
is defined via
$\braN{\mu} \tilde\rho \ketN{\nu} := 
\delta_{\mu\nu} \sum_\kappa \tilde u(\nu - \kappa) \rho^0_{\kappa\kappa}(0)$ with $\tilde u(n) := \sum_m u(n-m) u(m)$,
essentially amounting
to a  ``washed-out'' descendant of the so-called diagonal 
ensemble (long-time average of $\rho_0(t)$)
\cite{gog16,dal16,mor18}.
Following Deutsch \cite{deu91},
$\< A \>_{\!\tilde\rho}$ is thus 
commonly considered \cite{gog16,rei15,nat18} 
to closely approximate the 
microcanonical expectation value 
$\Ath$ (see below (\ref{22})).
Furthermore,
we can infer from ~\eqref{9}, \eqref{11}, 
and (\ref{15})
that
$g(t) \simeq \! \int \e^{\I n \lvsp t} \, u(n) \d n \! = \e^{-\Gamma \lvert t \rvert / 2}$,
i.e., we recover (\ref{s1}).
Finally, the last term in~\eqref{16} 
is given by
%takes the form
\begin{align}
\label{17}
	R(t) &:= \sum_{\mu,\nu} \rho^0_{\mu\mu}(0)
  \, A^0_{\nu\nu} \, r(\lvert t \rvert, \mu-\nu) \,, \\
\label{18}
	r(t, n) &:= \e^{-\Gamma t} \, \tilde u(n) \left\{ 1 - \cos(t n \lvsp) - \frac{\Gamma \sin(t n \lvsp)}{n \lvsp} \right\} .
\end{align}
Due to similar arguments as in the above approximation 
$\< A \>_{\!\tilde\rho}\simeq\Ath$,
this term is usually negligibly small.
For example, if the unperturbed system 
satisfies the eigenstate thermalization 
hypothesis (ETH), the $A^0_{\nu\nu}$ are well 
approximated by $\Ath$ 
%for all $n$ with $E_n \in I$ 
\cite{deu91,sre94,rig08,dal16}.
Observing that $\sum_n r(t, n)\! \simeq \! \int r(t, n) \d n \! =\! 0$ 
then immediately implies $R(t) \simeq 0$.
However, similar cancellations so that $R(t) \simeq 0$
can be shown to persist under much 
more general conditions \cite{dab20}
(some ETH violating examples are also
provided above and in \cite{suppl}).
Altogether, we thus obtain 
\begin{equation}
\label{19}
	\av{ \< A \>_{\!\rho(t)} }
		= \Ath
			+ \lvert g(t) \rvert^2 \left\{ \< A \>_{\!\rho_0(t)} - \Ath \right\}  ,
\end{equation}
see also Refs.~\cite{mor18,bor16,sre99,bar08,bar09a,gen13,san18,kni18,rei19,nat19}
for somewhat similar findings.

The advantage of (\ref{19}) over (\ref{16}) 
is that $\< A \>_{\!\tilde\rho}$ and $R(t)$ are often 
hard to determine in concrete examples.
For the rest -- 
especially if the above
approximations leading to (\ref{19}) 
might not apply -- one also could continue to 
employ (\ref{16}) 
(for an example, see Sec.~I\,B in \cite{suppl}).
%%%%%%%%%%%%%%%%%%%%%%%%%%%%%%%%%%%%%%%%

So far, we focused on sufficiently weak perturbations
as specified below (\ref{s1}).
Beyond this regime,
the solution $G(z)$ of (\ref{10})
and hence 
$g(t)$ from~\eqref{9}
will be different, yet we still recovered \cite{suppl}
the same final conclusion as in (\ref{19}).
In particular, for ``sufficiently strong''
perturbations as specified below~(\ref{s2})
one finds \cite{wig55,fyo96,suppl}
that $u(n)$ approaches a semicircular 
distribution with radius $\gamma/\lvsp$
and hence one recovers (\ref{s2}),
while in the intermediate regime,
$g(t)$ can still be readily
obtained by solving (\ref{10}) numerically.

Our final objective is to quantify the fluctuations
$\xi_V(t) := \langle A \rangle_{\!\rho(t)} - \avv{ \langle A \rangle_{\!\rho(t)} }$ 
about the average behavior 
in terms of the variance $\av{ \xi_V^2(t) }$.
In view of ~\eqref{4} and~\eqref{6}, 
averages over eight 
$U_{m\nu}$'s are thus required.
Referring to \cite{suppl} for their quite tedious evaluation, 
we finally obtain 
the estimate
\begin{equation}
\label{20}
	\av{ \xi_V^2(t) } \leq c \, \delta \, \norm{A}^2 \ ,
\end{equation}
where 
$\norm{A}$ indicates the operator norm
and $c$ is some positive real number, which does 
not depend on any further details of the considered system
and is at most on the order of $10^3$.
Exploiting Chebyshev's inequality, the probability that 
$\lvert \xi_V(t) \rvert \leq \delta^{1/3}\norm{A}$
when randomly 
sampling $V$'s from the ensemble 
can thus be lower bounded by 
$1 - \delta^{1/3} c$.
In view of (\ref{15}),
the deviations from the average behavior
(\ref{19}) are thus negligibly small for 
most $V$'s
and any preset $t$.
Moreover, one can show similarly as in \cite{rei16}
that for most $V$'s the deviations $\xi_V(t)$ 
must be negligibly small not only for any given 
$t$, but even for the vast majority of all $t$
within any given time interval $[t_1,t_2]$.
Overall, we thus recover our main result
from (\ref{22}).

%%%%%%%%%%%%%%%%%%%%%%%%%%%%%%%%%%%%%%%%%%%%%%%%%%%%%%%%
{\em Conclusions --} We derived an analytical prediction 
for the ensemble-averaged, time-dependent deviations 
of the perturbed from the unperturbed expectation 
values in isolated many-body quantum systems.
Moreover, we showed that nearly all members of 
the ensemble behave very similarly to the average.
Provided the ensemble has been chosen 
appropriately, the same behavior is thus typically 
expected to also apply when dealing with a specific 
physical model, resulting in our main
analytical prediction (\ref{22}).
As a validation, we demonstrated good agreement
with a variety of numerical and experimental
findings from the literature.
Technically speaking, substantial extensions 
of previously established, non-perturbative
supersymmetry methods were indispensable
to arrive at those results \cite{suppl}.
Related analytical \cite{fyo96,fyo95} and 
numerical \cite{rei19} studies 
suggest that 
such methods may in the future even be further 
extended to considerably more general 
ensembles than those we admitted here.
On the conceptual side, a better understanding
of when a given physical system is {\em not}
a typical member of any 
permitted
ensemble \cite{kni18,rei19a,ham18}
remains as yet another important issue 
for further studies.

%%%%%%%%%%%%%%%%%%%%%%%%%%%%%%%%%%%%%%%%%%%%%%%%%%%%%%
{\em Acknowledgments --}
Enlightening discussions with Jochen Gemmer
and Charlie Nation are gratefully acknowledged.
This work was supported by the 
Deutsche Forschungsgemeinschaft (DFG)
within the Research Unit FOR 2692
under Grant No. 397303734.
%%%%%%%%%%%%%%%%%%%%%%%%%%%%%%%%%%%%%%%%%%%%%%%%%%%%%%%%

\newpage

\clearpage
\newpage

\onecolumngrid
\begin{center}
{\bf\large SUPPLEMENTAL MATERIAL}
\\[1cm]
\end{center}
\twocolumngrid
\setcounter{equation}{0}
\setcounter{figure}{0}

Throughout this Supplemental Material, equations and figures
from the main paper are indicated by an extra letter ``m''. 
For example, ``Eq.~(m1)'' refers to Eq.~(1) in the main paper.

Sec.~\ref{sec:FurtherExamples}
provides
further examples for the applicability of 
our theoretical prediction~(\mref{22}).
In Sec.~\ref{supplSec1}, the arguments below~(\mref{2})
are elaborated in more detail.
Sec.~\ref{sec:RandMatEns} contains a detailed definition of the 
considered ensembles of perturbations 
along with some of their basic properties.
In the remaining Secs.~\ref{sec:OverlapsFromSupersymmetry}--\ref{s5}, 
we present the derivations of the various moments of eigenvector 
overlaps $U_{m\nu}:=\langle m | \nu \rangle_{\!0}$
underlying our
two key results~(\mref{10}) and~(\mref{20}):
the supersymmetry calculation for the second and fourth moments
in Sec.~\ref{sec:OverlapsFromSupersymmetry} 
[in particular~(\mref{10}) and (\mref{11})],
a generalized approximation for the fourth 
and higher such moments in Sec.~\ref{s4},
and the evaluation of the eighth moments
in Sec.~\ref{s5} [in particular~(\mref{20})].

%%%%%%%%%%%%%%%%%%%%%%%%%%%%%%%%%%%%%%%%%%%%%%%%%%%%
\section{Further examples}
\label{sec:FurtherExamples}
In this section, we compare our main result~(\mref{22}),
\begin{equation}
\label{suppl22}
	\< A \>_{\!\rho(t)}
		= \Ath
			+ \lvert g(t) \rvert^2 \left\{ \< A \>_{\!\rho_0(t)} - \Ath \right\} ,
\end{equation}
with
additional numerical examples, namely
two fermionic systems in 
Secs.~\ref{sec:FurtherExamples1} and
\ref{sec:FurtherExamples2},
and a sparse and banded random matrix model in 
Sec.~\ref{sec:FurtherExamples:RandomMatrixModel}.

As already pointed out in the main paper,
for the numerical results from the literature,
with which we compare our analytical prediction
(\ref{suppl22}), the necessary details to determine 
the function $g(t)$ can in principle 
be calculated numerically, but 
they unfortunately have not been provided
in the corresponding published works.
Therefore, we employ the approximation
$|g(t)|^2=\e^{-\Gamma t}$, which is 
theoretically predicted to apply, at least,
whenever the perturbation is sufficiently 
weak (see below~(\mref{s1})).
In the same vein, the quantitative value of
the ratio $\sigma_v^2/\lvsp$ in our theoretical 
prediction $\Gamma=2\pi\lambda^2\sigma_v^2/\lvsp$
from~(\mref{s1}) has not been published in 
the pertinent numerical works and is 
therefore treated as a fit parameter.

While the above considerations apply to 
all the examples from the main paper and also 
those in Secs.~\ref{sec:FurtherExamples1} and
\ref{sec:FurtherExamples2} below,
for the last example in Sec.~\ref{sec:FurtherExamples:RandomMatrixModel}
all relevant quantities to determine
$g(t)$ in (\ref{suppl22}) will be 
exactly known.

\subsection{Fermionic Hubbard model}
\label{sec:FurtherExamples1}

\begin{figure}
\includegraphics[scale=1]{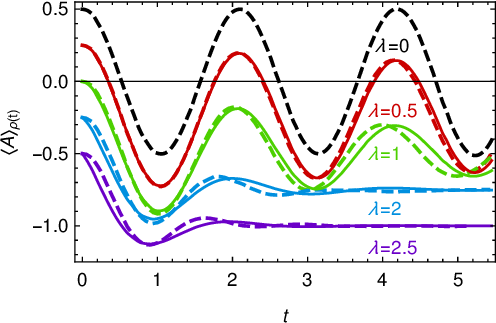}
\caption
{
{Fermionic Hubbard model.}
The Hamiltonian $H = H_0 + \lambda V$ and the 
observable $A$ are specified 
by~\eqref{eq:H:bal15} and~\eqref{eq:A:bal15}.
As detailed in~\cite{bal15},
the initial state consists of one fermion per 
site with opposing spins between connected sites.
Dashed: Dynamical mean-field theory results, 
adopted from Fig.~3(a) in~\cite{bal15} for 
$\lambda = 0, 0.5, 1, 2, 2.5$, and vertically 
shifted in steps of $-0.25$.
Solid: Corresponding theoretical 
prediction~\eqref{suppl22} with $\<A\>_{\mathrm{mc}} = 0$, 
$\lvert g(t) \rvert^2 = \e^{-\Gamma t}$, 
$\Gamma = 0.22\, \lambda^2$,
and using the numerical results for 
$\lambda=0$ 
as $\< A \>_{\!\rho_0\!(t)}$.}
\label{fig:ExBalzer}
\end{figure}

We consider the fermionic Hubbard model as studied 
by Balzer \etal\ in Ref.~\cite{bal15}, 
defined by the Hamiltonian $H = H_0 + \lambda V$ with
\begin{equation}
\label{eq:H:bal15}
	H_0 := -\sum_{\langle ij \rangle,\sigma} 
\hat c^\dagger_{i\sigma} \hat c_{j\sigma}, \quad
%\label{eq:V:bal15}
	V := \sum_{i} (\hat n_{i\sUp} - \tfrac{1}{2}) 
(\hat n_{i\sDn} - \tfrac{1}{2}),
\end{equation}
where $\hat c^\dagger_{i\sigma}$ 
($\hat c_{i\sigma}$) creates (annihilates) 
a fermion with spin $\sigma \in \{ \sUp, \sDn \}$ 
at site $i$, and 
$\hat n_{i\sigma} := \hat c^\dagger_{i\sigma} \hat c_{i\sigma}$.
The notation $\sum_{\langle ij \rangle}$ indicates the 
sum over all pairs of connected sites in a Bethe lattice 
of infinite coordination number.
This model was studied in~\cite{bal15} using dynamical 
mean-field theory.
The observable
\begin{equation}
\label{eq:A:bal15}
	A := \tfrac{1}{2} \langle \hat c^\dagger_k \hat c_{\bar k} + \hat c^\dagger_{\bar k} \hat c_k \rangle
\end{equation}
measures the correlation between conjugated momentum modes $k$ and $\bar k$ (see \cite{bal15} for details).

Similarly as for the examples presented in the 
main paper, we compare the numerics from~\cite{bal15} 
to our theoretical prediction~\eqref{suppl22} with 
$\lvert g(t) \rvert^2 = \e^{-\Gamma t}$.
Moreover, since the proportionality constant 
$\sigma_v^2/\lvsp$ between $\Gamma$ and 
$\lambda^2$ [cf.~Eq.~(\mref{s1})] is not 
available from the original publication 
\cite{bal15}, it is again treated as 
a fit parameter.
As Fig.~\ref{fig:ExBalzer} shows,
we find good agreement between 
numerics and theory up to quite large 
$\lambda$ values.
The remaining deviations for large $\lambda$ 
and small $t$ could hint at a banded structure of the 
perturbation matrix since the behavior in this regime 
is seemingly better described by a Bessel-like decay 
[see~(\mref{s2}) and the discussion between~(\mref{19}) and~(\mref{20}) in the main 
paper as well as the subsequent 
Sec.~\ref{sec:FurtherExamples:RandomMatrixModel} 
of this Supplemental Material].

\subsection{Spinless fermions}
\label{sec:FurtherExamples2}

\begin{figure}
\includegraphics[scale=1]{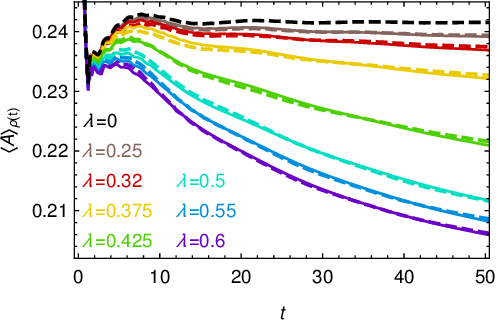}
\caption
{
{Spinless fermions.}
The model is described by the Hamiltonian~\eqref{eq:H:ber16}, 
the observable is the correlator~\eqref{eq:A:ber16}, 
and the initial state 
$\rho(0) \propto \exp[ 2 \sum_i (\hat c_i^\dagger \hat c_{i+1} 
+ \hat c_{i+1}^\dagger \hat c_{i}) ]$.
Dashed: Numerical equations-of-motion results
for various $\lambda$-values,
adopted from Fig.~3 in Ref.~\cite{ber16}.
Solid: Corresponding theoretical predictions,
as detailed in the main text.
}
\label{fig:ExBertini}
\end{figure}

Next, we turn to a model of spinless fermions 
in a periodic one-dimensional lattice, whose 
Hamiltonian $H = H_0 + \lambda V$ consists of
\begin{subequations}
\label{eq:H:ber16}
\begin{align}
\label{eq:H0:ber16}
	H_0 &:= -\sum_{i=1}^L \left\{
\!\left[ 1 + (-1)^i \delta \right] \! \left( \hat c_i^\dagger \hat c_{i+1} + \hat c_{i+1}^\dagger \hat c_{i} \right) 
%	\right.\notag \\ & \qquad\qquad\left. \vphantom{[ (-1)^i c_{i+1}^\dagger}
		+ U \hat n_i \hat n_{i+1} \!\right\}\!, \\
\label{eq:V:ber16}
	V & := -\sum_{i=1}^L \left( \hat c_i^\dagger \hat c_{i+2} + \hat c_{i+2}^\dagger \hat c_{i} \right)   .
\end{align}
\end{subequations}
Here, $c_i^\dagger$ ($c_i$) creates (annihilates) a spinless 
fermion at site $i$, and $\hat n_i := \hat c_i^\dagger \hat c_i$.
Bertini \etal\ \cite{ber16} investigated the effect of the 
next-nearest neighbor tunneling $V$ on the dynamics by means of 
equations-of-motion techniques for a lattice of $L = 320$ sites.
The observable is a tunneling correlation
\begin{equation}
\label{eq:A:ber16}
	A := \frac{1}{2} \left( \hat c_{L/2}^\dagger \hat c_{L/2-1} + \hat c_{L/2 - 1}^\dagger \hat c_{L/2} \right)
\end{equation}
and the system is prepared in a thermal state with 
respect to $H_0$ from~\eqref{eq:H0:ber16} with 
dimerization parameter $\delta = 0$ and nearest-neighbor 
interaction $U = 0$ at inverse temperature $2$.
Quenching to $\delta = 0.1$ and $U = 0.4$, the 
numerical result from~\cite{ber16} for the 
unperturbed dynamics ($\lambda =0$) is shown as 
a dashed black curve in Fig.~\ref{fig:ExBertini}.
Since the system is nearly integrable for
$\lambda=0$, this dashed black curve settles 
down to a nonthermal prethermalization plateau 
within the time scales shown \cite{ber16}.
Upon activating and increasing $\lambda$, 
the system departs farther and farther
from integrability, and the time-dependent 
expectation values approach the 
thermal value better 
and better for large times.

Yet, as already observed in Ref.~\cite{ber16}, 
the numerically approached long-time limit 
still appears to notably deviate from the thermal 
value for small $\lambda$, presumably 
due to the finite system size in combination 
with the still rather mild integrability breaking, 
and possibly also due to non-negligible
numerical inaccuracies at large times.
For this reason, instead of comparing 
the numerics from \cite{ber16} with the analytic 
approximation from~\eqref{suppl22}, we rather 
employed -- as announced in the main paper 
below Eq.~(\mref{19}) -- our more general 
original result from (\mref{16}).
In doing so, we utilized, as before, the 
approximations $R(t)=0$ and
$\lvert g(t) \rvert^2 = \e^{-\Gamma t}$,
and we fitted the ratio $\sigma_v^2 / \lvsp$, 
yielding $\Gamma = 0.098\, \lambda^2$.
Moreover, we again used for
the unperturbed behavior $\< A \>_{\!\rho_0\!(t)}$
in (\mref{16}) the numerical results
for $\lambda=0$ (dashed black line in Fig.~\ref{fig:ExBertini}).
But unlike before, we now also treated 
the long-time limit $\<A\>_{\!\tilde\rho}$ 
in Eq.~(\mref{16}) as a fit parameter.
The resulting agreement between numerics
and analytics in Fig.~\ref{fig:ExBertini}
is very good for all numerically available
$\lambda$-values.
The actually employed values of $\<A\>_{\!\tilde\rho}$ were 
$0.232,\, 0.230,\, 0.223,\, 0.207,\, 0.199,\, 0.199,\, 0.199$
for $\lambda=0.25,\, 0.32,\, 0.375,\, 0.425, \,0.5,\, 0.55,\, 0.6$,
respectively.
In particular, with increasing $\lambda$ the long-time
limit indeed seems to converge towards some constant
(presumably thermal) value.

\subsection{Sparse and banded random matrix model}
\label{sec:FurtherExamples:RandomMatrixModel}

%%%%%%%%%%%%%%%%%%%%%%%%%%%%%%%%%%%%
\begin{figure*}
\includegraphics[scale=1]{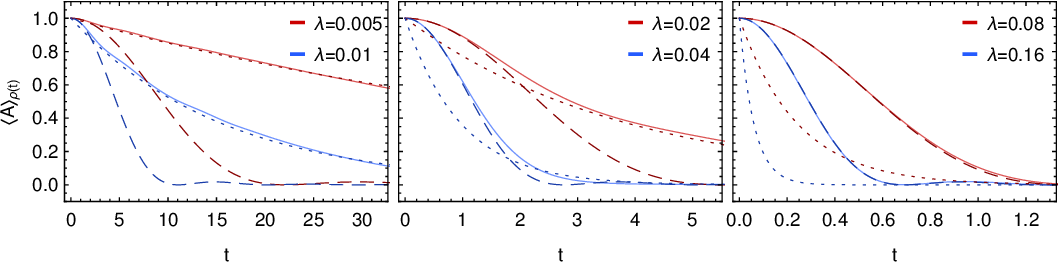}
\caption{{Sparse and banded random matrix model.}
The unperturbed Hamiltonian $H_0$ has eigenvalues $E^0_\mu = \lvsp \mu$ with $\lvsp = 1/512$ and $\mu = 1, \ldots, 16384$.
The perturbation $V^0_{\mu\nu}$ is a random matrix with sparsity $p = 0.8$ and Gaussian distributed nonvanishing entries with mean zero and variance $\av{ \lvert V^0_{\mu\nu} \rvert^2 } = \sigma_v^2 F(\lvert \mu-\nu \rvert)$, where $\sigma_v^2 = (1-p) = 0.2$ and 
$F(n) = \Theta(B - n)$
with $B = 750$.
The initial state and observable are $\rho(0) = A = \ketN{\mu} \braN{\mu}$ with $\mu = 8192$.
Accounting for the different time scales, the dynamics for smaller and larger
$\lambda$ values are presented in 
separate plots for better visibility.
Solid: Numerically determined $\langle A \rangle_{\!\rho(t)}$.
Dotted: Analytical prediction 
$\< A \>_{\!\rho(t)}=\lvert g(t) \rvert^2$ for the exponential decay~\eqref{eq:g:exp},
following from~\eqref{suppl22} by exploiting that
$\< A \>_{\!\rho_0\!(t)}=1$ and
$\Ath=1/N\simeq 0$.
Dashed:
Analytical prediction 
$\< A \>_{\!\rho(t)}=|g(t)|^2$
for the Bessel-like decay~\eqref{eq:g:Bessel}.
As theoretically predicted,
the numerically exact results 
(solid lines)
are well approximated
by the dotted lines for 
$\lambda\ll\lambda_c$
and by the dashed lines for 
$\lambda\gg\lambda_c$,
where the crossover value
$\lambda_c\simeq 0.05$
follows from
the condition $\Gamma\simeq\gamma$.
}
\label{fig:ExRandMatBanded}
\end{figure*}
%%%%%%%%%%%%%%%%%%%%%%%%%%%%%%%%%%%%

In order to have detailed control of all relevant parameters 
and to obviate any sort of fitting, we finally 
consider a model of sparse and banded 
random matrices, which constitutes a direct 
implementation of the 
%typicality 
approach from the main paper
(see also Sec.~\ref{sec:RandMatEns:Definition} below):
The unperturbed Hamiltonian $H_0$ consists of equally spaced 
eigenvalues $E^0_\mu = \lvsp \mu$ with level spacing $\lvsp = 1/512$.
The matrix elements $V^0_{\mu\nu} := \braN{\mu} V \ketN{\nu}$ of the 
perturbation $V$ in the eigenbasis of $H_0$ are independent random 
variables (apart from $(V^0_{\mu\nu})^* = V^0_{\nu\mu}$), distributed as follows:
With probability $p = 0.8$ we set $V^0_{\mu\nu} = 0$, so that the matrix 
is sparse with about $80\,\%$ of the matrix elements vanishing.
The nonzero elements, in turn, have Gaussian distributed real 
and imaginary parts with mean zero and variance 
$F(\lvert \mu - \nu \rvert)/2$ if $\mu \neq \nu$, while those 
with $\mu = \nu$ are purely real and of unit variance.
For the band profile $F(n)$, we choose a sharp cutoff 
at $n=B$, i.e.,
$F(n) := \Theta(B - n)$
with the Heaviside step function $\Theta(x)$.
Altogether, we thus obtain the first two moments $\av{ V^0_{\mu\nu} } = 0$ and
$\av{ \lvert V^0_{\mu\nu} \rvert^2 } = \sigma_v^2 \, F(\lvert \mu - \nu \rvert)$
with $\sigma_v^2 = 1-p = 0.2$, corresponding to~(\mref{7}) in the main paper.

In the numerical example presented in Fig.~\ref{fig:ExRandMatBanded}, 
the total Hilbert space dimension is $N = 2^{14} = 16384$.
For the initial state, we choose the eigenstate $\rho(0) := \ketN{\mu} \braN{\mu}$
of the unperturbed system with $\mu = N/2 = 8192$, i.e., from the middle 
of the spectrum.
As the observable, we choose $A = \rho(0)$, too, measuring 
the survival probability of the initially populated energy 
level $\ketN{\mu}$.
Consequently, the unperturbed time evolution is constant, 
$\langle A \rangle_{\!\rho_0(t)} = 1$, while the thermal 
expectation value is $\Ath = 1/N \simeq 0$.
The theory~\eqref{suppl22} thus predicts $\langle A \rangle_{\!\rho(t)} = \lvert g(t) \rvert^2$ 
for the perturbed dynamics, where,
as in~(\mref{9}),
\begin{equation}
\label{eq:g}
	g(t) := \sum_n \e^{\I n \lvsp t} \, u(n) 
\end{equation}
is the Fourier transform of 
$u(n)$ from~(\mref{8}).
As stated in the main paper and derived in 
Sec.~\ref{sec:OverlapsFromSupersymmetry:2ndMoment} below, 
the function $u(n)$ exhibits a crossover
from the Breit-Wigner form~(\mref{11}) for weak perturbations 
in comparison with the energy scale of the perturbation 
bandwidth $\lvsp B$,
\begin{equation}
\label{eq:u:BW}
	u(n) = \frac{1}{2\pi} \frac{\Gamma / \lvsp}{ (\Gamma / 2\lvsp)^2 + n^2}, \quad \Gamma := \frac{2\pi\lambda^2 \sigma_v^2}{\lvsp} \,,
\end{equation}
to the semicircular shape 
[cf.~above~(\mref{20})]
for stronger perturbations,
\begin{equation}
\label{eq:u:SC}
%	u(n) = \frac{2 \lvsp^2}{\pi \gamma^2} \sqrt{ \gamma^2 / \lvsp^2 - n^2} \, \Theta(\gamma^2 / \lvsp^2 - n^2) \,,
	u(n) = \frac{2 \lvsp^2}{\pi \gamma^2} \sqrt{ \frac{\gamma^2 }{\lvsp^2} \!-\! n^2} \, 
	\Theta\!\left(\!\frac{\gamma^2 }{\lvsp^2} - n^2\!\right) ,\,
	\gamma := \sqrt{8 B} \, \lambda \sigma_v \,.
\end{equation}
For the Fourier transform $g(t)$ 
from (\ref{eq:g}),
this implies a crossover from an exponential decay with rate $\Gamma/2$,
\begin{equation}
\label{eq:g:exp}
	g(t) = \e^{-\Gamma \lvert t \rvert / 2} \,,
\end{equation}
to a Bessel-like decay with rate $\gamma$,
\begin{equation}
\label{eq:g:Bessel}
	g(t) = 2 J_1(\gamma t)/\gamma t \, ,
\end{equation}
where $J_1(x)$ is a Bessel function of the first kind.

Choosing a fixed bandwidth $B = 750$, we compare these two 
theoretical predictions with 
numerical results
for a single realization 
$V$ of the 
above specified
random matrix ensemble in Fig.~\ref{fig:ExRandMatBanded}.
We find excellent agreement between the 
numerics and the exponential decay~\eqref{eq:g:exp} for 
sufficiently small
values of $\lambda$, and equally splendid agreement between 
the numerics and the Bessel-like decay~\eqref{eq:g:Bessel} 
for 
sufficiently large
values of $\lambda$.
Moreover, the crossover between the two regimes around 
$\Gamma \simeq \gamma$ is nicely illustrated.
Finally, we observe
that upon increasing $\lambda$,
deviations from the exponential behavior 
become pronounced at short times first.
This can be traced back to the tails of $u(n)$, 
which begin to decay faster than 
in the Breit-Wigner distribution (\ref{eq:u:BW})
when $n$ approaches the bandwidth $B$
\cite{fyo96, wig55}, 
i.e., for large $n$, and hence the deviations of the Fourier transform
(\ref{eq:g}) from the behavior in (\ref{eq:g:exp}) 
will be most pronounced for small $t$.

Remarkably enough,
the deviations for small $t$
appearing at larger perturbations strengths $\lambda$
in the above Fig.~\ref{fig:ExBalzer} as well as 
(less pronounced)
in Figs.~\mref{fig:ExFlesch}, \mref{fig:ExBarmettler}, 
and~\mref{fig:ExMallayya} of the main paper
seem to be of somewhat similar character as those in Fig.~\ref{fig:ExRandMatBanded}, 
suggesting that they might be caused by a finite bandwidth of the corresponding 
perturbation matrix $V^0_{\mu\nu}$.

%%%%%%%%%%%%%%%%%%%%%%%%%%%%%%%%%%%%%%%%%%%%%%%%%%%%
\section{Justification of Eq. ($\mbox{m}$10)}
\label{supplSec1}
In this section, additional details regarding
the line of reasoning below Eq.~(\mref{2}) and 
thus the justification for approximating $(E_n - E_m) t$ in~(\mref{2}) by $(n-m)\varepsilon t$, leading to Eq.~(\mref{6}),
are provided.

Our starting point is Eq.~(\mref{2}), i.e.,
\begin{equation}
	\langle A \rangle_{\!\rho(t)}
		= \sum_{m, n} \e^{\I (E_n - E_m) t} 
                  \rho_{mn}(0) A_{nm} \, .
\label{suppl2}
\end{equation}
As said below~(\mref{1}), we assume that only 
energies $E_{n}$ within some macroscopically 
small but microscopically large interval 
$I := [E, E + \Delta]$ exhibit non-negligible
populations $\rho_{nn}(0)$ of the corresponding
energy eigenstates $|n\rangle$.
Accordingly, we approximate all 
$\rho_{nn}(0)$ with $E_n\not\in I$ by zero,
and by exploiting the Cauchy-Schwarz inequality 
$|\rho_{mn}(0)|^2\leq \rho_{mm}(0) \rho_{nn}(0)$, 
we can conclude that only summands with 
$E_m,E_n\in I$ actually contribute 
in (\ref{suppl2}). 

Temporarily, we thus restrict ourselves 
to labels $m$ and $n$ with $E_m,E_n\in I$.
Moreover, we assume without loss of generality
that the $E_n$'s are ordered by magnitude.

Again as said below (\mref{1}), we furthermore focus on
intervals $I$ so that the (locally averaged) level 
spacings $E_{n+1}-E_n$ are approximately constant 
throughout $I$, and we denote this constant by
$\lvsp$.

Our next assumption is that 
the level spacings $E_{n+1}-E_n$ 
exhibit some well-defined statistical 
distribution, for instance a Poisson 
or a Wigner-Dyson statistics,
or some other (intermediate) statistics 
with qualitatively similar features.
It is well-established \cite{haa10,bro81} 
that this assumption can be taken for 
granted for practically any Hamiltonian $H$.
We furthermore assume that the level spacings
are statistically independent of each other.
Again, it is well-known \cite{haa10,bro81}
that this assumption is generically satisfied
in very good approximation, although neighboring
level spacings strictly speaking still exhibit
some very weak correlations, which further 
decrease very rapidly with increasing 
distance between the level pairs.
Such correlations could still be included in
the subsequent considerations without changing
the final conclusions, but for the sake of
simplicity we henceforth disregard them.

In other words, the level spacings
$E_{n+1}-E_n$ are considered as 
independent, identically distributed 
random variables with mean value $\lvsp$.
With respect to their variance, one
readily sees that, independently of 
whether we are dealing with a Poisson, 
a Wigner-Dyson, or some other (reasonable) 
statistics, the variance is always
on the order of
$\lvsp^2$.
Since the $E_{n+1}-E_n$ are independent and
identically distributed, it follows that
$E_n-E_{n_0}$ is a random variable
with mean value $(n-n_0)\lvsp$ and
variance of order 
$|n-n_0|\lvsp^2$ for an arbitrary but fixed
``reference index'' $n_0$ 
(with $E_{n_0}\in I$).
It follows that
$E_n+c+\epsilon_n=n\lvsp$, where
$c:=n_0\lvsp-E_{n_0}$ is an 
$n$-independent constant and 
$\epsilon_n:=E_{n_0}-E_n+(n-n_0)\lvsp$
is a random variable of mean zero and 
variance on the order of 
$|n-n_0|\lvsp^2$.
Furthermore,
since the expectation value in (\ref{suppl2})
remains unchanged if all energies 
$E_n$ are shifted by the same constant 
value, we can assume without loss of 
generality that $c=0$ and thus 
$E_n+\epsilon_n=n\lvsp$.

Next we exploit the following
rigorous result, derived in 
Appendix B of Ref.~\cite{rei19a}:
If each $E_n$ is changed
into $E_n+\epsilon_n$ then the 
corresponding change of the expectation 
value in~(\ref{suppl2}) is upper bounded 
(in modulus) by 
$\norm{A}\,|t|\max|\epsilon_n|$,
where $\norm{A}$ is the operator norm 
of $A$.
Combining this bound with the considerations in
the previous paragraphs, one sees that
$\max|\epsilon_n|$ will be of the same order as 
$\max |n-n_0|^{1/2}\lvsp$ and that
$\max |n-n_0|\lvsp$ can be upper bounded in
very good approximation by $\max|E_n-E_{n_0}|$
and thus by the width $\Delta$ of the 
interval $I$.
It follows that $\max|\epsilon_n|$ 
is (approximately) upper bounded by
$\sqrt{\lvsp\Delta}$.

Altogether, $(E_n - E_m)t$ can thus be 
replaced by $(n-m)\lvsp t$ without 
any significant change of the 
expectation value in (\ref{suppl2})
as long as $t\sqrt{\lvsp\Delta}\ll 1$.
At this point, abandoning our temporary
restriction to $E_m,E_n\in I$ is trivial.
Finally, since $\lvsp$ is exponentially small
in the systems's degrees of freedom
(see main paper), the latter condition 
is fulfilled for all $t$ of actual
interest later on, i.e., before the 
expectation value (\ref{suppl2}) very closely 
approaches its equilibrium 
(long-time average) value.

%%%%%%%%%%%%%%%%%%%%%%%%%%%%%%%%%%%%%%%%%%%%%%%%%%%%
%
\section{Perturbation ensembles}
\label{sec:RandMatEns}
In this section, we
specify and discuss in detail 
the ensembles of perturbation matrices
considered in the main paper and 
in the following sections.
We recall that we are dealing with 
isolated quantum many-body systems 
whose Hamiltonian takes the form
\begin{equation}
\label{eq:H}
	H = H_0 + \lambda V
\end{equation}
with a fixed unperturbed Hamiltonian $H_0$ and a random
perturbation $V$.
For the latter, we will define in Sec.~\ref{sec:RandMatEns:Definition}
an ensemble of operators in terms of their matrix elements
in the basis of $H_0$,
\begin{equation}
\label{eq:V}
	V^0_{\mu\nu} := \braN{\mu} V \ketN{\nu}
	\ ,
\end{equation}
whose distribution is motivated by the generic form of ``real'' perturbations in physical systems,
as well as by certain invariance properties of such systems  
(Sec.~\ref{sec:RandMatEns:InvarianceProperty}).
Finally, we will explain in Sec.~\ref{sec:RandMatEns:SmallParameter} 
that the parameter
\begin{equation}
\label{eq:delta}
	\delta := \lvsp / \Gamma
	\ ,
\end{equation}
as already introduced below 
Eq.~(\mref{11})
in the main paper, is generically exponentially small in the system's 
degrees of freedom [see also Eq. (\mref{15})], 
a crucial prerequisite for the applicability of our 
theory to many-body systems.

%%%%%%%%%%%%%%%%%%%%%%%%%%%%%%%%%%%%%%%%%%%%%%%%%%%%
%
\subsection{Definition}
\label{sec:RandMatEns:Definition}
We consider statistical ensembles 
of perturbations $V$,
where each member of the ensemble
is given by a formally
infinite-dimensional random matrix~\eqref{eq:V}.
The distribution of the matrix elements $V^0_{\mu\nu}$ should be chosen such that the main properties of the ``true'' perturbation are emulated reasonably well.
As explained in the main paper, the distribution 
should allow for potentially (but not necessarily) sparse and/or banded matrices 
with real- or complex-valued entries \cite{bro81, bor16, fyo96, fla97, deu91, fei89, wig55, gen12}.
As already emphasized, e.g., in Ref.~\cite{fyo96},
this random matrix approach should be contrasted 
with the much more common modeling of a Hamiltonian 
such as $H$ in~\eqref{eq:H} in terms of a Gaussian 
orthogonal or unitary ensemble 
(GOE or GUE \cite{haa10, bro81}), which would 
be entirely inappropriate in the present 
context.

From a technical point of view, the central 
objects are the overlaps
\begin{equation}
\label{eq:U}
	U_{m\nu} := \langle m | \nu \rangle_{\!0}
\end{equation}
[see~(\mref{3})] between the 
eigenvectors
of $H_0$ and $H$.
More precisely, the essential quantities 
underlying our central
results~(\mref{16}) and~(\mref{20}) are averages over four and eight factors of such $U_{m\nu}$, respectively, which in turn build on the average of two such overlaps.
This raises the question which of the above exemplified specific features of the $V$ ensemble are relevant with respect to those averages over several $U$ matrix elements.

The simplest connection between the $V$'s and $U$'s is obtained by treating $H$ in~\eqref{eq:H} in terms of elementary (Rayleigh-Schr\"odinger) perturbation theory.
Quantitatively, such an approach may only lead to useful analytical approximations for exceedingly small $\lambda$ values due to the extremely dense energy eigenvalues of $H_0$ (the level spacing decreases exponentially with the system's degrees of freedom \cite{gol10a, lan70}) and the concomitant small denominators.
Hence, nonperturbative methods will be indispensable for our present purposes.
However, qualitatively it is quite plausible on the basis of such a perturbative approach that beyond those extremely small $\lambda$ values, a large number of $V^0_{\mu\nu}$'s will appreciably contribute to any given $U_{m\nu}$ or products thereof.
It is thus reasonable to expect that some generalized kind of central limit theorem (CLT) may apply, so that only a few basic statistical properties of the $V$'s will be actually relevant for the $U$'s.
We point out that all these abstract considerations will become concrete in Sec.~\ref{sec:OverlapsFromSupersymmetry}, in particular Sec.~\ref{sec:OverlapsFromSupersymmetry:2ndMoment}.

Notably, quite significant correlations between the $V^0_{\mu\nu}$ may still be admitted, but like in the CLT, they should be largely irrelevant with respect to the $U$'s.
In this spirit, we take for granted 
that all matrix elements $V^0_{\mu\nu}$ can be treated as statistically
independent random variables, apart from the trivial 
constraint $(V_{\mu\nu}^0)^\ast=V_{\nu\mu}^0$ since $V$ is Hermitian.
Furthermore, we recall that we are working within a macroscopically small but microscopically large energy interval $I := [E, E + \Delta]$, so that the spectra of $H_0$ and $H$ are approximately homogeneous throughout $I$ [see also the discussion below~(\mref{1})].
Consequently, for the dynamics, only states with $E^0_\mu \in I$ are actually relevant.
Therefore, we can simply assume an infinite homogeneous spectrum by suitably extending the levels beyond the energy interval $I$.
This in turn implies that the statistics of the $V^0_{\mu\nu}$ can only depend on the difference $\mu-\nu$, and similarly for the $U_{m\nu}$ [see also~(\mref{8}) and the discussions above~(\mref{7})].

Altogether, it is thus reasonable to expect (and will be further corroborated later)
that the most important properties of the $V$ ensemble are its first two 
moments. Indicating ensemble averages as in the 
main paper by the symbol $\av{\,\cdots}$, we moreover
expect them to be expressible as [cf.~Eq.~(\mref{7})]
\begin{subequations}
\label{eq:MeanVarV}
\begin{align}
\label{eq:MeanV}
	\av{ V^0_{\mu\nu} } &= 0 \,, \\
\label{eq:VarV}
	\av{ \lvert V^0_{\mu\nu} \rvert^2 } &= \sigma_v^2 \, F(\lvert \mu - \nu \rvert) =: \sigma_{\mu\nu}^2 \, .
\end{align}
\end{subequations}
Here, the ``band profile'' function $F(n)$ 
admits the possibility of banded $V$ matrices.
As discussed below Eq.~(\mref{7}) in the main paper,
$\sigma_v^2$ sets the overall scale, while
$F(n)$ should be slowly varying with $n$, normalized 
so as to approach unity for small $n$, 
and upper bounded by a constant of order unity for all $n$.
Notably, however, $F(n)$ is not required to approach zero for 
large $n$, i.e., the $V$ matrices generally may, but need not 
be banded, nor is $F(n)$ required to monotonically decrease.

For the rest, the distribution of the $V^0_{\mu\nu}$ 
will indeed turn out (as anticipated above)
to be rather arbitrary provided that the dimension of the pertinent energy interval is sufficiently large (see Sec.~\ref{sec:OverlapsFromSupersymmetry:2ndMoment} in particular).
We therefore
stipulate the probability distributions of
the matrix elements $V^0_{\mu\nu}$ to be of the explicit general form
\begin{eqnarray}
	P_{\mu\nu}(v) & := & \av{\delta(V^0_{\mu\nu}-v)}=\tilde P_{|\mu-\nu|}(v)
	\ \ \ (\mu \leq \nu)
	\ ,\ 
\label{eq:PDFH:Entries'}
	\\
	\tilde P_j(v) & = & p_j \, \delta(v) 
	+ (1-p_j) \, f_j \!\left( \frac{v} {\tilde\sigma_j 
	\sqrt{F(j)} } \right)
\label{eq:PDFH:Entries}
\end{eqnarray}
with $j\in\NN_0$, $0\leq p_j<1$, 
$\tilde \sigma_j := \sigma_v 
/ \sqrt{1-p_j}$, 
and where $f_j(z)$ are generic probability 
distributions with vanishing mean and unit 
variance.
For $\mu=\nu$, the argument $v$ in~(\ref{eq:PDFH:Entries'}) and~(\ref{eq:PDFH:Entries})
is understood to be real.
For $\mu < \nu$, the argument may be complex or real,
depending on whether we are dealing with an
ensemble of complex or purely real matrices
$V^0_{\mu\nu}$
(see also Secs.~\ref{s42} and ~\ref{s43} below
for additional formal details). 
Moreover,
we generally take the 
corresponding $f_j(z)$ with $j\geq 1$
to depend on the absolute value 
$\lvert z \rvert$ only, an assumption we will elaborate on in the 
subsequent Sec.~\ref{sec:RandMatEns:InvarianceProperty} 
in more detail.
Finally, the restriction to indices 
with $\mu\leq \nu$ in~(\ref{eq:PDFH:Entries'})
is justified by the fact that $(V_{\mu\nu}^0)^\ast=V_{\nu\mu}^0$.

Note that from (\ref{eq:PDFH:Entries'}), (\ref{eq:PDFH:Entries}),
and the properties of $f_j(z)$
one readily recovers the relations \eqref{eq:MeanVarV}
for the first two moments. [This is one main reason behind the somewhat
involved structure of (\ref{eq:PDFH:Entries}).]
While the first two moments~\eqref{eq:MeanVarV}
are thus directly built into the formal definition of the
probability distributions (\ref{eq:PDFH:Entries'}) 
and (\ref{eq:PDFH:Entries}), all their remaining
statistical properties are still very general.

In particular, the possibility of sparse matrices 
is accounted for by the $\delta$-peak in~\eqref{eq:PDFH:Entries}.
In most cases, one actually expects
that the diagonal elements are
non-sparse, i.e., $p_0=0$,
and that all off-diagonal elements
exhibit (approximately) the same
sparsity, i.e., $p_j=p_1$ for all 
$j\geq 1$.
In this context, we also recall the definition of the effective bandwidth
\begin{equation}
\label{eq:B}
	B := \sum_{n=1}^\infty F(n)
\end{equation}
introduced below~(\mref{7}), 
implying $B=\infty$ if the matrices are not (or only very weakly) 
banded.

From the distributions of the individual entries~\eqref{eq:PDFH:Entries'}, 
we then obtain the total distribution of the matrix $V$ in the $H_0$ 
eigenbasis,
\begin{equation}
\label{eq:PDFH}
\begin{aligned}
	P(V) 
		&:=  \prod_{\mu \leq \nu} P_{\mu\nu}(V^0_{\mu\nu}) 
		\ .
\end{aligned}
\end{equation}
Averaging over the ensemble of perturbations means an integral over the 
probability measure $[\d V] \, P(V)$, where
\begin{equation}
\label{eq:dV}
	[\d V] := \left[ \prod_\mu \d V^0_{\mu\mu} \right] \left[ \prod_{\mu < \nu} \d V^0_{\mu\nu} \, \d V^{0*}_{\mu\nu} \right]
\end{equation}
is the Lebesgue measure of all independent entries of $V$ and
$\d V^0_{\mu\nu} \, \d V^{0*}_{\mu\nu}$ is understood as a shorthand
notation for $2 \, \d (\textrm{Re}\, V^0_{\mu\nu}) \, \d (\textrm{Im}\, V^{0}_{\mu\nu})$,
and where we tacitly focused on complex-valued $v$ for
$\mu\neq\nu$ in (\ref{eq:PDFH:Entries'}).
Symbolically, taking averages over the $V$ 
ensemble may thus be written in the two 
equivalent forms
 $\av{\,\cdots}=\int \,[\d V] \,\cdots \,P(V)$.

%%%%%%%%%%%%%%%%%%%%%%%%%%%%%%%%%%%%%%%%%%%%%%%%%%%%
%
\subsection{Physical motivation}
\label{sec:RandMatEns:InvarianceProperty}
In the previous section we specified the random matrix 
ensembles that will be admitted in our subsequent 
calculations. 
There, as well as in the main paper, we also argued 
why and how the investigation of such ensembles may provide 
insight with respect to our actual objective, namely to 
describe the effects of some given (non-random) 
perturbation of the ``true'' physical system of actual 
interest.

Here, we further elaborate on such considerations.
Specifically, we are concerned with the physical
motivation for our assumptions regarding
the functions $f_j(z)$ in (\ref{eq:PDFH:Entries}).
For this purpose,
the true (non-random) perturbation of actual interest
will be temporarily denoted as $\hat V$, while the 
symbol $V$ is reserved for the members of the 
considered random ensemble.

Our first observation is that if one takes for granted 
that the probability distributions in (\ref{eq:PDFH:Entries'}) 
indeed depend only on the differences $|\mu-\nu|$,
then those distributions can be estimated from the 
``empirical distribution'' of the true matrix elements  
$\hat V^0_{\mu\nu}$ by considering subsets with 
identical values of $|\mu-\nu|$ within the energy interval $I$.
Similarly, the various parameters and functions
appearing in (\ref{eq:PDFH:Entries}) can be 
(approximately) determined.
Along this line, the appropriate choice of the random matrix
ensemble is thus uniquely fixed by the given $\hat V$.
Analogously, also some of the further assumptions 
about this ensemble might be ``empirically checked'',
for instance the statistical independence of the matrix 
elements, or the features of $F(n)$.
Next we specifically address the assumed 
properties of the functions $f_j(z)$.

According to textbook quantum mechanics,
multiplying each eigenvector
$\ketN{\nu}$ of the unperturbed 
Hamiltonian $H_0$ 
by a factor of the form
$\e^{\I\phi_\nu}$ with an arbitrary but 
fixed phase $\phi_\nu$ does not entail 
any physically observable changes.
On the other hand, by doing so, 
each matrix element
$\hat V^0_{\mu\nu}$
acquires an extra factor
$\e^{\I(\phi_\nu-\phi_\mu)}$.
If we now choose a basis with an
arbitrary but fixed set of randomly
generated phases $\phi_\nu$,
it is quite plausible that the corresponding
``empirical distributions'' from above
will be well approximated for
any given index pair $\mu\neq\nu$
by a probability distribution $P_{\mu\nu}(v)$ in 
(\ref{eq:PDFH:Entries'}) which does not depend 
separately on the real and imaginary parts 
of $v$, but only on the absolute value 
$|v|$ (see also Sec.~\ref{s43} below).
Hence, the same must apply to the functions
$f_j(z)$ with $j\geq 1$, appearing on the 
right-hand side of (\ref{eq:PDFH:Entries}).

Note that in the above argument we tacitly
assumed that after randomizing the phases of the basis 
vectors $\ketN{\nu}$, a 
reasonable description of the so-obtained 
model within the general framework of our 
present random matrix approach 
is possible at all.

Once the property that $P_{\mu\nu}(v)$
only depends on $|v|$ is established for one
(random) choice of the phases $\phi_\nu$,
the same property of $P_{\mu\nu}(v)$ (and thus 
of $f_j(z)$) can be readily recovered for any
other (random or non-random) choice
of the $\phi_\nu$. Moreover, the underlying
ensemble of random matrices $V^0_{\mu\nu}$ 
must exhibit the following invariance: 
all its statistical properties remain 
unchanged when each random matrix 
element $V^0_{\mu\nu}$
of every member of the ensemble
is multiplied by a factor 
$\e^{\I(\phi_\nu-\phi_\mu)}$
for any arbitrary but fixed set of
phases $\phi_\nu\in[0,2\pi]$.

Essentially, we thus can conclude that 
if a random matrix approach is possible at all, 
then focusing on such $V$-ensembles
does not amount to any significant loss 
of generality (see also \cite{rei19,gen12,rei15}).

Finally, the above invariance of the 
$V$-ensemble can also be employed to once 
again recover the relation
(\ref{eq:MeanV}) for any $\mu\neq\nu$.
For a direct numerical verification
of this relation by means of a 
concrete physical model system, 
see also Ref.~\cite{beu15}.

As an aside we note that even in case the
matrix $\hat V^0_{\mu\nu}$ of the true perturbation
happens to be purely real in some specific
basis, one may conjecture that a suitably chosen
ensemble with complex-valued off-diagonal elements
might still do the job as well as a purely real-valued 
ensemble (this issue will be continued in 
Sec.~\ref{s42} below).

Turning to the so far excluded 
case $\mu=\nu$,
we first observe that we can always 
add a trivial constant to the true 
perturbation $\hat V$ such that all the 
$\hat V^0_{\nu\nu}$ with the property 
$E_\nu^0\in I$ [cf. below~(\mref{1})] 
are zero on average.
A similar argument as before
may then be invoked:
If the true perturbation can be 
reasonably described within a random 
matrix approach at all,
then only random ensembles which exhibit 
the property $\av{V^0_{\nu\nu}}=0$
seem appropriate to faithfully 
emulate the actual system 
at hand.
One thus can infer that $f_0(v)$ 
in (\ref{eq:PDFH:Entries}) must be a 
probability distribution with real valued 
argument $v$ and vanishing first 
moment.
(Note that in the case $j=0$ it is not 
required that $f_j(z)$ should only depend on 
$|z|$.)

We finally remark that the statistics of the diagonal 
random matrix elements $V^0_{\nu\nu}$ 
is in fact well-known to be of minor relevance 
anyway \cite{rei19,fyo95}.

%%%%%%%%%%%%%%%%%%%%%%%%%%%%%%%%%%%%%%%%%%%%%%%%%%%%
%
\subsection{Small parameter}
\label{sec:RandMatEns:SmallParameter}
A general feature of all the subsequent
calculations is that they amount 
to approximations in which the quantity 
\begin{eqnarray}
s:=\max_n u(n)
\label{x10}
\end{eqnarray}
plays the role of a small parameter,
i.e., we always assume that
$s\ll 1$.
Here $u(n)$ 
is given by (\mref{8}), i.e.,
\begin{equation}
\label{eq:AvgU2}
	u(m-n) := \av{ \lvert U_{mn} \rvert^2 } \,.
\end{equation}
Our first remark is that
the small parameter 
$\delta$ from~\eqref{eq:delta} 
[see also below~(\mref{11})],
which measures the inverse full width at half maximum of $u(n)$ in the weak-perturbation limit,
is almost, but not exactly 
identical to $s$.
To be precise, we will find in Sec.~\ref{sec:OverlapsFromSupersymmetry:2ndMoment} that
the two small
parameters are related via
\begin{eqnarray}
\delta=(\pi/2)\, s
\ .
\label{eq:s}
\end{eqnarray}
Our second remark is that
in most cases $u(n)$ is found 
to assume its maximum at $n=0$ 
and hence 
$s=u(0)$. However, exceptions
with $s\not=u(0)$ are still
possible, as exemplified in
Fig. 1 of Ref. \cite{fyo96}.

Our next objective is to quantify the 
``smallness'' of $s$ and hence 
of $\delta$ somewhat more precisely.
Quite obviously, the actual value of $s$
depends on many details of the specific 
model under consideration, hence somewhat
more general statements are only possible
in terms of nonrigorous arguments and
rough estimates.

As mentioned earlier, the mean level spacing
$\lvsp$ is exponentially small in the 
system's degrees of freedom \cite{lan70,gol10a}.
For a typical macroscopic system with, 
say, $f=10^{23}$ degrees of freedom, 
$\lvsp$ is thus an unimaginably small 
number 
(on the corresponding natural energy 
scale, which usually will be very 
roughly speaking on the order of
$10^x$ Joule with $|x|\ll f$).
Also for ``mesoscopic'' systems with 
considerably less extreme $f$ values
(arising for example in cold-atom experiments 
or in numerical simulations), the 
level spacing $\lvsp$ is usually still
expected to be 
extremely small
on the natural energy scale
of the system at hand.

Since $\sum_n u(n) = 1$ due to~\eqref{eq:AvgU2}, 
and given~\eqref{eq:s},
it seems reasonable to expect, 
and will be confirmed by all our particular 
examples below, that $u(n)$ is generically 
a rather slowly varying function 
of its argument $n\in\ZZ$,
decaying on a characteristic scale of the order 
of $s^{-1}$ from its maximal value at $n=0$
towards its asymptotic value zero for 
$|n|\to\infty$.
Moreover, we recall that its Fourier transform $g(t)$ from~\eqref{eq:g}
governs the modification of the
temporal relaxation in~(\mref{2})
due to the perturbation $\lambda V$.
We thus can infer from~\eqref{eq:g} that
the characteristic time 
scale of this modification will be 
of the order of $s/\lvsp$,
where 
we are still working in time units
with $\hbar=1$ [cf.\ below~(\mref{1})].
Since $\lvsp$ is exponentially small 
in the degrees of freedom $f$,
also $s$ is expected to be exponentially
small in $f$, at least if one disregards
extremely weak perturbations, which
entail modifications of the
relaxation which are themselves 
exponentially slow in $f$.
All these considerations
may thus be symbolically summarized 
like in~(\mref{15}) as
\begin{eqnarray}
s \approx \exp\{-\ord(f)\}\ll 1
\ .
\label{eq:SmallParam}
\end{eqnarray}

%%%%%%%%%%%%%%%%%%%%%%%%%%%%%%%%%%%%%%%%%%%%%%%%%%%%%
\section{Eigenvector correlations from supersymmetry}
\label{sec:OverlapsFromSupersymmetry}
In this section, we derive the results~(\mref{10}), (\mref{11}), and~\eqref{eq:u:SC} [see also below~(\mref{19})]
for the ``second moments''~(\ref{eq:AvgU2}), 
as well as the related results
for the ``fourth moments''
$\av{U_{m\mu_1}U_{n\mu_2}U^*_{m\nu_1}U^*_{n\nu_2}}$,
solving the pertinent random matrix
problem by means of supersymmetry methods.
We lay out the general procedure in Sec.~\ref{sec:OverlapsFromSupersymmetry:Outline} 
before presenting the explicit calculations for the second and 
fourth moments in Secs.~\ref{sec:OverlapsFromSupersymmetry:2ndMoment} and~\ref{sec:OverlapsFromSupersymmetry:4thMoment}, 
respectively.

Throughout this section we tacitly restrict ourselves 
to matrices with complex-valued 
off-diagonals 
$V^0_{\mu\nu}$ [see also the discussion below 
Eq.~(\ref{eq:PDFH:Entries})].
The case of purely real matrices is 
recovered by combining the results 
from \cite{fyo96} with the approach 
from Sec.~\ref{s4} below.

%%%%%%%%%%%%%%%%%%%%%%%%%%%%%%%%%%%%%%%%%%%%%%%%%%%%
%
\subsection{Outline of the method}
\label{sec:OverlapsFromSupersymmetry:Outline}
In this subsection, we sketch the general 
methodology on which the more detailed 
calculations in the subsequent 
Secs.~\ref{sec:OverlapsFromSupersymmetry:2ndMoment} and~\ref{sec:OverlapsFromSupersymmetry:4thMoment}
are based.

During most of the actual calculation, we choose a 
Hilbert space of large, but finite dimension $N$ 
so that $H_0$, $V$, and $H$ in~(\ref{eq:H}) amount to 
$(N\times N)$ matrices (in the eigenbasis of $H_0$).
Eventually, we will let $N\to\infty$ while keeping both the 
perturbation strength $\lambda$ and the average level 
spacing $\lvsp$ fixed.
In the considered ensemble of these Hamiltonians~(\ref{eq:H}), $H_0$ is fixed and $V$ is a 
Hermitian and possibly sparse and/or banded random 
matrix distributed according to~\eqref{eq:PDFH}.

\paragraph{Eigenvector overlaps from resolvents.}
As pointed out in the main paper and in Sec.~\ref{sec:RandMatEns:Definition} 
above, the key task is to evaluate the ensemble average 
over products of $U$ matrix elements~\eqref{eq:U}.
Such products can be expressed in terms of the 
\emph{resolvent} or Green's function of the Hamiltonian 
$H = H_0 + \lambda V$, defined as the operator 
\begin{equation}
\RsvG(z) := (z-H)^{-1}
\,.
\label{eq:DefG}
\end{equation}
To do so, we choose an arbitrary but fixed
energy level $E_n$ of $H$ and consider the matrix 
element $\RsvG^0_{\nu\mu}(E_n \pm \I\eta) := \braN{\nu} \RsvG(E_n \pm \I\eta) \ketN{\mu}$, yielding
\begin{equation}
\label{eq:ResolventH0}
	\RsvG^0_{\nu\mu}(E_n \pm \I \eta) = \sum_m \frac{E_n - E_m \mp \I\eta}{(E_n - E_m)^2 + \eta^2 } 
	U_{m\mu} U^*_{m\nu} \,.
\end{equation}

Next we adopt the assumption that the ensemble-averaged second moments $\av{U_{m\mu} U^*_{m\nu}}$ 
are (for arbitrary but fixed $\mu$ and $\nu$)
sufficiently slowly varying with $m$, so that the sum over $m$
in (\ref{eq:ResolventH0}) can be approximated by an integral, 
i.e.,
\begin{equation}
\label{eq:EigvecProdFromRsv}
\begin{aligned}
	\av{ U_{n\mu} U^*_{n\nu} } 
		&= \frac{\lvsp}{2\pi\I} \lim_{\eta\to 0+} \av{ \RsvG^0_{\nu\mu}(E_n - \I\eta) - \RsvG^0_{\nu\mu}(E_n + \I\eta) } \\
%		&\LD{= \frac{\lvsp}{\pi} \lim_{\eta\to 0+} \, \av{ ( \operatorname{Im} \RsvG(E_n - \I\eta) )^0_{\nu\mu} }, }
\end{aligned}
\end{equation}
and similarly for higher-order moments.
Our justification of this assumption is threefold:
(i) Exploiting the assumption, we will later determine an explicit
(approximate) expression for $\av{U_{m\mu} U^*_{m\nu}}$,
which indeed turns out to be slowly varying with
$m$.
Taking for granted that the exact solution for
$\av{U_{m\mu} U^*_{m\nu}}$ is unique and 
that our approximation is not fundamentally 
flawed, it follows that we have found a close
approximation to the unique exact 
$\av{U_{m\mu} U^*_{m\nu}}$.
(ii) In numerical examples we always found
that $\av{U_{m\mu} U^*_{m\nu}}$ indeed 
satisfies the assumption.
(iii) In random matrix theory, such assumptions are
routinely taken for granted without any further comment, 
while to our knowledge no counterexample has ever
been observed.

According to (\ref{eq:EigvecProdFromRsv}), we thus
can express the ensemble-averaged product of two eigenvector overlaps in 
terms of the \emph{advanced} and \emph{retarded} resolvents 
$\RsvG^0_{\nu\mu}(E_n - \I\eta)$ and $\RsvG^0_{\nu\mu}(E_n + \I\eta)$, respectively.
Analogously, for the fourth-order moments, we will later combine two such second-order 
expressions (details will be given in Sec.~\ref{sec:OverlapsFromSupersymmetry:4thMoment}).

The advantage of the resolvent formalism is that the matrix elements of $\RsvG(z)$ 
can be expressed as a Gaussian integral with kernel $\RsvG^{-1}(z) = z-H$.
Introducing the abbreviation $z^\pm := E_n \pm \I\eta$ with $E_n \in \mathbb R$ 
and $\eta > 0$ for the argument of the retarded and advanced resolvents,
the resolvent matrix element can be written as
\begin{equation}
\label{eq:RsvGaussInt}
\begin{aligned}
	\RsvG^0_{\nu\mu}(z^\pm) &= \frac{ \mp \I \, \det (z^\pm-H) }{(\pm2\pi\I)^N } \int \left[ \prod\nolimits_\alpha \d x_\alpha \d x_\alpha^* \right] x_\nu x^*_\mu \\
	&\quad \times \exp \left\{ \pm \I \sum\nolimits_{\alpha, \beta} x^*_\alpha \left( z^\pm \, \delta_{\alpha\beta} - H^0_{\alpha\beta} \right) x_\beta \right\} ,
\end{aligned}
\end{equation}
where $H^0_{\alpha\beta} := \braN{\alpha} H \ketN{\beta}$, and the choice of the sign in the exponent ensures convergence since $\eta > 0$.
The normalization factor $\det(z^\pm - H)$, and even more so its average, are 
in general hard to compute because $z^\pm - H$ is a high-dimensional matrix.
Nevertheless, it can be expressed conveniently by extending the Gaussian 
integral to anticommuting numbers.

\paragraph{Supersymmetry method.}
Our method of choice for the evaluation of average resolvents uses 
so-called supersymmetry techniques \cite{efe83, zuk96, mir00, ber10, haa10}.
The underlying concept of graded algebras and vector spaces 
introduces a set of \emph{anticommuting} or \emph{Grassmann numbers} 
$\chi_1, \chi_2, \ldots$ with the defining property that 
$\chi_i \chi_j = -\chi_j \chi_i$ for any two such elements.
The above cited references provide an introduction to 
the linear algebra and calculus on the resulting superspaces.

The crucial observation is that for a Gaussian integral 
similar to~\eqref{eq:RsvGaussInt}, if performed over 
Grassmannian, anticommuting numbers $\chi_\alpha$ and $\chi_\alpha^*$, we obtain 
\begin{equation}
\label{eq:GaussIntAnti}
\begin{aligned}
	&\int \left[ \prod\nolimits_\alpha \d\chi_\alpha \d\chi_\alpha^* \right] \exp \left\{ \I \sum\nolimits_{\alpha, \beta} \chi^*_\alpha \left( z^\pm \, \delta_{\alpha\beta}  - H^0_{\alpha\beta} \right) \chi_\beta \right\} \\
	&\; = \I^N \det (z^\pm - H) \,
\end{aligned}
\end{equation}
where our normalization for the 
Grassmann integral is 
$\int \d\chi \d\chi^* \, \chi^* \chi = 1$.
Combining~\eqref{eq:RsvGaussInt} and~\eqref{eq:GaussIntAnti}, we can thus write
\begin{equation}
\label{eq:RsvSuperGaussInt0}
\begin{aligned}
	\RsvG^0_{\nu\mu}(z^\pm) &= \frac{\mp\I}{(\mp 2\pi)^N} \int \left[ \prod\nolimits_\alpha \d x_\alpha \d x_\alpha^* \d \chi_\alpha \d\chi_\alpha^* \right] x_\nu x^*_\mu \\
		& \quad \times \exp \left\{ \I \sum\nolimits_{\alpha, \beta} \left[ \pm x^*_\alpha \left( z^\pm \, \delta_{\alpha\beta} - H^0_{\alpha\beta} \right) x_\beta \right. \right. \\
		&\qquad\qquad\qquad \left. \vphantom{\sum\nolimits_{\alpha, \beta}} \left. + \chi^*_\alpha \left( z^\pm \, \delta_{\alpha\beta} - H^0_{\alpha\beta} \right) \chi_\beta \right] \right\} .
\end{aligned}
\end{equation}
To condense the notation, we introduce a 
supervector $X := (X_1 \; \cdots \; X_N)^\trans$ 
with
\begin{equation}
\label{eq:Supervec2}
	X_\alpha := \lmat x_\alpha \\ \chi_\alpha \rmat .
\end{equation}
In the following, we will refer to the commuting component 
$X_{\alpha\bos} := x_\alpha$ as the \emph{bosonic} part, 
and to the anticommuting component $X_{\alpha\fer} := \chi_\alpha$ 
as the \emph{fermionic} part of the supervector $X_\alpha$.
Similarly, for a supermatrix $M$, we use the notation $M_{\bos\bos}$ to 
address the bosonic-bosonic sector, and consequently $M_{\bos\fer}$ for 
the bosonic-fermionic, $M_{\fer\bos}$ for the fermionic-bosonic, and 
$M_{\fer\fer}$ for the fermionic-fermionic sectors.
Using the above definition of $X$, we can then 
express~\eqref{eq:RsvSuperGaussInt0} as
\begin{equation}
\label{eq:RsvSuperGaussInt}
	\RsvG^0_{\nu\mu}(z^\pm) =  \mp\I \int \frac{\left[ \d X \d X^* \right]}{(\mp 2\pi)^N} \, x_\nu \, x^*_\mu \, \e^{ \I X^\dagger \left[ \left( z^\pm - H  \right) \otimes L^\pm \right] X } .
\end{equation}
Here, $L^\pm := \mathrm{diag}(\pm 1, 1)$ 
and $[\d X \d X^*] 
:= \left[ \prod_\alpha \d x_\alpha \d x_\alpha^* \d \chi_\alpha \d\chi_\alpha^* \right]$ 
for short.
The Kronecker product in the exponent thus consists of a ``Hilbert space'' factor $(z^\pm - H)$ and a ``superspace'' factor $L^\pm$.
In a slight abuse of notation, we will omit the Kronecker products in the following and identify $(z^\pm - H)$ and $L^\pm$ with their respectively lifted counterparts $(z^\pm - H) \otimes \id$ and $\id \otimes L^\pm$ on the combined Hilbert-superspace, and similarly for other operators acting trivially on either of the two factor spaces.
To obtain the average over all perturbations, we then have to integrate over the distribution~\eqref{eq:PDFH} of $V$, i.e.
\begin{equation}
\label{eq:AvgRsvSuperGaussInt}
\begin{aligned}
	\av{ \RsvG^0_{\nu\mu}(z^\pm) }
	&= \mp\I \int \left[ \d V \right] \int \frac{\left[ \d X \d X^* \right]}{(\mp 2\pi)^N} \,  P(V) \, x_\nu x^*_\mu\, \\
	&\qquad\qquad \times \e^{ \I X^\dagger L^\pm  \left( z^\pm - H_0 - \lambda V \right) X }.
\end{aligned}
\end{equation}
The necessary modifications for the computation 
of the fourth moments
$\av{U_{m\mu_1}U_{n\mu_2}U^*_{m\nu_1}U^*_{n\nu_2}}$,
involving products of two resolvents, 
will be further discussed in Sec.~\ref{sec:OverlapsFromSupersymmetry:4thMoment}.

\paragraph{Outline of the algorithm.}
The general procedure to evaluate expressions like 
Eq.~\eqref{eq:AvgRsvSuperGaussInt} involves the following steps:
First, 
we integrate over the matrix elements $V^0_{\alpha\beta}$ of 
the perturbation $V$, which can be carried out straightforwardly 
as the integrals are of Gaussian type due to a (generalized) 
central limit theorem.
The remaining superintegral then involves an exponent of fourth order in 
the supervector $X$.
Therefore, second, we invoke a Hubbard-Stratonovich transformation 
\cite{str57, hub59, efe83} by introducing an auxiliary integral over a supermatrix.
Thereby, we reduce the dependence on $X$ in the exponent to second order.
This allows us, third,
to perform the (now Gaussian) integral over the supervector $X$.
Fourth and last,
we evaluate the remaining integral over the Hubbard-Stratonovich supermatrix 
by means of a saddle-point approximation, making use of the large dimensionality 
$N$ of the considered Hilbert space.
In the end, we thus obtain explicit expressions for (products of) averaged resolvents, 
which give us access to the averaged eigenvector overlap products according 
to Eq.~\eqref{eq:EigvecProdFromRsv}.

% -------------------------------------------------------------

\subsection{Second moment}
\label{sec:OverlapsFromSupersymmetry:2ndMoment}

In this subsection, we derive the nonlinear integral 
equation~(\mref{10}), whose solution 
$G(z)$ gives access to the function $u(n)$ from
\eqref{eq:AvgU2}.
In addition, we solve the equation explicitly in the limiting cases of weak and stronger perturbations [compared to the scale $\lvsp B$ set by the perturbation bandwidth, cf.~\eqref{eq:B}], yielding expressions~\eqref{eq:u:BW} 
and~\eqref{eq:u:SC}, respectively, for $u(n)$.
To this end, we compute the average of the second 
moment~\eqref{eq:EigvecProdFromRsv} by calculating the 
average resolvents~\eqref{eq:AvgRsvSuperGaussInt}.

\paragraph{Ensemble average.}
Evaluating the ensemble average in Eq.~\eqref{eq:AvgRsvSuperGaussInt} 
amounts to computing the average
\begin{widetext}
\begin{equation}
\label{eq:AvgRsvExp}
	\avv{ \e^{-\I\lambda \, X^\dagger L^\pm V X} }
		= \avv{ \e^{ -\I \lambda \left[ \sum_\alpha X^\dagger_\alpha L^\pm X_\alpha V^0_{\alpha\alpha} + \sum_{\alpha < \beta} ( X^\dagger_\alpha L^\pm X_\beta + X^\dagger_\beta L^\pm X_\alpha ) (\operatorname{Re} V^0_{\alpha\beta}) + \sum_{\alpha < \beta} ( X^\dagger_\alpha L^\pm X_\beta - X^\dagger_\beta L^\pm X_\alpha ) (\I \operatorname{Im} V^0_{\alpha\beta}) \right] } } .
\end{equation}
\end{widetext}
Recalling the definition of $\sigma_{\alpha\beta}$
from~\eqref{eq:VarV}, and introducing
$Y_\alpha := \sigma_{\alpha\alpha} X^\dagger_\alpha L^\pm X_\alpha/2$ 
and $Y_{\alpha\beta} := \sigma_{\alpha\beta} X^\dagger_\alpha L^\pm X_\beta$, 
the exponent $X^\dagger L^\pm V X$ on the left-hand side 
of (\ref{eq:AvgRsvExp}) can be rewritten as $Z+Z^\ast$, where
\begin{equation}
\label{eq:AvgRsvExpLinComb}
\begin{aligned}
%	& X^\dagger L^\pm V X 
%		= \sum_\alpha Y_\alpha \frac{V_{\alpha\alpha}}{\sigma_{\alpha\alpha}} 
%			+ \sum_{\alpha < \beta} \left( Y_{\alpha\beta} 
%			\frac{ V^0_{\alpha\beta}}{\sigma_{\alpha\beta}} + Y_{\alpha\beta}^{*} \frac{ V^{0*}_{\alpha\beta}}{\sigma_{\alpha\beta}} \right)
%			\ .
& Z 
:= 
\sum_\alpha Y_\alpha \frac{V^0_{\alpha\alpha}}{\sigma_{\alpha\alpha}} 
+ \sum_{\alpha < \beta} Y_{\alpha\beta} 
			\frac{ V^0_{\alpha\beta}}{\sigma_{\alpha\beta}} 
			\ .
\end{aligned}
\end{equation}
In other words, $Z$ amounts to
a weighted sum of $N(N-1)/2$ independent 
random variables $V^0_{\alpha\beta}/\sigma_{\alpha\beta}$
of zero mean and unit variance.
According to the central limit theorem (cf.\ also the discussion 
in Sec.~\ref{sec:RandMatEns:Definition}),
we can thus conclude that
$Z$ approaches, for large $N$,
a Gaussian distribution with mean zero and variance 
$Y^2 := \sum_\alpha Y_\alpha^2 + \sum_{\alpha < \beta} 
\lvert Y_{\alpha\beta} \rvert^2 $,
and analogously for $X^\dagger L^\pm V X=Z+Z^\ast$.

Since all random variables $V^0_{\alpha\beta}$ only occur in 
combinations of the form~\eqref{eq:AvgRsvExpLinComb} in the 
average resolvent~\eqref{eq:AvgRsvSuperGaussInt}, only the 
limiting distribution of 
$Z$ actually matters.
Hence, we may approximate the distributions $P_{\mu\nu}(v)$ 
from Eq.~\eqref{eq:PDFH:Entries'} by any other distributions 
with the same values of the mean and variance
(since they lead,
for a given $X$, 
to the same limiting distribution for 
$Z$).
In particular, we may use
\begin{equation}
\label{eq:PDFH:Entries:Normal}
	P_{\alpha\alpha}(v) = \frac{ \e^{-v^2 / 2 \sigma^2_{\alpha\alpha}} }{ \sqrt{2 \pi} \, \sigma_{\alpha\alpha} }
	\; \text{and} \;
	P_{\alpha\beta}(v) = \frac{ \e^{-\lvert v \rvert^2 / \sigma^2_{\alpha\beta}} }{ \pi (\sigma_{\alpha\beta})^2 } \; (\alpha < \beta),
\end{equation}
approximating each matrix element by a normal distribution
(with real argument $v$ for $\alpha=\beta$ and complex $v$ for $\alpha < \beta$).
Evaluating the integrals over all $V^0_{\alpha\beta}$ ($\alpha \leq \beta$) 
in~\eqref{eq:AvgRsvSuperGaussInt}, which factorize because all matrix 
elements are independent, we then obtain
\begin{widetext}
\begin{equation}
\label{eq:AvgRsvAfterHInt}
\begin{aligned}
	\av{ \RsvG^0_{\nu\mu}(z^\pm) }
		&= \mp\I \int \frac{[\d X \d X^*]}{(\mp 2\pi)^N} \, x_\nu x_\mu^* \, \exp\left\{ -\tfrac{\ptst^2}{2} \sum\nolimits_{\alpha, \beta}  (\sigma_{\alpha\beta})^2 \, \str \left[ X_\alpha  X_\alpha^\dagger L^\pm X_\beta X_\beta^\dagger L^\pm \right] 
		 + \I \sum\nolimits_\alpha (z^\pm - E^0_\alpha) X_\alpha^\dagger L^\pm X_\alpha \right\} ,
\end{aligned}
\end{equation}
\end{widetext}
where
`$\str$' denotes the supertrace, i.e., $\str M = \tr M_{\bos\bos} - \tr M_{\fer\fer}$ 
for a supermatrix $M$.

For later use, we also note here that the integrand is invariant under a transformation 
$X \mapsto T X$, $X^\dagger \mapsto X^\dagger T^\dagger$ for any
(pseudo)unitary $T$ which satisfies $T^\dagger L^\pm T = L^\pm$.

\paragraph{Hubbard-Stratonovich transformation.}
Our next step is a supersymmetric Hubbard-Stratonovich transformation~\cite{str57, hub59, efe83} to get rid of the fourth-order term in $X$ in the exponent of~\eqref{eq:AvgRsvAfterHInt}.
To do so, we introduce a set of $(2\times 2)$ supermatrices $R_\alpha$,
\begin{equation}
\label{eq:HSMat}
	R_\alpha := \lmat r_{\alpha 1} & \rho_\alpha \\ \rho^*_\alpha & \I r_{\alpha 2} \rmat ,
\end{equation}
with real numbers $r_{\alpha 1}$, $r_{\alpha 2}$ and anticommuting $\rho_\alpha$, $\rho^*_\alpha$.
The choice of an imaginary $\fer\fer$ entry ensures convergence of the following integral.
Namely, the Hubbard-Stratonovich transformation is then based on the identity~\cite{haa10}
%\begin{widetext}
\begin{equation}
\label{eq:HS}
\begin{aligned}
	&\int \frac{[\d R]}{(2\pi)^N} \exp\left\{ -\sum\nolimits_{\alpha, \beta} \tfrac{ (\sigma^{-2})_{\alpha\beta} }{2 \lambda^2} \str\left( R_\alpha R_\beta \right)  \right. \\
	&\qquad\qquad\qquad\qquad \left. \vphantom{ \sum\nolimits_{\alpha, \beta} \tfrac{ (\sigma^{-2})_{\alpha\beta} }{2 \lambda^2} } + \I \sum\nolimits_\alpha \str \left( R_\alpha X_\alpha X_\alpha^\dagger L^\pm \right) \right\} \\
	&\quad = \exp\left\{ -\sum\nolimits_{\alpha, \beta} \tfrac{\lambda^2 (\sigma_{\alpha\beta})^2}{2} \str\left( X_\alpha X^\dagger_\alpha L^\pm X_\beta X^\dagger_\beta L^\pm \right) \right\} \,,
\end{aligned}
\end{equation}
where $\sigma^{-2}$ is the inverse of the Hilbert space matrix $\sigma^2$ 
with entries $(\sigma^2)_{\alpha\beta} := (\sigma_{\alpha\beta})^2$.
Moreover, $[\d R] := \prod_\alpha \d R_\alpha$ denotes the 
collective measure of 
all $R_\alpha$ with $\d R_\alpha = \d r_{\alpha 1} \d r_{\alpha 2} \d\rho_\alpha \d\rho^*_\alpha$.
Substituting~\eqref{eq:HS} into~\eqref{eq:AvgRsvAfterHInt}, we obtain
\begin{widetext}
\begin{equation}
\label{eq:AvgRsvAfterHS}
\begin{aligned}
	\av{ \RsvG^0_{\nu\mu}(z^\pm) } 
	&= \mp\I \int\frac{[\d R]}{(2\pi)^N} \int \frac{[\d X \d X^*]}{(\mp 2\pi)^N} \, x_\nu x^*_\mu \exp\left\{ -\sum\nolimits_{\alpha, \beta} \!\! \frac{(\sigma^{-2})_{\alpha\beta}}{2 \ptst^2} \str( R_\alpha R_\beta ) 
	 + \I \sum\nolimits_\alpha X^\dagger_\alpha L^\pm \left( R_\alpha + z^\pm - E^0_\alpha \right) X_\alpha \right\} .
\end{aligned}
\end{equation}
Comparing~\eqref{eq:AvgRsvAfterHInt} and~\eqref{eq:AvgRsvAfterHS}, we observe that the Hubbard-Stratonovich transformation essentially identifies $R_\alpha \sim X_\alpha X^\dagger_\alpha L^\pm$.
The integral over the supervector $X$ in~\eqref{eq:AvgRsvAfterHS} now has a Gaussian structure.
Therefore, we can evaluate it straightforwardly to find
\begin{equation}
\label{eq:AvgRsvAfterSuperInt}
\begin{aligned}
	\av{ \RsvG^0_{\nu\mu}(z^\pm) } 
		&= \delta_{\mu\nu} \int\frac{[\d R]}{(2\pi)^N} \, \left[ \left( R_\mu + z^\pm - E^0_\mu \right)^{-1} \right]_{\bos\bos} 
		 \exp\left\{ -\str\left[ \sum\nolimits_{\alpha, \beta} \tfrac{(\sigma^{-2})_{\alpha\beta}}{2 \ptst^2} R_\alpha R_\beta + \sum\nolimits_{\alpha} \ln \left( R_\alpha + z^\pm - E^0_\alpha \right) \right] \right\}.
\end{aligned}
\end{equation}
\end{widetext}

\paragraph{Saddle-point approximation.}
Keeping in mind the huge dimension $N$ of the considered Hilbert space and the 
fact that we will eventually take the limit $N\to\infty$, the remaining integral over the 
supermatrices $R_\alpha$ can be evaluated using a saddle-point approximation 
\cite{ben99, mir00, haa10}.
The reason is that the exponent in~\eqref{eq:AvgRsvAfterSuperInt} is a sum of $N$ 
very similar terms, so that the exponential is strongly peaked around the global 
maximum of the real part of its argument and/or highly oscillatory except at points 
where its imaginary part becomes stationary.
Hence the dominant contributions arise from the steepest saddle points in the 
complex, multidimensional $R$ plane, where the first variation of the exponent 
vanishes.
Computing this first variation and multiplying by the matrix $\ptst^2 \sigma^2$, 
we obtain the saddle-point equation
\begin{equation}
\label{eq:U2SaddleEq}
	R_\mu + \ptst^2 \sum_\alpha (\sigma_{\mu\alpha})^2 \left( R_\alpha + z^\pm - E^0_\alpha \right)^{-1} = 0 \,.
\end{equation}
To find the saddle points, we first look for diagonal solutions of this equation and 
consider one component $\hat R(E^0_\mu, z^\pm)$ of $R_\mu$, explicitly 
indicating the dependence of the solution on both the unperturbed and 
perturbed energy levels.
Any further solutions can be generated from diagonal ones by means of 
the (pseudo)unitary symmetry transformation introduced below~\eqref{eq:AvgRsvAfterHInt} \cite{haa10}.
Next, we substitute the variances $(\sigma_{\alpha\beta})^2$
from~\eqref{eq:VarV} with $F(n)$ as specified
there,
and we make use of the assumption
that the energy levels are very 
dense and approximately uniformly distributed, so that the 
density of states is essentially $\lvsp^{-1}$ [cf.\ discussion below~(\mref{1})].
Provided that the summands on the left-hand side of~\eqref{eq:U2SaddleEq} are slowly varying with $\alpha$, which -- as will become clear below -- is satisfied for $\delta \ll 1$ [cf.~Eq.~(\mref{15}) and Sec.~\ref{sec:RandMatEns:SmallParameter}],
we can approximate the sum by an integral and arrive at
\begin{equation}
	\hat R(E^0_\mu, z^\pm) + \int \frac{\d E}{\lvsp} \, \frac{ \ptst^2 \sigma_v^2 \, F(\lvert E/\lvsp - \mu \rvert) }{ z^\pm - E + \hat R(E, z^\pm) } = 0 \,.
\end{equation}
In fact, given the assumed homogeneity of the spectrum, the 
solution will only depend on the energy difference 
$z^\pm - E^0_\mu$, i.e., $\hat R(E_\mu^0, z^\pm)$ can be rewritten
in the form $\hat R(z^\pm - E^0_\mu)$.
This simplifies the equation further, resulting in the convolution-type relation
\begin{equation}
\label{eq:U2SaddleEqInt}
	\hat R(z^\pm) + \int \frac{\d E}{\lvsp} \, \frac{ \ptst^2 \sigma_v^2 \, F(\lvert E \rvert/\lvsp) }{ z^\pm - E + \hat R(z^\pm - E) } = 0 \,.
\end{equation}
Note that the sign of the imaginary part of $\hat R(z^\pm)$ has to match the sign of the imaginary part of $z^\pm$, $\operatorname{sgn} \operatorname{Im} \hat R(z^\pm) = \pm 1$, for otherwise the integrand in~\eqref{eq:U2SaddleEqInt} develops a pole.
Anticipating the final structure of $\av{ \RsvG^0_{\nu\mu}(z^\pm) }$ from~\eqref{eq:AvgRsvAfterSuperInt}, we introduce the function
\begin{equation}
\label{eq:G}
	G(z) := [z + \hat R(z)]^{-1} \,,
\end{equation}
for which, consequently, $\operatorname{sgn} \operatorname{Im} G(z^\pm) = \mp 1$ must hold.
Substituting into~\eqref{eq:U2SaddleEqInt}, we find that it satisfies
\begin{equation}
\label{suppl10}
	G(z^\pm) \left[ z^\pm - \frac{\lambda^2 \sigma_v^2}{\lvsp} \int \d E \; G(z^\pm - E) \, F(\lvert E \rvert / \lvsp) \right] = 1 \,,
\end{equation}
which is Eq.~(\mref{10}) from the main paper.
Assuming that $G(z)$ solves~\eqref{suppl10}, we immediately find a diagonal solution
\begin{equation}
\label{eq:U2Saddle}
	R_{* \mu} := \left[ G(z^\pm - E^0_\mu)^{-1} - (z^\pm - E^0_\mu) \right] \id
\end{equation}
of the saddle-point equation~\eqref{eq:U2SaddleEq}.
As mentioned below~\eqref{eq:U2SaddleEq}, any further solutions of 
the saddle-point equation can be generated from the diagonal one by means 
of symmetry transformations $T$ satisfying $T^\dagger L^\pm T = L^\pm$.
From the Hubbard-Stratonovich transformation~\eqref{eq:HS}, we 
understand that the supermatrix $R$ transforms as $R \mapsto T R T^{-1}$ 
under this symmetry.
But since the solution $R_{*\mu}$ from~\eqref{eq:U2Saddle} 
is proportional to the unit matrix, all transformed solutions collapse back 
onto the diagonal one, so that this is indeed the only solution of~\eqref{eq:U2SaddleEq}.
We remark that this will be manifestly different when computing 
the fourth moment in Sec.~\ref{sec:OverlapsFromSupersymmetry:4thMoment}.
Furthermore, one also verifies straightforwardly that the second 
variation of the exponent in~\eqref{eq:AvgRsvAfterSuperInt} with 
respect to $R_\mu$ is proportional to the unit matrix upon substitution of $R_{*\mu}$.
Therefore, its superdeterminant is unity and the saddle-point 
approximation of the integral~\eqref{eq:AvgRsvAfterSuperInt} is 
obtained from its integrand by plugging in the 
solution~\eqref{eq:U2Saddle} for $R_\mu$, so that
\begin{equation}
\label{eq:AvgRsv}
	\av{ \RsvG^0_{\nu\mu}(z^\pm) } = \delta_{\mu\nu} \, G(z^\pm - E^0_\mu) \,.
\end{equation}
Hence the function $G(z)$ is appropriately described as the scalar average resolvent.
Using $E_n - E^0_\mu \simeq (n-\mu)\lvsp$ [see below~(\mref{1})] and~\eqref{eq:EigvecProdFromRsv}, and 
remembering that $z^\pm = E_n \pm \I\eta$, we then find
\begin{equation}
\label{eq:AvgU2LargeBand}
	\av{ U_{n\mu} U^*_{n\nu} } = \delta_{\mu\nu} \, \ldos(n-\mu) 
\end{equation}
with
\begin{equation}
\label{eq:uFromG}
\begin{aligned}
	u(n) :=&\, \frac{\lvsp}{2\pi\I} \lim_{\eta\to 0+} \left[ G(n\lvsp - \I\eta) - G(n\lvsp + \I\eta) \right] \\
		=&\,  \frac{\lvsp}{\pi} \lim_{\eta\to 0+} \operatorname{Im} G(n\lvsp - \I\eta)  \,.
\end{aligned}
\end{equation}
Here, the last equality follows from $\RsvG(z^*) = \RsvG(z)^\dagger$
[see (\ref{eq:DefG})],
and hence, due to~\eqref{eq:AvgRsv}, $G(z^*) = G(z)^*$.
This establishes the results around Eq.~(\mref{10}) 
from the main paper.

The remaining task is to solve the defining equation~\eqref{suppl10} for the scalar average resolvent $G(z)$. 
Unfortunately, no general solution to this nonlinear integral equation is known to us.
However, analytic solutions are available in the limiting cases of weak and 
strong perturbations compared to the scale $\lvsp B$ of its bandwidth~\eqref{eq:B}.

\paragraph{Weak perturbations.}
Provided that $G(z)$ in 
Eq.~\eqref{suppl10} decays much faster than $F(\lvert E \rvert / \varepsilon)$ (which we will verify self-consistently in the end), 
we can neglect the band profile in the integral altogether and approximate $F(\lvert E \rvert / \lvsp) \approx 1$ (by normalization), leading to
\begin{equation}
\label{eq:IntEqG:Weak:Integral}
\int \d E \; G(z^\pm - E) F(\lvert E \rvert / \lvsp) \approx \int \d E \; G(E \pm \I\eta) =: C^\pm \,.
\end{equation}
From the full equation~\eqref{suppl10}, 
we then obtain
\begin{equation}
\label{eq:G:BW:UnknkownConst}
	G(z^\pm) = \frac{1}{z^\pm - \lambda^2 \sigma_v^2 C^\pm / \lvsp} \,.
\end{equation}
Next, we substitute this functional form back into~\eqref{eq:IntEqG:Weak:Integral} to determine the constant $C^\pm$.
The resulting integral needs to be evaluated in the principal-value sense, indicated by the symbol ``$\mathrm{PV}$'' in the following.
For $\eta > 0$, this leads to
\begin{equation}
	C^\pm = \mathrm{PV} \int \d E \; G(E \pm \I\eta) = \mp \I\pi \,.
\end{equation}
With~\eqref{eq:G:BW:UnknkownConst}, we thus find that the average scalar resolvent in the weak-perturbation or large-bandwidth limit is given by
\begin{equation}
\label{eq:G:BW}
	G(z^\pm) = \frac{1}{z^\pm \pm \I\Gamma / 2}
\end{equation}
with $\Gamma = 2\pi \lambda^2 \sigma_v^2 / \lvsp$ from~\eqref{eq:u:BW}.
Together with~\eqref{eq:uFromG}, this proves the result~\eqref{eq:u:BW} and hence~(\mref{11}).
The self-consistency condition for the weak-perturbation approximation, meaning that $G(z)$ 
is decaying much faster than the band profile $F(\lvert E \rvert / \lvsp)$,
becomes $\rateExp \ll \lvsp B $ as stated below Eq.~(\mref{s1}),
provided that $F(n)$ remains of order unity for $n \lesssim B$.

Finally, as a technical remark, we note that the solution~\eqref{eq:U2Saddle} with $G(z)$ from~\eqref{eq:G:BW} 
does not lie on the original contour of integration 
in~\eqref{eq:HS}, see also~\eqref{eq:HSMat}.
However, the contour can be adjusted accordingly by shifting 
$r_1 \mapsto r_1 \pm \I \rateExp/2$, which is allowed because 
the poles in the integrand at $r_1 = E^0_\alpha - z^\pm$ all 
lie on the same side of the real axis, below it for `$+$' 
and above it for `$-$', respectively.

\paragraph{Stronger perturbations.}
In the opposite limiting case, when the band profile $F(\lvert E \rvert / \lvsp)$ decays much faster than the solution $G(z)$ of~\eqref{suppl10},
we can approximate $G(z^\pm - E)$ in the integrand by its central value $G(z^\pm)$.
Using the definition~\eqref{eq:B} of the effective bandwidth, we then find
\begin{equation}
\label{eq:IntEqG:Strong}
	\gamma^2 \, G(z^\pm)^2 / 4 - z^\pm \, G(z^\pm) + 1 = 0 \,,
\end{equation}
where $\gamma = \sqrt{8 B} \, \lambda \sigma_v$ was defined as in~\eqref{eq:u:SC}.
Solving this algebraic equation, we immediately obtain
\begin{equation}
	G(z^\pm) = \frac{2}{\gamma^2} \left( z^\pm \mp \I \sqrt{\gamma^2 - (z^\pm)^2} \right) ,
\end{equation}
where we already incorporated the requirement that the imaginary parts of $G(z^\pm)$ and $z^\pm$ should have opposite signs [see below~\eqref{eq:G}].
Together with~\eqref{eq:uFromG}, we then recover~\eqref{eq:u:SC} and hence the result stated below~(\mref{19}) for $u(n)$ in the limit of a dominating overlap scale compared to the bandwidth.
More precisely, this latter assumption is verified self-consistently if
$\gamma \gg \lvsp B$, as stated below~(\mref{s2}) in the main paper.

% --------------------------------------------------------
\subsection{Fourth moment}
\label{sec:OverlapsFromSupersymmetry:4thMoment}

In this subsection, we evaluate the fourth moments
$\av{U_{m\mu_1}U_{n\mu_2}U^*_{m\nu_1}U^*_{n\nu_2}}$
of eigenvector overlaps when the perturbation strength 
is weak compared to its bandwidth, which is the physically 
most relevant case for generic many-body systems.
More precisely speaking, we will derive the
following results:
\begin{subequations}
\label{eq:AvgU4}
\begin{equation}
\label{eq:AvgU4:tot}
\begin{aligned}
	&\av{ U_{m \mu_1} U_{n \mu_2} U^*_{m \nu_1} U^*_{n \nu_2} } \\
		&\;= \delta_{\mu_1 \nu_1} \delta_{\mu_2 \nu_2} \, d^{mn}_{\mu_1 \mu_2}
			+ \delta_{\mu_1 \nu_2} \delta_{\mu_2 \nu_1} \left( \delta_{mn} \, d^{mn}_{\mu_1 \mu_2} + f^{mn}_{\mu_1 \mu_2} \right)
\end{aligned}
\end{equation}
with
\begin{align}
\label{eq:AvgU4:d}
	d^{mn}_{\mu_1 \mu_2} &:= \ldos(m -\mu_1) \, \ldos(n - \mu_2) \ ,
 \displaybreak[0] \\
\label{eq:AvgU4:f}
	f^{mn}_{\mu_1 \mu_2} & := - \left( \tfrac{\rateExp / \lvsp}{4\pi} \right)  \ldos(m - \mu_1) \, \ldos(n - \mu_2) \\
		& \hspace{-20pt} \times \frac{(\rateExp/\lvsp)^2 + (\mu_1 \!-\! \mu_2)^2 + (m \!-\! n)^2 - (\mu_1 \!+\! \mu_2 \!-\! m \!-\! n)^2 }{ \left[ (m - \mu_2)^2 + (\rateExp/2\lvsp)^2 \right] \left[ (n - \mu_1)^2 + (\rateExp/2\lvsp)^2 \right] } \notag
\end{align}
\end{subequations}
to leading order in the small parameter $s$ from~\eqref{x10}.
As some of the steps are straightforward generalizations of the calculation of the second moment in Sec.~\ref{sec:OverlapsFromSupersymmetry:2ndMoment}, the presentation will be tightened in these places.
Moreover, we immediately assume a normal distribution~\eqref{eq:PDFH:Entries:Normal} for the perturbation matrix elements $V^0_{\mu\nu}$ because the general case of sparse and banded perturbations can be reduced to this effective model as sketched in Sec.~\ref{sec:OverlapsFromSupersymmetry:2ndMoment}.
Focusing, as announced above, on the weak-perturbation/large-bandwidth limit, which
essentially amounts to a constant band profile (cf.\ Sec.~\ref{sec:OverlapsFromSupersymmetry:2ndMoment}),
we will assume right from the beginning that the variance $(\sigma_{\mu\nu})^2 = \sigma_v^2 = 1$ for all $V^0_{\mu\nu}$ and control the interaction strength by the parameter $\ptst$.

%%%%%%%%%%%%%%%%%%%%%%%%%%%%%%%%%%%%%%%
\begin{figure*}
\includegraphics[scale=1]{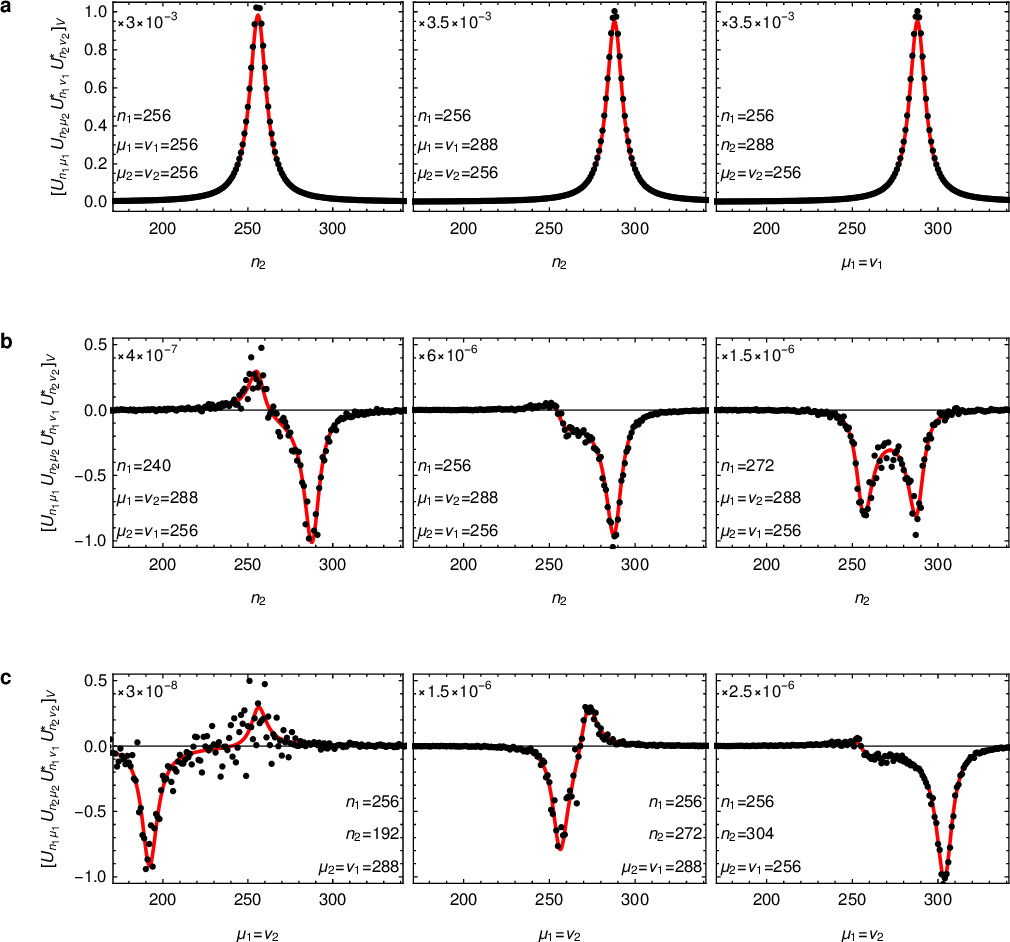}
\caption{
Fourth moments
$\av{ U_{n_1 \mu_1} U_{n_2 \mu_2} U^*_{n_1 \nu_1} U^*_{n_2 \nu_2} }$
for various combinations of fixed and variable indices as 
indicated in the respective panels.
Solid: Analytical approximation from ~\eqref{eq:AvgU4}.
Dots: Numerical averages over $10^5$ randomly 
sampled perturbation matrices $V^0_{mn}$.
The considered random matrix model is of the 
form (\ref{eq:H}), where $H_0$ exhibits $N=512$ 
equidistant levels with level spacing 
$\lvsp=1$.
The matrix elements $V_{mn}^0$ 
are independent random variables 
(apart from $(V_{mn}^0)^\ast=V_{nm}^0$).
The diagonal elements $V^0_{nn}$
are Gaussian distributed with zero mean and
unit variance.
For $m<n$, the real and imaginary parts 
of $V_{mn}^0$ are independent Gaussian 
distributed random variables
with mean zero and variance $1/2$.
The perturbation strength is $\ptst = 1.33$, 
implying $\sigma_v^2 = 1$, $\rateExp \simeq 11$,
and $s\simeq 0.06$
according to (\ref{eq:VarV}), (\ref{x10}), (\ref{eq:G:BW}), and (\ref{eq:u:BW}).
The $y$-axes are scaled as specified in the top-left corner of each panel.
In (a), the $d^{n_1 n_2}_{\mu_1 \mu_2}$ branch 
on the right hand side of ~\eqref{eq:AvgU4} is probed as a function 
of $n_2$ or $\mu_1 = \nu_1$.
Likewise, in (b) and (c) the $f^{n_1 n_2}_{\mu_1 \mu_2}$ branch 
is probed as a function of  $n_2$ and $\mu_1 = \nu_2$, respectively.
}
\label{fig:U4}
\end{figure*}
%%%%%%%%%%%%%%%%%%%%%%%%%%%%%%%%%%%%%%%%%%%%%%%%%%%

Before turning to its derivation, we demonstrate by means of Fig.~\ref{fig:U4}
that the analytical leading-order approximation~\eqref{eq:AvgU4} 
captures the actual and quite nontrivial behavior with great accuracy.
Concretely, Fig.~\ref{fig:U4} provides a detailed comparison
of the result~\eqref{eq:AvgU4} with a numerical 
computation of $\av{ U_{n_1 \mu_1} U_{n_2 \mu_2} U^*_{n_1 \nu_1} U^*_{n_2 \nu_2} }$ 
for an ensemble of random matrices as specified in the figure caption.
Notably, the approximation seems to work surprisingly well already for relatively 
moderate values of the small parameter $s$ or, equivalently, for relatively moderate matrix dimensions $N$.
%%%%%%%%%%%%%%%%%%%%%%%%%%%%%%%%%%%%%%%%%%%%%%%%%%%%%%%%

\paragraph{Properties of the fourth moment.}
The averaged fourth-order eigenvector overlap
\begin{equation}
\label{eq:AvgU4Raw}
	\av{ U_{n_1 \mu_1} U_{n_2 \mu_2} U^*_{n_1 \nu_1} U^*_{n_2 \nu_2} }
\end{equation}
involves the components $\mu_1$, $\mu_2$, $\nu_1$, and $\nu_2$ of two perturbed eigenvectors $\ket{n_1}$, $\ket{n_2}$ in the basis of the unperturbed system.
We begin by collecting a few properties and symmetries that follow directly from its general structure and the unitarity of the overlap matrix $U_{n\mu}$.
First, there are two symmetry properties that are obtained from interchanging indices in Eq.~\eqref{eq:AvgU4Raw}.
Namely, the result should
\begin{enumerate}[label=(\roman*)]
	\item \label{lst:AvgU4Swap12}
		be invariant under the exchange of labels $1$ and $2$, i.e.\ $n_1 \leftrightarrow n_2$, $\mu_1 \leftrightarrow \mu_2$, $\nu_1 \leftrightarrow \nu_2$;
	\item \label{lst:AvgU4SwapMuNu}
		be complex conjugated upon $\mu_1 \leftrightarrow \nu_1$ and $\mu_2 \leftrightarrow \nu_2$.
\end{enumerate}
Second, the fourth moment must relate to the second moment~\eqref{eq:AvgU2LargeBand} when tracing out eigenvectors of the unperturbed or perturbed Hamiltonians.
Hence the result should
\begin{enumerate}[resume, label=(\roman*)]
	\item \label{lst:AvgU4TraceH}
		reduce to the second moment when summing over $n_1$ or $n_2$, i.e.
	\begin{equation}
		\sum\nolimits_{n_1} \av{ U_{n_1 \mu_1} U_{n_2 \mu_2} U^*_{n_1 \nu_1} U^*_{n_2 \nu_2} } = \delta_{\mu_1 \nu_1} \, \av{ U_{n_2 \mu_2} U^*_{n_2 \nu_2} } \,;
	\end{equation}
	\item \label{lst:AvgU4TraceH0}
		reduce to the second moment when summing over $\mu_1 = \nu_2$ or $\mu_2 = \nu_1$, i.e.
	\begin{equation}
	\label{eq:AvgU4TraceH0}
	\begin{aligned}
		&\sum\nolimits_{\mu_1, \nu_2} \!\!\!\!\delta_{\mu_1 \nu_2} \, \av{ U_{n_1 \mu_1} U_{n_2 \mu_2} U^*_{n_1 \nu_1} U^*_{n_2 \nu_2} }
		\\&\quad
		= \delta_{n_1 n_2} \, \av{ U_{n_1 \mu_2} U^*_{n_1 \nu_1} } \,.
	\end{aligned}
	\end{equation}
\end{enumerate}
Furthermore, the Gaussian distribution of the perturbation matrix elements together 
with the Isserlis (or Wick) theorem 
(see also Secs. \ref{s41} and \ref{s43} below)
imply that we always need to pair up factors of $U$ and $U^*$.
This will become obvious in the explicit calculations below.
In any case, it already affirms the general structure~\eqref{eq:AvgU4:tot} with $d^{n_1 n_2}_{\mu_1 \mu_2}$ and $f^{n_1 n_2}_{\mu_1 \mu_2}$ to be determined. 

To compute the correlator~\eqref{eq:AvgU4Raw}, we proceed as for the second moment and aim at extracting it from the product of average resolvents similarly to~\eqref{eq:EigvecProdFromRsv}.
As we are dealing with a product of four overlap matrices now, we have to consider products of two resolvents.
A crucial observation is that we need to distinguish the cases $n_1 = n_2$ (single eigenvector) and $n_1 \neq n_2$ (two distinct eigenvectors).

\paragraph{Resolvent approach: single eigenvector.}
In the first case, $n_1 = n_2 = n$, we perform similar steps as those leading to Eq.~\eqref{eq:EigvecProdFromRsv} to express the fourth-order overlap in terms of resolvents, yielding
\begin{widetext}
\begin{equation}
\label{eq:U4SameFromRsv}
\begin{aligned}
	& -\frac{8 \pi^2}{\lvsp^2} \av{ U_{n \mu_1} U_{n \mu_2} U^*_{n \nu_1} U^*_{n \nu_2} } \\
		&\;= \lim_{\eta\to 0+}
		\left\{
			\av{ \RsvG^0_{\nu_1 \mu_1}(z^+) \, \RsvG^0_{\nu_2 \mu_2}(z^+) + \RsvG^0_{\nu_1 \mu_2}(z^+) \, \RsvG^0_{\nu_2 \mu_1}(z^+) }
		 + \av{ \RsvG^0_{\nu_1 \mu_1}(z^-) \, \RsvG^0_{\nu_2 \mu_2}(z^-) + \RsvG^0_{\nu_1 \mu_2}(z^-) \, \RsvG^0_{\nu_2 \mu_1}(z^-) } \right. \\
	&\qquad\qquad\qquad \left. - \av{ \RsvG^0_{\nu_1 \mu_1}(z^+) \, \RsvG^0_{\nu_2 \mu_2}(z^-) + \RsvG^0_{\nu_1 \mu_2}(z^+) \, \RsvG^0_{\nu_2 \mu_1}(z^-) } 
		- \av{ \RsvG^0_{\nu_1 \mu_1}(z^-) \, \RsvG^0_{\nu_2 \mu_2}(z^+) + \RsvG^0_{\nu_1 \mu_2}(z^-) \, \RsvG^0_{\nu_2 \mu_1}(z^+) } \right\} .
\end{aligned}
\end{equation}
\end{widetext}
Here $z^\pm := E_n \pm \I\eta$ as in Sec.~\ref{sec:OverlapsFromSupersymmetry:2ndMoment}.
The symmetrization in the Greek indices becomes necessary because such symmetrized combinations of resolvents are the only quantities that can be computed as Gaussian integrals if the argument of both factors is the same.
Indeed, the corresponding integral is essentially the same as in Eq.~\eqref{eq:RsvSuperGaussInt} except that we have four factors of $x$ in the integrand now to account for the fourth moment.
Therefore,
\begin{equation}
\label{eq:RsvRRSameSuperGaussInt}
\begin{aligned}
	&\int \frac{\left[ \d X \d X^* \right]}{(+2\pi)^N} \, x_{\nu_1} x_{\nu_2} \, x^*_{\mu_1} x^*_{\mu_2} \, \e^{ \I X^\dagger L^+ \left( z^+ - H  \right)  X } \\ 
		&\; = \RsvG^0_{\nu_1 \mu_1}(z^+) \RsvG^0_{\nu_2 \mu_2}(z^+) + \RsvG^0_{\nu_1 \mu_2}(z^+) \RsvG^0_{\nu_2 \mu_1}(z^+) \,,
\end{aligned}
\end{equation}
and similarly for $z^-$, where the right-hand side was obtained using the Isserlis/Wick theorem.
Averaging Eq.~\eqref{eq:RsvRRSameSuperGaussInt} over the ensemble of perturbations works analogously to the calculation of the second moment in Sec.~\ref{sec:OverlapsFromSupersymmetry:2ndMoment}, except that we now have four factors $x_{\nu_1} x_{\nu_2} x^*_{\mu_1} x^*_{\mu_2}$ in the integrand instead of two.
When going from the equivalent of Eq.~\eqref{eq:AvgRsvAfterHS} to the equivalent of Eq.~\eqref{eq:AvgRsvAfterSuperInt}, we need to use the Isserlis/Wick theorem again and obtain two inverse matrix elements in the remaining superintegral over the Hubbard-Stratonovich matrix $R$,
\begin{widetext}
\begin{equation}
\begin{aligned}
	&\av{ \RsvG^0_{\nu_1 \mu_1}(z^\pm) \RsvG^0_{\nu_2 \mu_2}(z^\pm) + \RsvG^0_{\nu_1 \mu_2}(z^\pm) \RsvG^0_{\nu_2 \mu_1}(z^\pm) } \\
		&\;= \left( \delta_{\mu_1 \nu_1} \delta_{\mu_2 \nu_2} + \delta_{\mu_1 \nu_2} \delta_{\mu_2 \nu_1} \right) \int \frac{\d R}{2\pi} \, (R \!+\! z^\pm \!-\! E^0_{\mu_1})^{-1}_{\;\;\;\bos\bos}\; (R \!+\! z^\pm \!-\! E^0_{\mu_2})^{-1}_{\;\;\;\bos\bos}
		 \exp\left\{ -\str\left[ \frac{ R^2}{2\lambda^2} + \sum\nolimits_\alpha \!\! \ln (R \!+\! z^\pm \!-\! E^0_\alpha) \right] \right\} .
\end{aligned}
\end{equation}
Note that due to the constant variance $(\sigma_{\mu\nu})^2 = 1$ of the perturbation matrix elements assumed in this subsection, a single Hubbard-Stratonovich matrix $R$ of the form~\eqref{eq:HSMat} suffices in the transformation~\eqref{eq:HS}.
Computing the $R$-integral by means of a saddle-point approximation works completely analogously, so that we eventually find
\begin{subequations}
\label{eq:AvgRsvRRSame}
\begin{align}
	&\av{ \RsvG^0_{\nu_1 \mu_1}(z^+) \RsvG^0_{\nu_2 \mu_2}(z^+) + \RsvG^0_{\nu_1 \mu_2}(z^+) \RsvG^0_{\nu_2 \mu_1}(z^+) } 
	=  \frac{\delta_{\mu_1 \nu_1} \delta_{\mu_2 \nu_2} + \delta_{\mu_1 \nu_2} \delta_{\mu_2 \nu_1}}{ \left( z^+ - E^0_{\mu_1} + \I \rateExp/2 \right) \left( z^+ - E^0_{\mu_2} + \I \rateExp/2 \right) } , \displaybreak[0] \\
	&\av{ \RsvG^0_{\nu_1 \mu_1}(z^-) \RsvG^0_{\nu_2 \mu_2}(z^-) + \RsvG^0_{\nu_1 \mu_2}(z^-) \RsvG^0_{\nu_2 \mu_1}(z^-) } 
	= \frac{\delta_{\mu_1 \nu_1} \delta_{\mu_2 \nu_2} + \delta_{\mu_1 \nu_2} \delta_{\mu_2 \nu_1}}{ \left( z^- - E^0_{\mu_1} - \I \rateExp/2 \right) \left(  z^- - E^0_{\mu_2} - \I \rateExp/2 \right) } .
\end{align}
\end{subequations}
\end{widetext}
In the remaining terms in Eq.~\eqref{eq:U4SameFromRsv}, the resolvents are evaluated at distinct arguments $z^+$ and $z^-$ in the products, e.g., $\RsvG^0_{\nu_1 \mu_1}(z^+) \RsvG^0_{\nu_2 \mu_2}(z^-)$.
In this case, the calculation parallels the one for distinct eigenvectors, to which we will turn next.

\paragraph{Resolvent approach: distinct eigenvectors.}
If the perturbed eigenvectors $\ket{n_1}$ and $\ket{n_2}$ in Eq.~\eqref{eq:AvgU4Raw} are distinct, $n_1 \neq n_2$, the product of overlap matrices can be expressed in terms of the retarded and advanced resolvents by multiplying two expressions of the form~\eqref{eq:EigvecProdFromRsv}.
Defining $z^\pm_k := E_{n_k} \pm \I \eta$, we obtain
\begin{equation}
\label{eq:U4DistFromRsv}
\begin{aligned}
	&-\frac{4\pi^2}{\lvsp^2}  \av{ U_{n_1 \mu_1} U_{n_2 \mu_2} U^*_{n_1 \nu_1} U^*_{n_2 \nu_2} } \\
		&\;= \lim_{\eta\to 0+} \!\!\left\{
			\av{ \RsvG^0_{\nu_1 \mu_1}(z^+_1) \RsvG^0_{\nu_2 \mu_2}(z^+_2) + \RsvG^0_{\nu_1 \mu_1}(z^-_1) \RsvG^0_{\nu_2 \mu_2}(z^-_2) } \right. \\
			&\qquad \left. - \av{ \RsvG^0_{\nu_1 \mu_1}(z^+_1) \RsvG^0_{\nu_2 \mu_2}(z^-_2) + \RsvG^0_{\nu_1 \mu_1}(z^-_1) \RsvG^0_{\nu_2 \mu_2}(z^+_2) }
		\right\} .
\end{aligned}
\end{equation}
There is a decisive difference between the terms in the second line and the ones in the third line of this relation.
In the second line, the imaginary parts of the argument of the two factors in each resolvent product are the same, so that we have a product of two retarded or two advanced resolvents.
In contrast, the terms in the third line each involve one retarded and one advanced resolvent.

We inspect the first kind with energies shifted to the same side of the real axis first.
Since $z_1 \neq z_2$, we now need two supervectors $X^{(1)}$ and $X^{(2)}$ of the form~\eqref{eq:Supervec2} in order to write the product of resolvents as a Gaussian integral,
\begin{equation}
\label{eq:RsvRRDistFromSuperInt}
\begin{aligned}
	&\RsvG^0_{\nu_1 \mu_1}(z_1^\pm) \, \RsvG^0_{\nu_2 \mu_2}(z_2^\pm) \\
	&\;=  \int \frac{\left[ \d X^{(1)} \d X^{(1)*} \right]}{(\mp 2\pi)^N} \frac{\left[ \d X^{(2)} \d X^{(2)*} \right]}{(\mp 2\pi)^N} x^{(1)}_{\nu_1} \, x^{(1)*}_{\mu_1} \, x^{(2)}_{\nu_2} \, x^{(2)*}_{\mu_2} \\
			&\qquad\qquad \times \e^{ \I X^{(1)\dagger} L^\pm \left( z_1^\pm - H  \right)  X^{(1)} + \I X^{(2)\dagger} L^\pm \left( z_2^\pm  - H  \right) X^{(2)} } .
\end{aligned}
\end{equation}
Averaging this expression over the ensemble of perturbations, the entire calculation for the second moment from Sec.~\ref{sec:OverlapsFromSupersymmetry:2ndMoment} carries over, except that we have essentially two copies of the same integral \cite{haa10}.
Most importantly, the resulting saddle-point equation still has a single solution proportional to the unit matrix.
The upshot is that the averaged product of two advanced or two retarded resolvents factorizes into the product of two averages, i.e.
\begin{subequations}
\label{eq:AvgRsvRRDist}
\begin{align}
	\av{ \RsvG^0_{\nu_1 \mu_1}(z_1^+) \, \RsvG^0_{\nu_2 \mu_2}(z_2^+) }
		&= \av{ \RsvG^0_{\nu_1 \mu_1}(z_1^+) } \,\, \av{ \RsvG^0_{\nu_2 \mu_2}(z_2^+) } \,, \displaybreak[0] \\
	\av{ \RsvG^0_{\nu_1 \mu_1}(z_1^-) \, \RsvG^0_{\nu_2 \mu_2}(z_2^-) }
		&= \av{ \RsvG^0_{\nu_1 \mu_1}(z_1^-) } \,\, \av{ \RsvG^0_{\nu_2 \mu_2}(z_2^-) } \,,
\end{align}
\end{subequations}
where the single averages were given in~\eqref{eq:AvgRsv}.

The situation is manifestly different for the average product of one retarded and one advanced resolvent, which is the form of the remaining terms both in Eq.~\eqref{eq:U4DistFromRsv} for two distinct eigenvectors and in Eq.~\eqref{eq:U4SameFromRsv} for a single eigenvector.
There, the saddle-point equation obtained after averaging over the ensemble of $V$'s does not have a unique diagonal solution anymore, but instead exhibits a whole manifold of degenerate saddles.

Expressed as a Gaussian integral, the product of two different resolvents takes the form
\begin{equation}
\label{eq:RsvRAFromSuperGaussInt0}
\begin{aligned}
	& \RsvG^0_{\nu_1 \mu_1}(z_1^+) \, \RsvG^0_{\nu_2 \mu_2}(z_2^-) \\
	&\;=  \int \frac{\left[ \d X^{(1)} \d X^{(1)*} \right]}{(\mp 2\pi)^N} \frac{\left[ \d X^{(2)} \d X^{(2)*} \right]}{(\mp 2\pi)^N} x^{(1)}_{\nu_1} \, x^{(1)*}_{\mu_1} \, x^{(2)}_{\nu_2} \, x^{(2)*}_{\mu_2} \\
			&\qquad\qquad \times \e^{ \I X^{(1)\dagger} L^+  \left( z_1^+ - H  \right) X^{(1)} + \I X^{(2)\dagger} L^- \left( z_2^-  - H  \right) X^{(2)} } ,
\end{aligned}
\end{equation}
where the matrices $L^\pm$ were defined below Eq.~\eqref{eq:RsvSuperGaussInt}.
To compress notation, we introduce a collective supervector $X := (X_1 \; \cdots \; X_N)^\trans$ with
\begin{equation}
	X_\alpha := \lmat x^{(1)}_\alpha & \chi^{(1)}_\alpha & x^{(2)}_\alpha & \chi^{(2)}_\alpha \rmat^{\mathsf T} .
\end{equation}
In the following, we will refer to the first two components (superscript 1) as the retarded and the last two components (superscript 2) as the advanced sector.
Furthermore, we define the abbreviations
\begin{equation}
	\bar z := \frac{z_1^+ + z_2^-}{2}
	\quad \text{and} \quad
	\Delta_z := z_1^+ - z_2^-
\end{equation}
as well as the diagonal matrices
\begin{equation}
	L := \mathrm{diag}(1,1,-1,1)
	\quad \text{and} \quad
	\Lambda := \mathrm{diag}(1,1,-1,-1) \,.
\end{equation}
Note that $\mathrm{Im}\, \bar z = 0$ and $\mathrm{Im}\, \Delta_z = 2\eta$.
With these definitions, we can write Eq.~\eqref{eq:RsvRAFromSuperGaussInt} in the more compact form
\begin{equation}
\label{eq:RsvRAFromSuperGaussInt}
\begin{aligned}
	\RsvG^0_{\nu_1 \mu_1}(z_1^+) \, \RsvG^0_{\nu_2 \mu_2}(z_2^-)
	&=  \int \frac{\left[ \d X \d X^* \right]}{(2\pi\I)^{2N}}  x^{(1)}_{\nu_1} \, x^{(1)*}_{\mu_1} \, x^{(2)}_{\nu_2} \, x^{(2)*}_{\mu_2} \\
	&\; \times \exp\left\{ \I\, X^\dagger L \left( \bar z + \tfrac{\Delta_z}{2} \Lambda - H \right) X \right\} .
\end{aligned}
\end{equation}
To compute an explicit expression for the average of this equation, we then follow the 
recipe outlined at the end of Sec.~\ref{sec:OverlapsFromSupersymmetry:Outline}.

\paragraph{Ensemble average and symmetries.}
In the first step, we integrate Eq.~\eqref{eq:RsvRAFromSuperGaussInt} over the ensemble of perturbations specified by the distribution $P(V)$ from Eqs.~\eqref{eq:PDFH} and~\eqref{eq:PDFH:Entries:Normal}.
Similarly as in Sec.~\ref{sec:OverlapsFromSupersymmetry:2ndMoment}, this leads to
\begin{widetext}
\begin{equation}
\label{eq:AvgRsvRAAfterH}
\begin{aligned}
	&\av{ \RsvG^0_{\nu_1\mu_1}(z_1^+) \, \RsvG^0_{\nu_2\mu_2}(z_2^-) } \\
		&\; = \int \frac{[\d X \d X^{*}]}{(2\pi\I)^{2N}} \, x^{(1)}_{\nu_1} \, x^{(1)*}_{\mu_1} \, x^{(2)}_{\nu_2} \, x^{(2)*}_{\mu_2}  
		\exp\left\{ -\tfrac{\lambda^2}{2} \sum\nolimits_{\alpha, \beta} \str\left( X_\alpha X^\dagger_\alpha L X_\beta X^\dagger_\beta L \right)
			  + \I \sum\nolimits_\alpha X^\dagger_\alpha L \left( \bar z + \Delta_z \Lambda / 2 - E^0_\alpha \right) X_\alpha \right\} .
\end{aligned}
\end{equation}
\end{widetext}
Before invoking the Hubbard-Stratonovich transformation, we inspect the symmetries of the integrand in this equation.
To this end, we need to consider the relative location of the considered eigenstates $\ket{n_1}$ and $\ket{n_2}$ with respect to each other, quantified by the parameter $\Delta_z$.
On the one hand, if the energy difference is on the order of the mean level spacing, $\Delta_z \sim \lvsp \ll \rateExp$, the two eigenvectors belong to close-by levels in the spectrum,
corresponding to the regime where significant correlations due to the orthonormality constraint are expected.
As $\Delta_z$ is small compared to the typical scale $\rateExp$ of eigenvector correlations in this case, we can neglect the term to leading order and the integrand in Eq.~\eqref{eq:AvgRsvRAAfterH} becomes invariant under pseudounitary transformations $X \mapsto TX$, $X^\dagger \mapsto X^\dagger T^\dagger$ with $T^\dagger L T = L$.
Hence the integrand has an approximate pseudounitary symmetry for small $\Delta_z$.

On the other hand, if $\Delta_z \gg \rateExp$, the $\Delta_z$ term is not negligible and the approximate symmetry breaks down.
In this case, the approximation collapses to the result that would be obtained if we treated the eigenvectors as independent random variables, see also Sec.~\ref{s4}.
In the intermediate regime, the situation is much more subtle and will be discussed in more detail below, after we have derived a result for small $\Delta_z$.

\paragraph{Hubbard-Stratonovich transformation.}
The Hubbard-Stratonovich identity applicable to transform Eq.~\eqref{eq:AvgRsvRAAfterH} looks similar to Eq.~\eqref{eq:HS} for the second moment,
\begin{equation}
\label{eq:HS4}
\begin{aligned}
	&\int \frac{\d R}{(2\pi)^2} \exp\left\{ - \str\left( \frac{R^2}{2 \ptst^2} \right) + \I \sum\nolimits_\alpha \str \left( R X_\alpha X_\alpha^\dagger L \right) \right\} \\
	&\quad = \exp\left\{ -\sum\nolimits_{\alpha, \beta} \frac{\lambda^2 }{2} \str\left( X_\alpha X^\dagger_\alpha L X_\beta X^\dagger_\beta L \right) \right\} \,,
\end{aligned}
\end{equation}
except that a single $(4\times 4)$ supermatrix $R$ suffices due to the constant variance $(\sigma_{\mu\nu})^2 = 1$.
It is conveniently parameterized as \cite{zuk96, mir00, haa10}
\begin{equation}
	R = T \lmat P_1 - \I \delta_0 & 0 \\
		        0 & P_2 + \I \delta_0 \rmat T^{-1} .
\end{equation}
The $(2\times 2)$ supermatrices $P_1$ and $P_2$ are Hermitian, and the parameter $\delta_0 > 0$ will be adapted so that the integration contour passes through the saddle points \cite{mir00}.
The block-diagonalizing transformation matrix $T$ satisfies $T^\dagger L T = L$ and thus belongs to the (approximate) pseudounitary symmetry group of the integrand in~\eqref{eq:AvgRsvRAAfterH}.
Applying the transformation~\eqref{eq:HS4} to the integral~\eqref{eq:AvgRsvRAAfterH}, we obtain
\begin{widetext}
\begin{equation}
\label{eq:AvgRsvRAAfterHS}
\begin{aligned}
	&\av{ \RsvG^0_{\nu_1\mu_1}(z_1^+) \, \RsvG^0_{\nu_2\mu_2}(z_2^-) }
%	\\ &\;
		= \int \frac{\d R}{(2\pi)^2} \frac{[\d X \d X^{*}]}{ (2\pi\I)^{2N}} \, x^{(1)}_{\nu_1} \, x^{(1)*}_{\mu_1} \, x^{(2)}_{\nu_2} \, x^{(2)*}_{\mu_2}  
		\exp\left\{ -\str( \tfrac{R^2}{2 \ptst^2}) + \I \sum\nolimits_\alpha X^\dagger_\alpha L \left( R + \bar z + \tfrac{\Delta_z}{2} \Lambda - E^0_\alpha \right) X_\alpha \right\} .
\end{aligned}
\end{equation}
Using the Isserlis/Wick theorem, we can then perform the Gaussian integration over $X$, leading to
\begin{equation}
\label{eq:AvgRsvRAAfterX}
\begin{aligned}
	\av{ \RsvG^0_{\nu_1\mu_1}(z_1^+) \, \RsvG^0_{\nu_2\mu_2}(z_2^-) } 
		&= \int \frac{\d R}{(2\pi)^2} \exp\left\{ -\str \left[ \frac{R^2}{2 \ptst^2} + \sum\nolimits_\alpha \ln \left( R + \bar z + \Delta_z \Lambda / 2 - E^0_\alpha \right) \right] \right\} \\
		&\quad	\times \left\{ - \delta_{\mu_1 \nu_1} \delta_{\mu_2 \nu_2}  \left[ \left( R + \bar z + \Delta_z \Lambda / 2 - E^0_{\nu_1} \right)^{-1} \right]_{1 \bos, 1 \bos} \left[ \left( R + \bar z + \Delta_z \Lambda / 2 - E^0_{\nu_2} \right)^{-1} \right]_{2 \bos, 2 \bos} \right. \\
		&\quad\qquad\left.		- \delta_{\mu_1 \nu_2} \delta_{\mu_2 \nu_1} \left[ \left( R + \bar z + \Delta_z \Lambda / 2 - E^0_{\nu_1} \right)^{-1} \right]_{1 \bos, 2 \bos} \left[ \left( R + \bar z + \Delta_z \Lambda / 2 - E^0_{\nu_2} \right)^{-1} \right]_{2 \bos, 1 \bos} \right\} .
\end{aligned}
\end{equation}
\end{widetext}
As before, the indices $1$ and $2$ for the supermatrix elements refer to the retarded and advanced components (corresponding to $z_1^+$ and $z_2^-$), and $\bos$ and $\fer$ denote the bosonic and fermionic sectors of superspace, respectively.

\paragraph{Saddle-point approximation.}
To evaluate the remaining superintegral over $R$, we employ a saddle-point approximation again.
Upon variation of the exponent in~\eqref{eq:AvgRsvRAAfterX}, the resulting saddle-point equation reads
\begin{equation}
\label{eq:U4SaddleEq}
	R  + \ptst^2 \sum_\alpha \left( R + \bar z + \Delta_z \Lambda/2 - E^0_\alpha \right)^{-1} = 0 \,.
\end{equation}
Following the calculation to solve Eq.~\eqref{eq:U2SaddleEq} for the second moment, we readily conclude that a diagonal solution of~\eqref{eq:U4SaddleEq} is given by
\begin{equation}
\label{eq:U4SaddleDiag}
	R_* := \I \, \rateExp \Lambda / 2
\end{equation}
with $\rateExp = 2 \pi \ptst^2 / \lvsp$ as before.
The crucial difference to the calculation for the second moment is that this diagonal solution is no longer proportional to the unit matrix.
As observed below Eq.~\eqref{eq:AvgRsvRAAfterH}, the integral and hence also the saddle-point equation~\eqref{eq:U4SaddleEq} has a pseudounitary symmetry for $\Delta_z \sim \lvsp$, mediated by linear transformation matrices $T$ with $T^\dagger L T = L$.
Given the solution $R_*$ from Eq.~\eqref{eq:U4SaddleDiag}, all matrices $T R_* T^{-1}$ with pseudounitary $T$ are also (approximate) saddle points,
and they all contribute equally because the exponent of Eq.~\eqref{eq:AvgRsvRAAfterX} is invariant when $\Delta_z$ is negligible.
Therefore, we have to sum the contributions from all these saddles, i.e., we need to integrate over the symmetry group of transformation matrices $T$.

Evaluating the integrand at the saddle points is straightforward and works analogously to the calculation for the second moment.
Substituting $R = T R_* T^{-1}$ and introducing a new integration variable $Q := T \Lambda T^{-1}$, the remaining integral over the manifold of degenerate saddles then reads
\begin{widetext}
\begin{equation}
\label{eq:U4SaddleInt}
\begin{aligned}
	\av{ \RsvG^0_{\nu_1\mu_1}(z_1^+) \, \RsvG^0_{\nu_2\mu_2}(z_2^-) } 
		&\; = -\int \d\mu(Q) \; \exp\left[-  \str \sum\nolimits_\alpha \ln \left( \I\rateExp Q / 2 + \bar z + \Delta_z \Lambda/2 - E^0_\alpha \right) \right] \\
			&\;\qquad \times \left[ \delta_{\mu_1 \nu_1} \delta_{\mu_2 \nu_2} \left( \I\rateExp Q / 2+ \bar z + \Delta_z \Lambda/2 - E^0_{\nu_1} \right)^{-1}_{\;\;1 \bos, 1 \bos}  \left( \I\rateExp Q / 2 + \bar z + \Delta_z \Lambda/2 - E^0_{\nu_2} \right)^{-1}_{\;\;2 \bos, 2 \bos} \right. \\
			 &\;\qquad\quad \left. + \delta_{\mu_1 \nu_2} \delta_{\mu_2 \nu_1} \left( \I\rateExp Q / 2 + \bar z + \Delta_z \Lambda/2 - E^0_{\nu_1} \right)^{-1}_{\;\;1 \bos, 2 \bos} \left( \I\rateExp Q / 2 + \bar z + \Delta_z \Lambda/2 - E^0_{\nu_2} \right)^{-1}_{\;\;2 \bos, 1 \bos} \right] ,
\end{aligned}
\end{equation}
\end{widetext}
where $\d\mu(Q)$ denotes the integration measure, which we will specify in a moment after fixing a convenient parameterization for $Q$.
Note that we used $Q^2 = \id$ and hence $\str(Q^2) = 0$ in the exponent.

\paragraph{Setting up the saddle-point integral.}
We observe that $Q$ can be split into a product of matrices that are block-diagonal in either the retarded-advanced or the boson-fermion decomposition \cite{zir86, fyo94, haa10}.
To this end,
define $(2\times 2)$ supermatrices $A$ and $B$ as
\begin{equation}
\label{eq:PseudoUTBlockAB}
	A := \exp \lmat 0 & -\alpha^* \\ \alpha & 0 \rmat
	\quad \text{and} \quad
	B := \exp \lmat 0 & -\I\beta^* \\ \I\beta & 0 \rmat
\end{equation}
with anticommuting $\alpha$, $\alpha^*$, $\beta$, $\beta^*$,
so that $A$ is unitary ($A^\dagger A = \id$), while $B$ is pseudounitary ($B^\dagger k B = k$
with $k = \operatorname{diag}(1, -1)$).
With these and the diagonal matrix $\tilde\tau := \operatorname{diag}(\tau_\bos, \tau_\fer)$, the pseudounitary transformation matrix $T$ can be expressed as \cite{haa10, zir86, mir00}
\begin{equation}
\label{eq:PseudoUTBlockDiag}
	T = \lmat A & 0 \\ 0 & B \rmat
		\lmat \sqrt{1 + k \lvert \tilde\tau \rvert^2} & k \tilde\tau^* \\
			  \tilde\tau & \sqrt{1 + k \lvert \tilde\tau \rvert^2} \rmat
		\lmat A^{-1} & 0 \\ 0 & B^{-1} \rmat ,
\end{equation}
where the left and right matrices are block-diagonal in the boson-fermion decomposition, and the middle matrix is block-diagonal in the retarded-advanced decomposition.
Writing the bosonic and fermionic eigenvalues as $\tau_{\bos, \fer} = r_{\bos, \fer} \, \e^{\I\phi_{\bos,\fer}}$ with $r_\bos \in [0, \infty)$, $r_\fer \in [0, 1]$, $\phi_\bos, \phi_\fer \in [0, 2\pi)$, the measure $\d\mu(T)$ takes the form \cite{haa10}
\begin{equation}
	\d\mu(T) = \frac{\d r_\bos \, \d\phi_\bos \, r_\bos}{\pi} \frac{\d r_\fer \, \d\phi_\fer \, r_\fer}{\pi} \d\alpha \, \d\alpha^* \, \d(\I\beta) \, \d(\I\beta^*) \,.
\end{equation}
Substituting~\eqref{eq:PseudoUTBlockDiag} into the definition of $Q = T \Lambda T^{-1}$ and introducing new integration variables $\ell_\bos := 1 + 2 r_\bos^2 \in [0, \infty)$ and $\ell_\fer := 1 - 2 r_\fer^2 \in [-1, 1]$, we find
\begin{subequations}
\label{eq:PseudoUQ}
\begin{align}
\label{eq:PseudoUQ1}
	Q =& \lmat A & 0 \\ 0 & B \rmat
		\tilde Q
		\lmat A^{-1} & 0 \\ 0 & B^{-1} \rmat 
\end{align}
with
\begin{widetext}
\begin{align}
\label{eq:PseudoUQTilde}
	\tilde Q :=& \lmat \ell_\bos & 0 & -\sqrt{\ell_\bos^2 - 1} \, \e^{-\I \phi_\bos} & 0 \\
		      0 & \ell_\fer & 0 & \sqrt{1-\ell_\fer^2} \, \e^{-\I\phi_\fer} \\
		      \sqrt{\ell_\bos^2 - 1} \, \e^{\I\phi_\bos} & 0 & -\ell_\bos & 0 \\
		      0 & \sqrt{1 - \ell_\fer^2} \, \e^{\I\phi_\fer} & 0 & -\ell_\fer \rmat ,
\end{align}
\end{widetext}
\end{subequations}
and the measure takes the form
\begin{equation}
\label{eq:PseudoUQMeasure}
	\d\mu(Q) = - \frac{\d\ell_\bos \, \d\phi_\bos \, \d\ell_\fer \, \d\phi_\fer}{ (2\pi)^2 (\ell_\bos - \ell_\fer)^2 } \d\alpha \, \d\alpha^* \, \d\beta \, \d\beta^* \,.
\end{equation}
This completes the parameterization of the integrand in Eq.~\eqref{eq:U4SaddleInt}.
To evaluate the integral, we need explicit expressions for the $\bos\bos$-submatrix  of $(\I\rateExp Q/2 + \bar z + \Delta_z \Lambda / 2 - E^0_{\nu_i} )^{-1}$,
\begin{subequations}
\label{eq:U4SaddleIntRationals}
\begin{align}
	&\left(  \I\rateExp Q / 2 + \bar z + \Delta_z \Lambda / 2 - E^0_{\nu_i} \right)^{-1}_{\;\;1\bos, 1\bos} \notag \\
		&\;= K_{\nu_i}(\ell_\bos) (z_2^- - E^0_{\mu_i} - \I\rateExp \ell_\bos / 2) (1 + \alpha \alpha^*) \notag \\
		&\qquad - K_{\nu_i}(\ell_\fer) (z_2^- - E^0_{\nu_i} - \I\rateExp \ell_\fer / 2) \, \alpha \alpha^* \,, \displaybreak[0] \\
	&\left(  \I\rateExp Q / 2 + \bar z + \Delta_z \Lambda / 2 - E^0_{\nu_i} \right)^{-1}_{\;\;1\bos, 2\bos}\notag \\
		&\;= \I\rateExp K_{\nu_i}(\ell_\bos) \sqrt{\ell_\bos^2 - 1} \, \e^{-\I\phi_\bos} \left(\tfrac{1}{2} + \tfrac{\alpha \alpha^* - \beta\beta^*}{4} - \tfrac{\alpha\alpha^*\beta\beta^*}{8} \right) \notag \\
			&\qquad + \rateExp K_{\nu_i}(\ell_\fer) \sqrt{1-\ell_\fer^2} \, \e^{-\I \phi_\fer} \, \alpha^* \beta / 2 \,, \displaybreak[0] \\
	&\left(  \I\rateExp Q / 2+ \bar z + \Delta_z \Lambda / 2 - E^0_{\nu_i} \right)^{-1}_{\;\;2\bos, 1\bos}\notag \\
		&\;= -\I\rateExp K_{\nu_i}(\ell_\bos) \sqrt{\ell_\bos^2 - 1} \, \e^{\I\phi_\bos} \left(\tfrac{1}{2} + \tfrac{\alpha \alpha^* - \beta\beta^*}{4} - \tfrac{\alpha\alpha^*\beta\beta^*}{8} \right) \notag \\
			&\qquad -\rateExp K_{\nu_i}(\ell_\fer) \sqrt{1 - \ell_\fer^2} \, \e^{\I\phi_\fer} \, \alpha \beta^* / 2 \,, \displaybreak[0] \\
	&\left(  \I\rateExp Q / 2 + \bar z + \Delta_z \Lambda / 2 - E^0_{\nu_i} \right)^{-1}_{\;\;2\bos, 2\bos}\notag \\
		&\;= K_{\nu_i}(\ell_\bos) (z_1^+ - E^0_{\nu_i} + \I \rateExp \ell_\bos / 2) (1-\beta\beta^*) \notag \\
			&\qquad + K_{\nu_i}(\ell_\fer) (z_1^+ - E^0_{\mu_i} + \I\rateExp \ell_\fer / 2) \, \beta \beta^* \,,
\end{align}
\end{subequations}
where
\begin{equation}
	K_\nu(\ell) := \left[ (z_1^+ - E^0_\nu) (z_2^- - E^0_\nu) - \I \rateExp \Delta_z \, \ell / 2 + \rateExp^2 / 4 \right]^{-1} .
\end{equation}
For the exponent of the integrand in Eq.~\eqref{eq:U4SaddleInt}, we recall that the degeneracy of saddles occurs for $\Delta_z \sim \lvsp$, so that we can expand it to first order in $\Delta_z$.
Using the saddle-point equation~\eqref{eq:U4SaddleEq}, we obtain
\begin{equation}
\label{eq:U4SaddleIntExp}
\begin{aligned}
	&-\str \sum_\alpha \ln \left( \I\rateExp Q / 2 + \bar z + \Delta_z \Lambda / 2 - E^0_\alpha \right) \\
%		= \tfrac{\I\pi\Delta}{\lvsp} \str(Q\Lambda) + \mc O(\Delta^2)
		&\;= \I\pi\Delta_z (\ell_\bos - \ell_\fer) / \lvsp + \mc O(\Delta_z^2) \,.
\end{aligned}
\end{equation} 
Substituting the measure~\eqref{eq:PseudoUQMeasure}, the rational part~\eqref{eq:U4SaddleIntRationals} of the integrand and the exponent~\eqref{eq:U4SaddleIntExp}, the integral over the saddle-point manifold thus reads
\begin{widetext}
\begin{equation}
\label{eq:U4SaddleIntSubs}
\begin{aligned}
	\av{ \RsvG^0_{\nu_1\mu_1}(z_1^+) \, \RsvG^0_{\nu_2\mu_2}(z_2^-) } 
		&= \int_1^\infty \!\! \d\ell_\bos \int_{-1}^1 \!\! \d\ell_\fer \frac{1}{(\ell_\bos - \ell_\fer)^2} \int_0^{2\pi} \frac{\d\phi_\bos}{2\pi} \int_0^{2\pi} \frac{\d\phi_\fer}{2\pi} \int \d\alpha \, \d\alpha^* \, \d\beta \, \d\beta^* \, \exp\left[ \tfrac{\I\pi \Delta_z}{\lvsp} (\ell_\bos - \ell_\fer) \right] \\
		&\;\qquad \times \left[ \delta_{\mu_1 \nu_1} \delta_{\mu_2 \nu_2} \left( D_{00} + D_{11} + D_{10} + D_{01} \right)
				+ \delta_{\mu_1 \nu_2} \delta_{\mu_2 \nu_1} \left( F_{00} + F_{11} + F_{\mathrm S} + F_{\mathrm C} \right) \right]
\end{aligned}
\end{equation}
with
\begin{subequations}
\label{eq:U4SaddleIntSubsTerms}
\begin{align}
	D_{00} &:= K_{\nu_1}(\ell_\bos) K_{\nu_2}(\ell_\bos) (z_1^+ - E^0_{\nu_2} + \tfrac{\I\rateExp}{2} \ell_\bos) (z_2^- - E^0_{\nu_1} - \tfrac{\I\rateExp}{2} \ell_\bos) \,, 
	\displaybreak[0] \\
	D_{11} &:= -\alpha\alpha^*\beta\beta^* \left[ K_{\nu_1}(\ell_\bos) (z_2^- - E^0_{\mu_1} - \tfrac{\I\rateExp}{2} \ell_\bos) - K_{\nu_1}(\ell_\fer) (z_2^- - E^0_{\nu_1} - \tfrac{\I\rateExp}{2} \ell_\fer) \right] \notag \\
		&\qquad\qquad\quad \times \left[ K_{\nu_2}(\ell_\bos) (z_1^+ - E^0_{\nu_2} + \tfrac{\I\rateExp}{2} \ell_\bos) - K_{\nu_2}(\ell_\fer) (z_1^+ - E^0_{\nu_2} + \tfrac{\I\rateExp}{2} \ell_\fer) \right]  \,, 
	\displaybreak[0]\\
	D_{10} &:= \alpha\alpha^* \left[ K_{\nu_1}(\ell_\bos) (z_2^- - E^0_{\nu_1} - \tfrac{\I\rateExp}{2} \ell_\bos) - K_{\nu_1}(\ell_\fer) (z_2^- - E^0_{\nu_1} - \tfrac{\I\rateExp}{2} \ell_\fer) \right]
	 K_{\nu_2}(\ell_\bos) (z_1^+ - E^0_{\nu_2} + \tfrac{\I\rateExp}{2} \ell_\bos) \,, 
	\displaybreak[0]\\
	D_{01} &:= -\beta\beta^* K_{\nu_1}(\ell_\bos) (z_2^- - E^0_{\nu_2} + \tfrac{\I\rateExp}{2} \ell_\bos) 
	 \left[ K_{\nu_2}(\ell_\bos) (z_1^+ - E^0_{\nu_2} + \tfrac{\I\rateExp}{2} \ell_\bos) - K_{\nu_2}(\ell_\fer) (z_1^+ - E^0_{\mu_2} + \tfrac{\I\rateExp}{2} \ell_\fer) \right] \,, 
	\displaybreak[0]\\
	F_{00} &:= \tfrac{\rateExp^2 }{4} \left[ K_{\nu_1}(\ell_\bos) K_{\nu_2}(\ell_\bos) (\ell_\bos^2 - 1) - K_{\nu_1}(\ell_\fer) K_{\nu_2}(\ell_\fer) (1- \ell_\fer^2) \right]  \,, 
	\displaybreak[0]\\
	F_{11} &:= -\alpha\alpha^*\beta\beta^* \frac{\rateExp^2}{4} \left[ K_{\nu_1}(\ell_\bos) K_{\nu_2}(\ell_\bos) (\ell_\bos^2 - 1) + K_{\nu_1}(\ell_\fer) K_{\nu_2}(\ell_\fer) (1 - \ell_\fer^2) \right] \,, 
	\displaybreak[0]\\
	F_{\mathrm S} &:= (\alpha\alpha^* - \beta\beta^*) \frac{\rateExp^2}{4} K_{\nu_1}(\ell_\bos) K_{\nu_2}(\ell_\bos) \,, 
	\displaybreak[0]\\
	F_{\mathrm C} &:= -\tfrac{\I \rateExp^3 }{8} \sqrt{\ell_\bos^2-1} \sqrt{1-\ell_\fer^2} \left[ \alpha \beta^* \, K_{\nu_1}(\ell_\bos) K_{\nu_2}(\ell_\fer) \, \e^{-\I(\phi_\bos - \phi_\fer)} 
	 + \alpha^* \beta \, K_{\nu_1}(\ell_\fer) K_{\nu_2}(\ell_\bos) \, \e^{\I (\phi_\bos - \phi_\fer)} \right] \,.
\end{align}
\end{subequations}
\end{widetext}
Hence we sorted the terms by their dependence on the Grassmann variables.

\paragraph{Evaluating the saddle-point integral.}
To calculate the integral~\eqref{eq:U4SaddleIntSubs}, we analyze the various terms~\eqref{eq:U4SaddleIntSubsTerms} individually.
Naively, the Grassmann integral over all terms that do not contain the full set $\alpha\alpha^* \beta\beta^*$ of anticommuting generators vanishes.
This would imply that the only contributions come from the $D_{11}$ and $F_{11}$ terms.
However, we observe that the integration measure is singular in the commuting variables at the boundary of the domain of integration where $\ell_\bos = \ell_\fer = 1$, corresponding to the ``origin'' $Q = \Lambda$, see Eq.~\eqref{eq:PseudoUQ}.
This singularity potentially leads to diverging integrals over the commuting variables, and the total integral of ``$0 \cdot \infty$'' type has to be dealt with using the Parisi-Sourlas-Efetov-Wegner (PSEW) theorem \cite{par79, mck80, efe83, weg83, zir86, con89, haa10},
telling us that
the value of the integral is given by the value of the integrand at the origin, i.e.\ for the ``superrotation'' matrix $T$ replaced by the identity.
For our current problem, this means that $Q = \Lambda$ or, equivalently, $\ell_\bos = \ell_\fer = 1$.

Inspecting the terms~\eqref{eq:U4SaddleIntSubsTerms}, the conditions of the PSEW theorem are met by the integrals of $D_{00}$ and $F_{00}$.
Consequently, we obtain
\begin{equation}
\label{eq:U4SaddleInt:D00}
	\int D_{00} = D_{00} |_{\ell_\bos = \ell_\fer = 1} = \frac{1}{ \left( z_1^+ \!-\! E^0_{\nu_1} \!+\! \tfrac{\I\rateExp}{2} \right) \left( z_2^- \!-\! E^0_{\nu_2} \!-\! \tfrac{\I\rateExp}{2} \right)}
\end{equation}
and
$\int F_{00} = 0$.
For the integrals of $D_{10}$ and $D_{01}$, the singularity in the measure is lifted by an additional factor $(\ell_\bos - \ell_\fer)$ in the integrand, rendering the commuting integrals convergent.
Due to the incomplete set of Grassmann variables in these terms, the total integral thus vanishes,
$\int D_{10} = \int D_{01} = 0$.
Similarly, the bosonic integral of $F_{\mathrm C}$ converges, so that
$\int F_{\mathrm C} = 0$.
For the integral of $F_{\mathrm S}$, we note that the terms proportional to $\alpha\alpha^*$ and $\beta\beta^*$, respectively, come with opposite signs, but are otherwise symmetric.
Hence this integral has to vanish, too,
$\int F_{\mathrm S} = 0$.
Finally, we are left with the integrals of $D_{11}$ and $F_{11}$.
The fermionic integral is evaluated straightforwardly, giving $\int \d\alpha \d\alpha^* \d\beta \d\beta^* \, \alpha\alpha^*\beta\beta^* = 1$.
Furthermore, the angular integrals over $\phi_\bos$ and $\phi_\fer$ both yield factors of $2\pi$ compensating the corresponding factors in the measure.
For the remaining integrals over $\ell_\bos$ and $\ell_\fer$, the following identities \cite{gra07, erd54} turn out useful:
 \begin{equation}
\label{eq:EiId:I}
	\mc I(v,c) := \int_{1}^\infty \d\ell \; \frac{\e^{\I v \ell}}{\ell + c} = - \e^{-\I v c} \, \Ei(\I v [c+1]) 
\end{equation}
if $\mathrm{Im}(v) > 0, \, \lvert \arg(1+c) \rvert < \pi$, and
\begin{equation}
\label{eq:EiId:J}
\begin{aligned}
	\mc J(v,c) :=& \int_{-1}^1 \d\ell \; \frac{\e^{-\I v \ell}}{\ell + c} \\
	=& \e^{\I v c} \left[ \Ei(-\I v [c+1]) - \Ei(-\I v[c-1]) \right]
\end{aligned}
\end{equation}
if $c \notin [-1, 1]$.
Here $\Ei(x)$ denotes the exponential integral 
function \cite{gra07},
\begin{equation}
\label{eq:Ei}
	\Ei(z) := - \mathrm{PV} \int_{-z}^\infty \!\! \d t \, \frac{\e^{-t}}{t} \,.
\end{equation}
The integrals of $D_{11}$ and $F_{11}$ can then be computed by decomposing the integrand into partial fractions in $\ell_\bos$ and $\ell_\fer$ and using identities~\eqref{eq:EiId:I} and~\eqref{eq:EiId:J} as well as derivatives thereof.
Denoting $a_{ij} := z_i - E^0_{\nu_j}$ and $c_k := 2\I a_{1 k} a_{2 k} / \rateExp \Delta_z + \I \rateExp / 2 \Delta_z$, we obtain
\begin{widetext}
\begin{equation}
\label{eq:U4SaddleInt:D11}
\begin{aligned}
&\int D_{11}
	= \frac{ (a_{11} a_{21} - a_{21} \Delta_z + \tfrac{\rateExp^2}{4}) ( a_{12} a_{22} + a_{12} \Delta_z + \tfrac{\rateExp^2}{4}) }{ (a_{11} a_{21} - a_{12} a_{22})^2 \Delta_z^2 } 
	\left[ \mc I(\tfrac{\pi\Delta_z}{\lvsp}, c_1) - \mc I(\tfrac{\pi\Delta_z}{\lvsp}, c_2) \right]  \left[ \mc J(\tfrac{\pi\Delta_z}{\lvsp}, c_1) - \mc J(\tfrac{\pi\Delta_z}{\lvsp}, c_2) \right]
\end{aligned}
\end{equation}
and
\begin{equation}
\label{eq:U4SaddleInt:F11}
\begin{aligned}
	\int F_{11}
		&= \frac{1}{(E^0_{\nu_1} - E^0_{\nu_2}) \Delta_z (E^0_{\nu_1} + E^0_{\nu_2} - z_1^+ - z_2^-)} \\
		&	\quad \times \left\{ \frac{ (a_{11} - \tfrac{\I\rateExp}{2}) (a_{21} + \tfrac{\I\rateExp}{2}) \, \mc J( \tfrac{\pi\Delta_z}{\lvsp}, c_1)}{\Delta_z} \left[ \frac{ 2 \pi (a_{11} + \tfrac{\I\rateExp}{2})(a_{21} -\tfrac{\I\rateExp}{2}) \, \mc I( \tfrac{\pi\Delta_z}{\lvsp}, c_1) }{ \rateExp \lvsp } - \e^{\I \pi \Delta_z / \lvsp} \right]  \right. \\
		& 	\qquad - \left. \frac{ (a_{12} - \tfrac{\I\rateExp}{2}) (a_{22} + \tfrac{\I\rateExp}{2}) \, \mc J( \tfrac{\pi\Delta_z}{\lvsp}, c_2)}{\Delta_z} \left[ \frac{ 2 \pi (a_{12} + \tfrac{\I\rateExp}{2})(a_{22} - \tfrac{\I\rateExp}{2}) \, \mc I( \tfrac{\pi\Delta_z}{\lvsp}, c_2) }{ \rateExp \lvsp } - \e^{\I \pi \Delta_z / \lvsp} \right]  \right. \\
		&	\qquad \left. - \I \rateExp \, \cos(\tfrac{\pi\Delta_z}{\lvsp}) \left[ \mc I( \tfrac{\pi\Delta_z}{\lvsp}, c_1) - \mc I( \tfrac{\pi\Delta_z}{\lvsp}, c_2) \right] + \rateExp \, \sin(\tfrac{\pi\Delta_z}{\lvsp}) \left[ c_1 \, \mc I( \tfrac{\pi\Delta_z}{\lvsp}, c_1) - c_2 \mc I( \tfrac{\pi\Delta_z}{\lvsp}, c_2) \right]
		\vphantom{\frac{ (a_{11} - \tfrac{\I\rateExp}{2}) (a_{21} + \tfrac{\I\rateExp}{2}) \, \mc J( \tfrac{\pi\Delta_z}{\lvsp}, c_1)}{\Delta_z}} \right\}
\end{aligned}
\end{equation}
\end{widetext}
To simplify these expressions further, we observe that the arguments entering the exponential integral functions $\mc I$ and $\mc J$ in both expressions depend on the large, real parameter $\rateExp / \lvsp = 2 / \pi s$ [cf.~Sec.~\ref{sec:RandMatEns:SmallParameter}].
Using the asymptotic expansion $\Ei(z) \sim \e^z / z$ as $\mathrm{Re}(z) \to \pm\infty$ (with a branch cut discontinuity of $2 \pi$ in the imaginary part along the negative real line), we can approximate the integrals as
\begin{equation}
\label{eq:U4SaddleInt:D11Approx}
\begin{aligned}
	&\int D_{11} \\ 
	&\;\approx \left(\frac{\rateExp \lvsp}{2\pi \Delta_z} \right)^2 \left[ \frac{ 
	\vphantom{\exp \left( \frac{2 \pi \I \Delta_z}{\lvsp} \right)}
	 (a_{12} - \tfrac{\I\rateExp}{2}) (a_{21} + \tfrac{\I\rateExp}{2}) }{ (a_{11} + \tfrac{\I\rateExp}{2})^2 (a_{12} + \tfrac{\I\rateExp}{2}) (a_{21} + \tfrac{\I\rateExp}{2}) (a_{22} - \tfrac{\I\rateExp}{2})^2 } \right. \\
			&\qquad\qquad\qquad \left. - \frac{ \exp \left( \frac{2 \pi \I \Delta_z}{\lvsp} \right) }{ (a_{11}^2 + \rateExp^2/4) (a_{22}^2 + \rateExp^2/4) } \right] \,.
\end{aligned}
\end{equation}
and
\begin{equation}
\label{eq:U4SaddleInt:F11Approx}
	\int F_{11} \approx 
	 \frac{\I\rateExp\lvsp / \pi\Delta_z}{ (a_{11} + \tfrac{\I\rateExp}{2}) (a_{12} + \tfrac{\I\rateExp}{2}) (a_{21} - \tfrac{\I\rateExp}{2}) (a_{22} - \tfrac{\I\rateExp}{2}) } .
\end{equation}

\paragraph{Collecting terms.}
The nonvanishing contributions to the averaged retarded-advanced resolvent $\av{ \RsvG^0_{\nu_1 \mu_1}(z_1^+) \, \RsvG^0_{\nu_2 \mu_2}(z_2^-) }$ as given in Eq.~\eqref{eq:U4SaddleIntSubs} are Eqs.~\eqref{eq:U4SaddleInt:D00}, \eqref{eq:U4SaddleInt:D11Approx}, and~\eqref{eq:U4SaddleInt:F11Approx}.
For the fourth moment of eigenvector overlaps, we also need the averaged advanced-retarded resolvent $\av{ \RsvG^0_{\nu_1 \mu_1}(z_1^-) \, \RsvG^0_{\nu_2 \mu_2}(z_2^+) }$ [see Eqs.~\eqref{eq:U4SameFromRsv} and~\eqref{eq:U4DistFromRsv}].
The contributions to this term are obtained from those of the retarded-advanced resolvent by exchanging all indices 1 and 2, i.e., $n_1 \leftrightarrow n_2$, $\mu_1 \leftrightarrow \mu_2$, and $\nu_1 \leftrightarrow \nu_2$.
Combining the two and letting $\eta\to 0$, we then find
\begin{widetext}
\begin{equation}
\label{eq:AvgRsvRAandAR}
\begin{aligned}
	& \lim_{\eta\to 0+} \av{ \RsvG^0_{\nu_1\mu_1}(z_1^+) \, \RsvG^0_{\nu_2\mu_2}(z_2^-) + \RsvG^0_{\nu_1\mu_1}(z_1^-) \, \RsvG^0_{\nu_2\mu_2}(z_2^+) } \\
		&\;= \delta_{\mu_1 \nu_1} \delta_{\mu_2 \nu_2} \left\{  \tfrac{1}{\left( E_{n_1} - E^0_{\nu_1} + \I\rateExp / 2 \right) \left( E_{n_2} - E^0_{\nu_2} - \I\rateExp / 2 \right) } 
			 + \tfrac{1}{ \left( E_{n_1} - E^0_{\nu_1} - \I\rateExp / 2 \right) \left( E_{n_2} - E^0_{\nu_2} + \I\rateExp / 2 \right) }  \right. \\
			&\;\qquad\qquad\qquad \left. + \tfrac{\rateExp^2}{ \left[ (E_{n_1} - E^0_{\nu_1})^2 + \rateExp^2 / 4 \right] \left[ (E_{n_2} - E^0_{\nu_2})^2 + \rateExp^2 / 4 \right] } \right.  \\
			&\;\qquad\qquad\qquad\qquad \left. \times
				 \left[ \operatorname{sinc}^2\left( \tfrac{\pi\Delta_z}{\lvsp} \right) 
				  - \left( \tfrac{\rateExp \lvsp}{2 \pi} \right)^2 \tfrac{ \left[ (E^0_{\nu_1})^2 + (E^0_{\nu_2})^2 + 2 E_{n_1} E_{n_2} - (E^0_{\nu_1} + E^0_{\nu_2}) (E_{n_1} + E_{n_2}) + \rateExp^2 / 2 \right]^2 }{ \left[ (E_{n_1}\!-\!E^0_{\nu_1})^2\!+\!\rateExp^2/4 \right] \left[ (E_{n_1}\!-\!E^0_{\nu_2})^2\!+\!\rateExp^2/4 \right]\left[ (E_{n_2}\!-\!E^0_{\nu_1})^2\!+\!\rateExp^2/4 \right]\left[ (E_{n_2}\!-\!E^0_{\nu_2})^2\!+\!\rateExp^2/4 \right] } \right]
		\right\} \\
		&\;\;\; + \delta_{\mu_1 \nu_2} \delta_{\mu_2 \nu_1} \tfrac{\rateExp^2 \lvsp}{\pi \left[ (E_{n_1} - E^0_{\nu_2})^2 + \rateExp^2/4 \right] \left[ (E_{n_2} - E^0_{\nu_1})^2 + \rateExp^2/4 \right]}  \\
		&\;\qquad \times \left[ \tfrac{\rateExp / 2}{(E_{n_1} - E^0_{\nu_1})^2 + \rateExp^2 / 4 } \left( 1 - \tfrac{E_{n_1} \!-\! E_{n_2}}{E^0_{\nu_1} \!+\! E^0_{\nu_2} \!-\! E_{n_1} \!-\! E_{n_2} } \right) 
			 + \tfrac{\rateExp / 2}{(E_{n_2} - E^0_{\nu_2})^2 + \rateExp^2/4 } \left( 1 + \tfrac{E_{n_1} \!-\! E_{n_2}}{E^0_{\nu_1} \!+\! E^0_{\nu_2} \!-\! E_{n_1} \!-\! E_{n_2} } \right) \right] ,
\end{aligned}
\end{equation}
where $\mathrm{sinc}(x) := (\sin x) / x$.
Using $(E_n-E_m) \simeq (E^0_n - E^0_m) \simeq (n-m)\lvsp$ [cf.\ discussion below~(\mref{1})] and combining Eqs.~\eqref{eq:AvgRsv}, \eqref{eq:AvgRsvRRDist} and~\eqref{eq:AvgRsvRAandAR} according to Eq.~\eqref{eq:U4DistFromRsv}, we obtain a first estimate for the fourth moment of the overlaps involving two distinct eigenvectors $\ket{n_1}$ and $\ket{n_2}$,
\begin{equation}
\label{eq:AvgU4DistNaive}
\begin{aligned}
	&\left. \av{ U_{n_1 \mu_1} U_{n_2 \mu_2} U^*_{n_1 \nu_1} U^*_{n_2 \nu_2} } \right|_{n_1 \neq n_2} \\
		&\;= \delta_{\mu_1 \nu_1} \delta_{\mu_2 \nu_2} \, \ldos(n_1 - \mu_1) \, \ldos(n_2 - \mu_2) 
			 \left\{ 1 + \mathrm{sinc}^2\left( \pi [n_1 - n_2] \right) 
			 	- \ldos(n_1 - \mu_1) \, \ldos(n_2 - \mu_1) \, \ldos(n_1 - \mu_2) \, \ldos(n_2 - \mu_2) 
			 	\vphantom{ \left[ (\mu_1\!\!-\!\!\mu_2)^2 \right]^2 } \right.  \\
			&\;\qquad\qquad\qquad\qquad\qquad\qquad\qquad\qquad \left.  \times \left(\tfrac{\pi \lvsp}
			{\rateExp} \right)^2 \left[ (\mu_1\!\!+\!\!\mu_2\!\!-\!\!n_1\!\!-\!\!n_2)^2\!+\!(\mu_1\!\!-\!\!\mu_2)^2\!-\!(n_1\!\!-\!\!n_2)^2\!+\! (\rateExp/\lvsp)^2 \right]^2 \right\} \\
		&\;\;\;- \delta_{\mu_1 \nu_2} \delta_{\mu_2 \nu_1} \left( \tfrac{\rateExp}{4\pi\lvsp} \right) \ldos(n_1 - \mu_1) \, \ldos(n_2 - \mu_2) 
			 \tfrac{ (\rateExp/\lvsp)^2 + \mu_1^2/2 +  \mu_2^2 /2 + n_1 n_2 - ( \mu_1 + \mu_2) ( n_1 + n_2 )/2 }{ \left[ (n_1 - \mu_2)^2 + (\rateExp/2\lvsp)^2 \right] \left[ (n_2 - \mu_1)^2 + (\rateExp/2\lvsp)^2 \right] } .
\end{aligned}
\end{equation}
Similarly, we combine Eqs.~\eqref{eq:AvgRsvRRSame} and~\eqref{eq:AvgRsvRAandAR} according to~\eqref{eq:U4SameFromRsv} to obtain an estimate of the fourth overlap moment of a single perturbed eigenvector $\ket{n}$,
\begin{equation}
\label{eq:AvgU4SameNaive}
\begin{aligned}
	\av{ U_{n \mu_1} U_{n \mu_2} U^*_{n \nu_1} U^*_{n \nu_2} } 
		&= \left( \delta_{\mu_1 \nu_1} \delta_{\mu_2 \nu_2} + \delta_{\mu_1 \nu_2} \delta_{\mu_2 \nu_1} \right)  \ldos(n - \mu_1) \, \ldos(n - \mu_2) 
			\left\{ \vphantom{\left( \tfrac{\pi}{\rateExp \lvsp} \right)^2} 1 + \tfrac{1}{2}\left[ \ldos(n - \mu_1) + \ldos(n - \mu_2) \right] \right. \\
			&\qquad\qquad\qquad\qquad \left. - \ldos(n - \mu_1)^2 \, \ldos(n - \mu_2)^2 \left( \tfrac{\pi \lvsp}{\rateExp} \right)^2 \left[ \tfrac{1}{2} \mu_1^2\!+\!\tfrac{1}{2} \mu_2^2 \!+\! n^{2}\!-\!(\mu_1\!+\!\mu_2) n\!+\!(\rateExp/\lvsp)^2 \right]^2 \right\} .
\end{aligned}
\end{equation}
\end{widetext}
Next, we can combine Eqs.~\eqref{eq:AvgU4DistNaive} and~\eqref{eq:AvgU4SameNaive} and write $\av{ U_{n_1 \mu_1} U_{n_2 \mu_2} U^*_{n_1 \nu_1} U^*_{n_2 \nu_2} }= (1 - \delta_{n_1 n_2}) \left. \av{ U_{n_1 \mu_1} U_{n_2 \mu_2} U^*_{n_1 \nu_1} U^*_{n_2 \nu_2} } \right|_{n_1 \neq n_2} + \delta_{n_1 n_2} \av{ U_{n_1 \mu_1} U_{n_1 \mu_2} U^*_{n_1 \nu_1} U^*_{n_1 \nu_2} }$.
In the following, we restrict ourselves to the leading-order contribution in the parameter $s = \pi \lvsp / 2 \rateExp$ from~\eqref{x10} or, equivalently, $\delta = \lvsp / \rateExp$ from~\eqref{eq:delta},
meaning that we
keep only terms up to order $s^3$.

We consider the terms in Eq.~\eqref{eq:AvgU4DistNaive} for $n_1 \neq n_2$ first.
The function $u(n)$ is of order $\ldos(n) \sim s$, see~(\ref{eq:u:BW}).
The $\mathrm{sinc}$ function acts as a Kronecker-$\delta$ and thus does not contribute for $n_1 \neq n_2$.
Keeping only terms up to order $s^3$, we then find
\begin{equation}
\begin{aligned}
	&\left. \av{ U_{n_1 \mu_1} U_{n_2 \mu_2} U^*_{n_1 \nu_1} U^*_{n_2 \nu_2} } \right|_{n_1 \neq n_2} \\
	&\;= \delta_{\mu_1 \nu_1} \delta_{\mu_2 \nu_2} \, d^{n_1 n_2}_{\mu_1 \mu_2} + \delta_{\mu_1 \nu_2} \delta_{\mu_2 \nu_1} \, \tilde f^{n_1 n_2}_{\mu_1 \mu_2} + \mc O(s^4)
\end{aligned}
\end{equation}
with
\begin{align}
	d^{mn}_{\mu_1 \mu_2} &:= \ldos(n_1 - \mu_1) \, \ldos(n_2 - \mu_2) \,, \displaybreak[0] \\
	\label{eq:AvgU4Naive:f}
	\tilde f^{mn}_{\mu_1 \mu_2} &:= - \left( \tfrac{\rateExp}{4\pi\lvsp} \right)  \ldos(n_1 - \mu_1) \, \ldos(n_2 - \mu_2) \notag \\
		&\quad \times \tfrac{ (\rateExp/\lvsp)^2 + \mu_1^2-2 + \mu_2^2 / 2 + n_1 n_2 - ( \mu_1 + \mu_2) ( n_1 + n_2 ) /2 }{ \left[ (n_1 - \mu_2)^2 + (\rateExp/2\lvsp)^2 \right] \left[ (n_2 - \mu_1)^2 + (\rateExp/2\lvsp)^2 \right] } .
\end{align}
Note that $d^{n_1 n_2}_{\mu_1 \mu_2}$ is already the same symbol as defined 
in~\eqref{eq:AvgU4:d} above.
For the single eigenvector correlations, we observe that they carry a prefactor $\delta_{n_1 n_2}$ in the combined expression.
This effectively decreases the order of these terms by a factor of $s$ because we eventually sum over all perturbed eigenvectors $\ket{n_1}$ and $\ket{n_2}$ [see, e.g., Eq.~(\mref{6})],
but the double sum reduces to a single sum for the single eigenvector correlations.
Hence we only keep terms of order $s^2$ in Eq.~\eqref{eq:AvgU4SameNaive}, so that
\begin{equation}
\begin{aligned}
	&\av{ U_{n \mu_1} U_{n \mu_2} U^*_{n \nu_1} U^*_{n \nu_2} } \\
	&\;= (\delta_{\mu_1 \nu_1} \delta_{\mu_2 \nu_2} + \delta_{\mu_1 \nu_2} \delta_{\mu_2 \nu_1}) d^{nn}_{\mu_1 \mu_2} + \mc O(s^3) .
\end{aligned}
\end{equation}
Altogether, we then obtain
\begin{equation}
\label{eq:AvgU4Naive}
\begin{aligned}
	&\av{ U_{n_1 \mu_1} U_{n_2 \mu_2} U^*_{n_1 \nu_1} U^*_{n_2 \nu_2} } \\
	&\;\simeq \delta_{\mu_1 \nu_1} \delta_{\mu_2 \nu_2} \, d^{n_1 n_2}_{\mu_1 \mu_2}
			+ \delta_{\mu_1 \nu_2} \delta_{\mu_2 \nu_1} \left( \delta_{n_1 n_2} \, d^{n_1 n_2}_{\mu_1 \mu_2} + \tilde f^{n_1 n_2}_{\mu_1 \mu_2} \right) .
\end{aligned}
\end{equation}

\paragraph{Symmetry restoration.}
Exchanging indices and/or approximating sums over the spectrum as integrals as before, one readily verifies that this result satisfies the Properties~\ref{lst:AvgU4Swap12}, \ref{lst:AvgU4SwapMuNu} and~\ref{lst:AvgU4TraceH} of the fourth moment collected in the beginning of this subsection below Eq.~\eqref{eq:AvgU4Raw}.
However, Property~\ref{lst:AvgU4TraceH0} is violated:
Substituting~\eqref{eq:AvgU4Naive} into~\eqref{eq:AvgU4TraceH0}, we get
%\begin{widetext}
\begin{equation}
\label{eq:AvgU4NaiveVioTraceH0}
\begin{aligned}
	&\sum_{\mu_1, \nu_2} \delta_{\mu_1 \nu_2} \, \av{ U_{n_1 \mu_1} U_{n_2 \mu_2} U^*_{n_1 \nu_1} U^*_{n_2 \nu_2} } 
	\\
		&= \delta_{\mu_2 \nu_1} \left[ \vphantom{\tfrac{n_1}{n_1}} \delta_{n_1 n_2} \, \ldos(n_1 - \mu_2) 
\right. \\ & \left.	\quad	 
		 + \ldosTilde(n_1 - n_2) \, \ldos(n_1 - \mu_2) \left( \tfrac{ (n_1 - n_2) (n_1 - 3 n_2 + 2 \mu_2) }{ 2 (n_2 - \mu_2)^2 + \rateExp^2 / 2 } - 1 \right) \right] ,
\end{aligned}
\end{equation}
%\end{widetext}
where $\ldosTilde(n)$ was defined below~(\mref{16}),
\begin{equation}
\label{eq:uTilde}
	\ldosTilde(n) := \sum_m u(n - m) u(m) \,.
\end{equation}
The first term on the right-hand side is the expected reduction, but the additional, nonvanishing second term spoils the symmetry.
Notably, this violation is not an artifact of the leading-order approximation.
Rather, it can apparently be
traced back to the approximate character of the saddle-point degeneracy discussed below Eq.~\eqref{eq:U4SaddleDiag}.
If the pseudounitary symmetry $T$ with $T^\dagger L T = L$ were perfect, all saddles $\I\rateExp Q / 2 = T R_* T^{-1}$ obtained from the diagonal solution by pseudounitary rotations would contribute equally.
However, as mentioned above, the degeneracy becomes exact only in the limit of close-by eigenvectors (i.e.\ $n_1 = n_2$),
and the approximation is still good for small $\Delta_z$ on the order of the mean level spacing $\lvsp$,
but breaks down when $\lvert \Delta_z \rvert \gg \rateExp$.
This latter limit is correctly reflected in Eq.~\eqref{eq:AvgU4Naive}, too, because then the term $\tilde f^{n_1 n_2}_{\mu_1 \mu_2}$ becomes negligible compared to $d^{n_1 n_2}_{\mu_1 \mu_2}$.
However, in the intermediate regime with $\lvsp \ll \lvert \Delta_z \rvert \lesssim \rateExp$, where the symmetry is neither perfect nor completely broken, the situation is much more subtle and, unfortunately, hardly analytically tractable.
Fortunately, though, the symmetry property~\ref{lst:AvgU4TraceH0} can be restored \emph{a posteriori}.

To this end, we look for a correction term $c^{n_1 n_2}_{\mu_1 \mu_2}$ such that Property~\ref{lst:AvgU4TraceH0} is retrieved when replacing $\tilde f^{n_1 n_2}_{\mu_1 \mu_2} \longrightarrow \tilde f^{n_1 n_2}_{\mu_1 \mu_2} + c^{n_1 n_2}_{\mu_1 \mu_2}$.
At the same time, the corrected term should still possess Properties~\ref{lst:AvgU4Swap12} through~\ref{lst:AvgU4TraceH}.
Properties~\ref{lst:AvgU4Swap12} and~\ref{lst:AvgU4SwapMuNu} limit the possible form of the correction to a function of the invariants
\begin{equation}
\label{eq:AvgU4CorrInv}
	\rateExp/\lvsp, \;
	(\mu_1 + \mu_2), \;
	(n_1 + n_2), \;
	(\mu_1 - \mu_2)^2, \;
	(n_1 - n_2)^2 \,.
\end{equation}
In addition, the correction should be of the same order in $s$.
Anticipating a structural similarity to the preliminary result~\eqref{eq:AvgU4Naive:f}, we thus make an ansatz of the form
\begin{subequations}
\begin{equation}
\begin{aligned}
	c^{n_1 n_2}_{\mu_1 \mu_2}
		&:= \left( \tfrac{\rateExp}{4\pi\lvsp} \right)  \ldos(n_1 - \mu_1) \, \ldos(n_2 - \mu_2) \\
		&\quad \times \frac{ A(c_0, c_1, c_2, c_3, c_4, c_5)}{ \left[ (n_1 - \mu_2)^2 + (\rateExp/2\lvsp)^2 \right] \left[ (n_2 - \mu_1)^2 + (\rateExp/2\lvsp)^2 \right] }
\end{aligned}
\end{equation}
with
\begin{equation}
\begin{aligned}
	&A(c_0, c_1, c_2, c_3, c_4, c_5) \\
		&\;:= c_0 (\rateExp/\lvsp)^2 + c_1 (\mu_1 + \mu_2)(n_1 + n_2) + c_2 (\mu_1 - \mu_2)^2 \\
		&\qquad + c_3 (n_1 - n_2)^2 + c_4 (\mu_1 + \mu_2)^2 + c_5 (n_1 + n_2)^2
\end{aligned}
\end{equation}
\end{subequations}
and $c_0, \ldots, c_5$ being constants, independent of all variables in~\eqref{eq:AvgU4CorrInv}.
Property~\ref{lst:AvgU4TraceH} requires that $\sum_{n_1} c^{n_1 n_2}_{\mu_1 \mu_2} = 0$.
Solving for $c_5$ and using the constraints that all $c_i$ should be constants, we find that $c_5 = -c_4 = c_3 = c_1/2$ and $c_2 = c_0 = 0$.
Hence the only free variable is, e.g., the coefficient $c_1$.
Substituting this ansatz into the defining equation~\eqref{eq:AvgU4TraceH0} of Property~\ref{lst:AvgU4TraceH0}, we find that $c_1 = -1$ solves the equation.
The correction term thus reads
\begin{equation}
\label{eq:AvgU4Corr}
\begin{aligned}
	c^{n_1 n_2}_{\mu_1 \mu_2}
		&= \tfrac{\rateExp}{2\pi\lvsp}  \ldos(n_1 - \mu_1) \, \ldos(n_2 - \mu_2) \\
		&\quad \times \frac{ (\mu_1 + \mu_2 - 2 n_1) (\mu_1 + \mu_2 - 2 n_2 )}{ \left[ (n_1 - \mu_2)^2 + (\rateExp/2\lvsp)^2 \right] \left[ (n_2 - \mu_1)^2 + (\rateExp/2\lvsp)^2 \right] } .
\end{aligned}
\end{equation}
Setting $f^{n_1 n_2}_{\mu_1 \mu_2} := \tilde f^{n_1 n_2}_{\mu_1 \mu_2} + c^{n_1 n_2}_{\mu_1 \mu_2}$, 
we finally recover our main result for the fourth 
moments from Eq.~\eqref{eq:AvgU4}.
We remark that this result also generalizes previous findings from Refs.~\cite{nat18, ith18},
which were obtained by means of completely 
different approximations and under quite 
substantial additional restrictions.
Yet another, and in fact more general,
such approach will be elaborated in 
the subsequent Sec.~\ref{s4}.

%%%%%%%%%%%%%%%%%%%%%%%%%%%%%%%%%%%%%%%%%%%%%%%%%%%%
%%%%%%%%%%%%%%%%%%%%%%%%%%%%%%%%%%%%%%%%%%%%%%%%%%%%
%
\section{Generalized Approximation}
\label{s4}
The central result~(\mref{16}) from the main paper was derived by averaging~(\mref{2}) over some suitably defined ensemble of perturbations $V$.
In doing so, the main task was to evaluate the average
over the products of four $U$ matrix elements on the
right-hand side of~(\mref{6}) [cf.~\eqref{eq:AvgU4Raw}], henceforth denoted as
\begin{eqnarray}
M_{mn}^{\mu_1\mu_2\nu_1\nu_2}
:=
\avv{U_{m \mu_1} U_{n \mu_2} U^*_{m \nu_1} U^*_{n \nu_2}}
\ . 
\label{s210}
\end{eqnarray}
This task was accomplished in the previous 
section by means of rather demanding and lengthy 
supersymmetry methods.
In the present section, an alternative way of 
evaluating such averages will be worked out.
It is based on an approximation
which considerably simplifies the
actual calculations and which will 
lead to practically the same final 
result as the supersymmetric approach.

The general idea is to approximate the 
``fourth moments'' on the right hand 
side of (\ref{s210}) solely in terms of the 
``second moments'' $u(n)$ from~(\ref{eq:AvgU2}) and then to exploit the
previously known results for 
$u(n)$.
In this way, we will be able to
treat, with relatively little effort,
also more general cases
than those which were explicitly
worked out in the previous section.

%%%%%%%%%%%%%%%%%%%%%%%%%%%%%%%%%%%%%%%%%%%%%%%%%%%%
%
\subsection{Symmetry consideration}
\label{s41}
Our starting point is the 
following observation: 
If a given random perturbation 
$V$ with matrix 
elements $V^0_{\mu\nu}$ results,
for the perturbed Hamiltonian $H$ 
in~(\ref{eq:H}), in a certain set of
eigenvectors $\ket{n}$ and
corresponding matrix elements
$U_{m\nu}$ according to (\ref{eq:U}),
then a perturbation $\tilde V$ with modified 
matrix elements
$\tilde V^0_{\mu\nu}:=\e^{\I(\phi_\nu-\phi_\mu)}V^0_{\mu\nu}$
leads to a modified basis transformation 
of the form 
$\tilde U_{m\nu}=\e^{\I(\phi_\nu-\psi_m)}U_{m\nu}$,
where the $\phi_\nu$ are arbitrary 
but fixed phases.
Likewise, the $\psi_m$ are in principle 
once again arbitrary but fixed phases, 
which however are actually irrelevant, 
since they pairwise cancel in the
fourfold products on the right 
hand side of (\ref{s210}).
Explicitly, the corresponding modification 
of (\ref{s210}) takes the form
\begin{eqnarray}
& &  
\tilde M_{mn}^{\mu_1\mu_2\nu_1\nu_2}
:=
\avv{\tilde U_{m \mu_1} \tilde U_{n \mu_2} \tilde U^*_{m \nu_1} \tilde U^*_{n \nu_2}}
=
\nonumber
\\
& & 
e^{i(\phi_{\mu_1}+\phi_{\mu_2}-\phi_{\nu_1}-\phi_{\nu_2})}
\,
\avv{U_{m \mu_1} U_{n \mu_2} U^*_{m \nu_1} U^*_{n \nu_2}}
\,. \ \ \
\label{s230}
\end{eqnarray}
According to Secs.~\ref{sec:RandMatEns:Definition} and~\ref{sec:RandMatEns:InvarianceProperty}, 
all statistical properties of
the $V$ ensemble are
independent of the phases $\phi_\nu$.
Therefore, the same must apply, in particular, 
to the average in (\ref{s230}).
It follows that the average in (\ref{s230}), 
and hence also the average in (\ref{s210}), 
must vanish except if the phases pairwise
cancel each other.
In other words, each of the first two
$U$ matrix elements (without the ``star'' symbol)
must have a ``partner'' among the 
last two $U$ matrix elements
(with ``stars''), where being partners
means that the second indices are equal.
The latter is the 
case if and only if one of the
following two conditions is fulfilled:
(i) $\mu_1=\nu_2$ and $\mu_2=\nu_1$;
(ii) $\mu_1=\nu_1$, $\mu_2=\nu_2$,
and $\mu_1\not=\mu_2$
(the extra condition  $\mu_1\not=\mu_2$ prevents double 
counting of the case $\mu_1=\nu_1=\mu_2=\nu_2$).
We thus can infer that
\begin{eqnarray}
M_{mn}^{\mu_1\mu_2\nu_1\nu_2}
& = &
\delta_{\mu_1\nu_2}\delta_{\mu_2\nu_1}
F^{mn}_{\mu_1\nu_1}
+
\delta_{\mu_1\nu_1}\delta_{\mu_2\nu_2}
G^{mn}_{\mu_1\nu_2}
, \ \  \ \ \ \ 
\label{s240}
\\
F^{mn}_{\mu\nu}
& := &
\avv{U_{m \mu} \, U^*_{m \nu} \,
U_{n \nu}\, U^*_{n \mu}}
\ ,
\label{s250}
\\
G^{mn}_{\mu\nu}
& := & 
\overline{\delta}_{\mu\nu}
\avv{\,|U_{m \mu}|^2 |U_{n \nu}|^2}
\ , 
\label{s260}
\\
\overline{\delta}_{mn} & := & 1-\delta_{mn}
\label{s265}
\ ,
\end{eqnarray}
where $\delta_{mn}$ is the Kronecker delta.

For later convenience, we can also conclude
along similar lines as below (\ref{s230})
that 
\begin{subequations}
\begin{eqnarray}
\av{U_{m\mu}} & = & 0
\label{s270}
\\
\av{U_{m\mu}U_{n\nu}} = \av{U^*_{m\mu}U^*_{n\nu}} & = & 0
\label{s280}
\\
\av{U_{m\mu}U_{n\nu}^*} = \av{U_{m\mu}^*U_{n\nu}} & = & 
\delta_{mn}\delta_{\mu\nu}
\, u(m-\mu)
\ \ \ \ \ \
\label{s290}
\end{eqnarray}
\end{subequations}
for arbitrary $m,\mu,n,\nu$,
where $u(n)$ is defined in (\ref{eq:AvgU2}).
Furthermore, taking into account 
(\ref{s210}) and  (\ref{s240}) we can infer from 
(\mref{4}) and (\mref{6}) that
\begin{eqnarray}
 \!\!
	\avv{\< A \>_{\!\rho(t)}} 
        \!\! & = & 
	\!\! \!\! \!\!  
        \sum_{m,n, \mu_1, \mu_2, \nu_1, \nu_2} 
        \!\!  \!\! 
	\rho^0_{\mu_1 \nu_2}(0) \, A^0_{\mu_2 \nu_1}
	\e^{\I (n-m)\lvsp t}
M_{mn}^{\mu_1\mu_2\nu_1\nu_2}
\nonumber
\\
        \!\! & = & 
	\sum_{m,n,\mu,\nu}
        \rho^0_{\mu \mu}(0) \, A^0_{\nu \nu} \,
        \e^{\I (n-m)\lvsp t}\,
         F^{mn}_{\mu\nu}
\nonumber 
\\
        \!\! & + &
	 \sum_{m,n,\mu,\nu} 
        \rho^0_{\mu \nu}(0) \, A^0_{\nu \mu} \,
        \e^{\I (n-m)\lvsp t} \,
         G^{mn}_{\mu\nu}
\ .\ \ \
\label{s291}
\end{eqnarray}

%%%%%%%%%%%%%%%%%%%%%%%%%%%%%%%%%%%%%%%%%%%%%%%%%%%%
%
\subsection{Real and complex random matrices}
\label{s42}
So far, it was always tacitly understood
that the off-diagonal matrix elements 
$V^0_{\mu\nu}$ are in general 
complex-valued.
On the other hand, it is well known from
textbook random matrix theory \cite{haa10,bro81}
that if the system exhibits a certain 
symmetry (related to time inversion)
then the eigenvectors can be chosen 
so that 
$(V^0_{\mu\nu})^\ast=V^0_{\mu\nu}$ and
$U^\ast_{m\nu}=U_{m\nu}$
for all $m$, $\mu$ and $\nu$.
Whether the considered system
exhibits this symmetry or not,
and thus the pertinent random 
matrix ensemble is purely real or not,
is known to be of great importance 
for instance with respect to the
level statistics \cite{haa10,bro81}.

With respect to the quantities which are
at the focus of our present work
[namely, the perturbed expectation 
values in (\mref{4}), the function 
$u(n)$ in (\ref{eq:AvgU2}), etc.], we found
that ensembles with real-valued matrix 
elements $V^0_{\mu\nu}$ behave practically
indistinguishable from their 
complex-valued counterparts.
For instance, in Sec.~\ref{sec:OverlapsFromSupersymmetry:2ndMoment} we derived
the result~\eqref{eq:u:BW} for certain
complex-valued ensembles, while the
same result was obtained in \cite{fyo96}
for the corresponding real-valued ensembles,
and likewise for the result 
in~\eqref{eq:u:SC}.
More precisely, this similarity between 
real- and complex-valued ensembles
is found to only hold true under our
usual assumption that $s$ 
in (\ref{x10}) is a small parameter 
(see (\ref{eq:SmallParam}))
and thus 
subleading terms in $s$ can be 
neglected.
Put differently, with respect to the
higher-order corrections, differences
between the real and complex 
ensembles 
may still be possible.

Throughout this Supplemental Material,
we confine ourselves to the complex 
case. 
Though omitted here, we also worked out 
the real case and always found 
identical final results in (\ref{s291})
apart from the higher-order 
corrections in $s$.
Unfortunately, we are not aware 
of a simple general argument
why this is 
so.
(The argument from Sec.~\ref{sec:RandMatEns:InvarianceProperty} 
may be considered as a first hint.)
We also remark that in contrast to the 
final results, the intermediate
steps often differ quite considerably.
For instance, in the real case only phases
$\phi_\nu\in\{0,\pi\}$ would be admitted in
(\ref{s230}), leading still to the same 
conclusion as in (\ref{s240})--(\ref{s270}) and (\ref{s290}),
whereas (\ref{s280}) becomes invalid, 
and also the evaluation
of $F^{mn}_{\mu_1\mu_2}$ and
$G^{mn}_{\mu_1\mu_2}$ 
in the following subsections
yields quantitatively different results.

%%%%%%%%%%%%%%%%%%%%%%%%%%%%%%%%%%%%%%%%%%%%%%%%%%%%
%
\subsection{Complex random variables}
\label{s43}
Complex random variables are well-established in probability
theory, yet it may be worthwhile to collect some issues of
particular interest in our present context.

A complex random variable $z$ is defined as a pair 
$(x,y)$ of real random variables via $z=x+\I y$.
Accordingly, its probability distribution amounts
to the joint distribution of the two real variables.
The corresponding probability density may
thus be written either in the form $\rho(z)$
or $\rho(x,y)$.
Given the probability density $\rho$, 
the expectation value $E[f(x,y)]$ 
of an arbitrary function $f(x,y)$
can be readily determined as usual, 
in particular arbitrary ``moments'' of 
the form $E[z^k (z^\ast)^l]$.

The distribution (or the random variable itself) 
is called circular symmetric if 
the probability density $\rho$
does not depend on $x$ and $y$ 
separately, but only on $|z|=\sqrt{x^2+y^2}$,
i.e., it can be written in yet another 
alternative form, namely as
$\rho(|z|)$.
It is called Gaussian if 
$\rho(|z|)=\exp\{-|z|^2/\sigma^2\}/\pi\sigma^2$
for some $\sigma$,
and non-Gaussian otherwise.

Similarly as in the discussion above (\ref{s230}),
one sees that $U_{m\nu}$ is a complex random 
variable in the above specified sense,
which must exhibit the same statistical 
properties as $\e^{\I \phi}U_{m\nu}$
for an arbitrary but fixed phase $\phi$. 
Hence, the 
distribution of $U_{m\nu}$ must be circular 
symmetric for any index pair $(m,\nu)$.
Moreover, one readily sees, similarly as 
above (\ref{eq:AvgU2}), that the distribution of 
$U_{m\nu}$ does
not depend on $m$ and $\nu$ separately,
but only on the difference $m-\nu$,
and likewise for all its moments 
$\av{\,|U_{m\nu}|^k}$.

If the distribution is furthermore 
Gaussian, then one readily verifies
that
\begin{eqnarray}
\avv{\, |U_{m\nu}|^4}=2 \avv{\, |U_{m\nu}|^2}
\ .
\label{c1}
\end{eqnarray}
Since the unitary $U$ must satisfy the
relation $\sum_m |U_{m\nu}|^2=1$ for any $\nu$,
we can conclude that $|U_{m\nu}|\leq 1$,
hence the random variable
$U_{m\nu}$ cannot be strictly Gaussian 
distributed.
In the following, we will therefore
{\em not} assume that the distribution 
is Gaussian. 
However, it will be taken for granted
that the deviations from a Gaussian 
distribution are not too extreme, 
so that the fourth moment
$\avv{\, |U_{m\nu}|^4}$ can be written in the form
\begin{eqnarray}
\avv{\, |U_{m\nu}|^4}=\gamma(\nu-m) \avv{\, |U_{m\nu}|^2}
\label{c2}
\end{eqnarray}
with some real-valued, non-negative function 
$\gamma(n)$ which remains of order unity 
for all $n\in\ZZ$. In other words, 
there exists a constant $C$ 
with the property that
\begin{eqnarray}
0\leq \gamma(n)\leq C=\ord(1)
\label{c3}
\end{eqnarray}
for all $n$.
(In principle, this requirement 
could still be considerably weakened).
For instance, probability densities $\rho$
with extremely slowly decaying 
tails are thus excluded.
Cases which are of interest to us but
disobey (\ref{c3}) have to our knowledge 
never been observed so far.

Analogously, two complex random variables 
$z_1=x_1+\I y_1$ and 
$z_2=x_2+\I y_2$ 
give rise to a joint probability density
$\rho(z_1,z_2)$, from
which arbitrary expectation values
$E[f(x_1,y_1,x_2,y_2)]$ can be 
deduced.
They are called uncorrelated if
$E[z_1z_2]=E[z_1]E[z_2]$ and
$E[z_1^\ast z_2]=E[z^\ast_1]E[z_2]$
(analogous relations for 
$E[z_1z^\ast_2]$ and 
$E[z^\ast_1z^\ast_2]$
then automatically follow).
\textit{A forteriori}, they are called independent
if $\rho(z_1,z_2)$ can be written
in the form $\rho_1(z_1)\rho_2(z_2)$.

Similar generalizations for more than
two random variables are straightforward.
In particular, if $(z_1,...,z_{2n})$ are 
multivariate Gaussian variables with 
zero mean, then the Isserlis
(or Wick) theorem reads
\begin{eqnarray}
E[z_1z_2\cdots z_{2n}]=\sum\prod E[z_jz_k]
\ ,
\label{c4}
\end{eqnarray}
where the notation $\sum\prod$ means summing over all 
distinct ways of partitioning $z_1,...,z_{2n}$ into pairs 
$z_j z_k$ and each summand is the product of the $n$ 
pairs. (The complementary relation 
$E[z_1z_2\cdots z_{2n-1}]=0$
is of less interest to us.)
In particular, one readily recovers
(\ref{c1}) as a special case.

Considering two arbitrary matrix elements 
$U_{m\mu}$ and $U_{n\nu}$ which 
are not identical, i.e., the index pairs $(m,\mu)$ 
and $(n,\nu)$ are different, 
it follows with (\ref{s270})--(\ref{s290}) 
that they are uncorrelated
in the above defined sense.
On the other hand,
they cannot be independent
for the following reason:
From the definition (\ref{eq:U}) of the unitary $U$,
one readily infers the usual orthonormality relations 
\begin{eqnarray}
\sum\nolimits_n U_{n\mu} U^*_{n\nu}
=
\delta_{\mu\nu}
\ .
\label{c5}
\end{eqnarray}
In particular, choosing $\mu$ 
arbitrary but fixed and focusing on
$\nu=\mu$,
it follows that
the value of $|U_{n\mu}|^2$ for one specific
$n=n_0$ is determined by the values of
$|U_{n\mu}|^2$ for all $n\not=n_0$.
Hence the random variable $U_{n_0\mu}$
cannot be independent from all the other
$U_{n\mu}$'s with $n\not=n_0$.

In some previous analytical investigations  
\cite{deu91,sre94,nat18}, it was 
taken for granted that the $U_{m\nu}$ are, 
at least in sufficiently good approximation, 
independent and/or Gaussian random 
variables,
and also in numerical examples we  observed
that these approximations seem to be fulfilled 
very well.
However, we have seen above that neither 
of the two properties can be strictly true.
Moreover, it has been observed for instance
in \cite{gen12} that the deviations from 
a Gaussian distribution may not 
necessarily be negligibly small.
Accordingly, one main point of our subsequent 
considerations is to still admit (small) deviations 
from strict Gaussianity and/or independence,
and to carefully keep track of their effects on
the final results in (\ref{s291}).

%%%%%%%%%%%%%%%%%%%%%%%%%%%%%%%%%%%%%%%%%%%%%%%%%%%%
%
\subsection{Simplest approximation}
\label{s44}
As seen above, the  complex random variables 
appearing in (\ref{s250}) and (\ref{s260}) are
uncorrelated but not strictly independent.
In this section, we investigate the consequences
of treating them as (approximately) independent
nevertheless (but still admitting deviations from
Gaussianity).

Assuming that the $U$ matrix elements
in (\ref{s250}) and (\ref{s260}) are independent,
and observing (\ref{s270})--(\ref{s290}), 
one can infer that (\ref{s250}) must 
be zero unless each of the 
two $U$ matrix elements without a 
``star'' symbol
has a complex conjugate counterpart 
among the two remaining $U$ matrix 
elements (with ``stars'').
This prerequisite can be fulfilled in two ways:
(i) $\mu=\nu$;
(ii) $\mu\not=\nu$ and $m=n$.
It follows that
\begin{eqnarray}
F^{mn}_{\mu\nu}
& = &
\delta_{\mu\nu} \, 
\av{\,|U_{m \mu}|^2\, |U_{n \mu}|^2}
\nonumber
\\
& + & 
\delta_{mn} \, \overline{\delta}_{\mu\nu} 
\av{\,|U_{m \mu}|^2 \,|U_{m \nu}|^2} 
\ .
\label{s300}
\end{eqnarray}
Exploiting the above 
independence assumption once more,
we can further conclude 
that $\av{\,|U_{m \mu}|^2\, |U_{n \mu}|^2}
=\av{\,|U_{m \mu}|^2}\, \av{\, |U_{n \mu}|^2}$
unless $m=n$, and likewise for the last term 
in (\ref{s300}),
yielding with (\ref{eq:AvgU2})
\begin{eqnarray}
F^{mn}_{\mu\nu}
& = &
\delta_{\mu\nu} \, \delta_{mn}\, 
\av{\,|U_{m \mu}|^4}
\nonumber
\\
& + & 
\delta_{\mu\nu}\, \overline{\delta}_{mn}\,
u(m-\mu) \, u(n-\mu)
\nonumber
\\
& + & 
\delta_{mn} \, \overline{\delta}_{\mu\nu} \,
u(m-\mu) \, u(m-\nu)
\ .
\label{s310}
\end{eqnarray}
Similarly, (\ref{s260}) takes the form
\begin{eqnarray}
G^{mn}_{\mu\nu}
& := & \overline{\delta}_{\mu\nu}
\, u(m-\mu) \, u(n-\nu)
\ .
\label{s320}
\end{eqnarray}
By means of (\ref{s310}) and (\ref{s320}) one thus can 
rewrite (\ref{s240}) as
\begin{eqnarray}
M_{mn}^{\mu_1\mu_2\nu_1\nu_2}
& = &
\delta_{\mu_1\nu_2}\delta_{\mu_2\nu_1}\delta_{mn}
\,u(m-\mu_1)\, u(m-\mu_2)
\nonumber
\\
& + &
\delta_{\mu_1\nu_1}\delta_{\mu_2\nu_2}
\, u(m-\mu_1)\, u(n-\mu_2)
\nonumber
\\
& + &
\delta_{\mu_1\nu_2}\delta_{\mu_2\nu_1}\delta_{\mu_1\mu_2}
\delta_{mn}\, y(m-\mu_1)
\ , 
\label{s321}
\end{eqnarray}
where we introduced the function
\begin{eqnarray}
y(m-\mu) 
& := & \av{\,|U_{m \mu}|^4} - 2\, u^2(m-\mu)
\nonumber
\\
& = & 
\{\gamma(m-\mu)- 2\}\, u^2(m-\mu)
\ ,
\label{s322}
\end{eqnarray}
and where we exploited (\ref{c2}) and (\ref{eq:AvgU2}) 
in the last step.
If the $U_{m\nu}$ were Gaussian random variables, 
then the last term in (\ref{s321}) would vanish 
(see (\ref{c1})), and the remaining first two 
terms could be readily recovered by exploiting 
the Isserlis theorem (\ref{c4}).

Introducing (\ref{s321}) into (\ref{s291})
yields, after a straightforward calculation,
the result
\begin{align}
\av{ \< A \>_{\!\rho(t)}}
& = 
\langle A \rangle_{\!\tilde\rho} + |g(t)|^2  \,  \< A \>_{\!\rho_0\!(t)} +  S \,,
\label{s330}
\end{align}
where $\tilde\rho$ is the state introduced below~(\mref{16}) with matrix elements
\begin{equation}
\label{eq:rhoTilde}
	\braN{\mu} \tilde\rho \ketN{\nu} := \delta_{\mu\nu} \sum_\kappa \ldosTilde(\nu-\kappa) \rho^0_{\kappa\kappa}(0)
\end{equation}
and $\ldosTilde(n)$ from~\eqref{eq:uTilde},
$g(t)$ from~\eqref{eq:g}, and
\begin{align}
S & :=  Y\, \sum_{m} \rho^0_{mm}(0)\, A^0_{mm} 
\,, \quad
\label{s370}
%\displaybreak[0] \\
Y :=  \sum_n y(n)
%=\sum_n \left\{\av{\, |U_{mn}|^4} - 2 u^2(m-n)\right\}
.
%\label{s375}
\end{align}
Note that the three terms on the right-hand side
of~(\ref{s330}) are the direct descendants of 
the three terms in (\ref{s321}).

If the $U_{m\nu}$ are approximated as
Gaussian random variables then the 
quantity $S$ in (\ref{s370}) is zero, as can be
concluded from the discussion below 
(\ref{s322}) .
More generally, one can readily
infer from (\ref{s370}) 
and  $\sum_{\mu} \rho^0_{\mu\mu}(0)=1$
that
\begin{eqnarray}
|S| \leq |Y|\, \norm{A}
\ ,
\label{s380}
\end{eqnarray}
where $\norm{A}$ denotes the operator norm of $A$
(largest eigenvalue in modulus).
Taking into account (\ref{c3}) and (\ref{s322}),
we can conclude from (\ref{s370}) that
\begin{eqnarray}
|Y| & \leq & (C+2)\, \sum_n u^2(n)
\label{s382}
\end{eqnarray}
Estimating $u^2(n)$ from above by
$u(n)\max_m u(m)$ and observing 
$\sum_n u(n) = 1$, it follows that
$|Y|=(C+2) \max_m u(m)$,
and with (\ref{x10}), (\ref{s380}) 
that 
\begin{eqnarray}
|S| \leq (C+2)\,s\,\norm{A}
\ .
\label{s383}
\end{eqnarray}
Recalling the discussion of $C$
above (\ref{c3}) and 
that $s$ is a small parameter
(see (\ref{eq:SmallParam})),
we can conclude that $S$ remains 
negligibly small in (\ref{s330})
even if the  $U_{m\nu}$ are not 
Gaussian distributed.

Altogether, we thus arrive at the 
general approximation
\begin{eqnarray}
\av{ \< A \>_{\!\rho(t)}}
& = & 
\langle A \rangle_{\!\tilde\rho} + |g(t)|^2  \,  \< A \>_{\!\rho_0\!(t)}
\ .
\label{s390}
\end{eqnarray}
Introducing the explicit result~\eqref{eq:u:BW} for $u(n)$ derived in 
the previous section, one readily recovers the 
approximation $|g(t)|^2=e^{-\Gamma \lvert t \rvert}$,
while~\eqref{eq:uTilde} becomes identical to
\begin{equation}
\label{eq:uTilde:BW}
	\ldosTilde(n) = \frac{1}{\pi} \frac{\Gamma /  \lvsp}{(\Gamma/\lvsp)^2 + n^2} \,.
\end{equation}
Hence (\ref{s390}) is seen to be already
somewhat similar to the more rigorous 
result~(\mref{16}), which followed from the supersymmetry calculations of 
the previous section.
Yet, the remaining differences 
are in fact quite serious, for instance:
(i) At time $t=0$ the system is always in the
same initial state $\rho(0)$, independent of
the perturbation $V$. It follows that both
the first and the last terms in (\ref{s390})
must be equal to $\< A \>_{\!\rho(0)}$.
Moreover, $\sum_n u(n) = 1$ and~\eqref{eq:g} imply 
$g(0)=1$.
Altogether, for $t=0$ the approximation (\ref{s390})
amounts to $\< A \>_{\!\rho(0)}=\langle A \rangle_{\!\tilde\rho}+\< A \>_{\!\rho(0)}$.
However, 
$\langle A \rangle_{\!\tilde\rho}$
is in general not a small quantity.
(ii) If one chooses $A=\id$ (identity operator),
then one readily sees that the left-hand side
of~(\ref{s390}) as well as $\langle A \rangle_{\!\tilde\rho}$ and 
$\< A \>_{\!\rho_0\!(t)}$ on the right-hand side 
must be unity, yielding $1=1+|g(t)|^2$.
As seen above, $|g(t)|^2$ is unity
for $t=0$ and usually is found to 
remain non-negligible also within
a notable interval of times $t>0$
(recall our standard example 
$|g(t)|^2=e^{-\Gamma \lvert t \rvert}$).
Even for the particularly simple observable 
$A=\id$ under consideration, 
(\ref{s390}) is thus far from 
being a satisfying approximation.
For the same reason, the result (\ref{s390})
does not exhibit the correct transformation
behavior if $A$ is replaced by $A+c\id$ 
with $c\in\RR$.

In conclusion, the idea
at the beginning of this subsection,
namely 
to approximate the matrix elements 
$U_{m\nu}$ in terms of independent 
(not necessarily Gaussian) random 
variables, does not work sufficiently
well for our present purposes.

%%%%%%%%%%%%%%%%%%%%%%%%%%%%%%%%%%%%%%%%%%%%%%%%%%%%
%
\subsection{Improved approximation}
\label{s45}
As seen in the previous
subsections, the 
matrix elements $U_{m\nu}$
are known to be pairwise uncorrelated,
but treating them as statistically 
independent is a too strong 
simplification.
The most immediate reason which comes
to mind is that such an assumption
leads to a violation of the
orthonormality conditions,
see the discussion below Eq.~(\ref{c5}).

In order to quantify the violation of
the orthonormality relations (\ref{c5}),
we note that those conditions are 
satisfied at least on average, 
i.e., $\av{\sum_n U_{n\mu} 
U^*_{n\nu}}=\delta_{\mu\nu}$,
as can be readily concluded 
from (\ref{eq:AvgU2}) and (\ref{s290}).
Assuming that the $U_{m\nu}$ are 
independent, one furthermore 
can show, similarly as in 
Secs. \ref{s43} and \ref{s44}, 
that
\begin{eqnarray}
\avv{
\Big(
\mbox{$\sum\limits_n$} 
U_{n\mu} 
U^*_{n\nu} - \delta_{\mu\nu}
\Big)^2
}
= \ord\Big(\ldosTilde(\mu-\nu)\Big)
\ ,
\label{e10}
\end{eqnarray}
where $\ldosTilde(n)$ was defined in~\eqref{eq:uTilde}.
Finally, one can show analogously as below
(\ref{s382}) that $\ldosTilde(\mu-\nu)\leq s$, 
where $s$ is our small parameter from
(\ref{x10}) and (\ref{eq:SmallParam}).

In other words, 
assuming that the $U_{m\nu}$ are 
independent leads to violations 
of the orthonormality relations 
(\ref{c5}) which, however,
are with very high probability 
very weak.
[A quantification of this qualitative
statement can be worked out along similar 
lines as below Eq.~(\mref{20}).]
Though treating the
$U_{m\nu}$ as independent might thus
seem to yield a very good 
approximation, 
the so-obtained final result 
(\ref{s390}) actually exhibits unacceptably 
large inaccuracies.
Intuitively, this may be understood 
as follows:
Though the committed error is 
with very high probability
very small for 
every single
summand on the right hand 
side of (\ref{s291}),
there are so many summands 
that the
resulting total error turns out 
to be non-negligible.

For the same reason, though the
$U_{m\nu}$ are expected to be nearly
Gaussian distributed, admitting small deviations
-- as it is done in our present approach --
seems a potentially relevant issue:
Even if each summand on the right-hand 
side of (\ref{s291}) entails with very high probability
a very small error when approximating the 
$U_{m\nu}$ as Gaussian distributed, 
the resulting total error in the final
result may potentially be non-negligible.

In the following, our main idea 
is to improve the approximation 
from the previous subsection 
so that the orthogonality 
relations (\ref{c5})
are even better fulfilled.

In order to work this out in detail,
let us focus as a first example
on the evaluation of the specific
average in (\ref{s250}).
Furthermore, we choose two arbitrary
indices $m$ and $n$ with
$m\not=n$, but then keep
them fixed.
Generalizing the setup from the
previous subsection, we consider
two sets (``vectors'' with 
components labeled by $\nu$) 
of complex random variables 
$y_\nu$ and $z_\nu$.
The distribution of each given 
random variable $y_\nu$ is assumed 
to be identical to the distribution 
of the corresponding random 
matrix element $U_{m\nu}$,
and analogously for $z_\nu$ and $U_{n\nu}$.
But unlike the ``true'' random variables
$U_{m\nu}$ and $U_{n\nu}$,
all the ``auxiliary'' random variables 
$y_\nu$ and $z_\nu$ are now assumed 
to be strictly independent of each other.
Indicating the corresponding averages
(or expectation values) by the symbol 
$\avx{...}$, we thus can
infer from (\ref{s270})--(\ref{s290}) 
that
\begin{subequations}
\begin{align}
\avx{y_\nu}=\avx{z_\nu}
& = 
0
\ ,
\label{s400}
\displaybreak[0] \\
\avx{y_\nu^2}=\avx{z_\nu^2}
& = 
0
\ ,
\label{s410}
\displaybreak[0] \\
\avvx{\,|y_\nu|^2}
& =
u(m-\nu)
\ ,
\label{s420}
\displaybreak[0] \\
\avvx{\,|z_\nu|^2}
& =
u(n-\nu)
\ ,
\label{s430}
\end{align}
\end{subequations}
and analogously for the higher 
moments (cf.~(\ref{c2})).
Similarly as before, the random
variables $y_\nu$ and
$z_\nu$ thus satisfy on average 
the orthonormalization conditions 
$\avx{\sum_\nu |y_\nu|^2}=1$,
$\avx{\sum_\nu |z_\nu|^2}=1$,
$\avx{\sum_\nu y^\ast_\nu z_\nu}=0$,
but the single realizations
do not strictly fulfill the corresponding 
(non-averaged) relations.
In order to fix this problem, 
we next define yet another set of 
random variables, namely
\begin{align}
v_\nu & :=  a\, y_\nu
\ ,
\label{s500}
\displaybreak[0] \\
w_\nu & :=  b\, (z_\nu - \epsilon\, y_\nu)
\ ,
\label{s510}
\displaybreak[0] \\
\epsilon & :=  a^2 \sum_\nu y_\nu^\ast z_\nu
\ ,
\label{s530}
\displaybreak[0] \\
a & :=  \bigg(\sum_\nu |y_\nu|^2\bigg)^{-1/2}
\ ,
\label{s520}
\displaybreak[0] \\
b & :=  \bigg(\sum_\nu
|z_\nu-\epsilon y_\nu|^2\bigg)^{-1/2}
\nonumber
\displaybreak[0] \\
& =  
\bigg(\sum_\nu
|z_\nu|^2-\big|\epsilon/a\big|^2\bigg)^{-1/2}
\ .
\label{s540}
\end{align}
It readily follows that the two vectors $v_\nu$ 
and $w_\nu$ are now properly
orthonormalized, i.e.,
\begin{subequations}
\begin{eqnarray}
\sum_\nu |v_\nu|^2 & = & 1
\ ,
\label{550}
\\
\sum_\nu |w_\nu|^2 & = & 1
\ ,
\label{560}
\\
\sum_\nu v^\ast_\nu w_\nu & = & 0
\ .
\label{570}
\end{eqnarray}
\end{subequations}
The central idea of the present section 
is to approximate the ``true''
random variables $U_{m\nu}$
and $U_{n\nu}$ in (\ref{s250})
for two arbitrary but fixed
indices $m\not=n$
by the above specified ``auxiliary'' 
random variables $v_\nu$ and $w_\nu$, 
i.e., we adopt the approximation
\begin{eqnarray}
F^{mn}_{\mu\nu}
& = &
\avx{v_{\mu} v^\ast_{\nu} w_{\nu} w^\ast_{\mu} }
\ \ \mbox{for $m\not=n$\,.}
\ 
\label{s580}
\end{eqnarray}
The evaluation of (\ref{s580}) turns out 
to be still an extremely tedious task.
Moreover, most of the effort is required
to find out that one could have adopted
right from the beginning the following
additional approximations in 
(\ref{s500})--(\ref{s540}) with only
negligibly small differences in the 
final results in (\ref{s291}):
(i) It is sufficient to consider
Gaussian random variables 
$y_\nu$ and $z_\nu$.
(ii) The factors $a$ and $b$ can be
approximated by unity.
In particular, one finds by exploiting 
(\ref{s400})--(\ref{s430}) that the 
relations
\begin{subequations}
\begin{eqnarray}
\avx{v_\nu}=\avx{w_\nu}
& = & 
0
\ ,
\label{s400'}
\\
\avvx{\,|v_\nu|^2}
& = &
u(m-\nu)
\ ,
\label{s420'}
\\
\avvx{\,|w_\nu|^2}
& = &
u(n-\nu)
\ ,
\label{s430'}
\end{eqnarray}
\end{subequations}
are satisfied in very good approximation,
i.e., up to higher order corrections
in the small parameter $s$ from 
(\ref{x10}), (\ref{eq:SmallParam}).
(Note that by adding suitable corrections of 
higher order in $s$ on the right-hand side of
(\ref{s420}{) and (\ref{s430}), it is even
possible to turn (\ref{s400'})--(\ref{s430'}) 
into exact equalities.)
The basic reason for (i) is similar as in
the (much simpler) example from Sec.~\ref{s44}: 
A possible non-Gaussianity may only show
up in terms which exhibit (at least) 
one ``extra Kronecker delta'' compared 
to the other terms, as exemplified by
the last summand in (\ref{s321}).
Therefore, the number of such terms
which contribute to the multiple sums
in (\ref{s291}) is relatively
small compared to the number of 
the other terms.
Their total contribution to the 
final result (\ref{s291}) thus amounts 
to a higher-order 
correction, as exemplified around 
(\ref{s382}).
An intuitive explanation of (ii) is
as follows: Similarly as in (\ref{e10})
one finds that the mean values of
$a$ and $b$ are very close to unity
and that the variances are very small.
Hence, approximating them by unity
amounts to very small corrections
on the right-hand side of 
(\ref{s500})--(\ref{s530}).
Since each of those terms is already
very small in itself, these 
corrections are even much smaller, 
hence their contribution to the 
final result in (\ref{s291})
is negligible.
The same applies to the factor $a^2$ 
in (\ref{s530}).
A better understanding of why approximating
$a$ and $b$ by unity is fundamentally
different from approximating $\epsilon$ 
by zero will only be possible after having 
seen the way in which $\epsilon$ 
acts in the following calculations.

Taking for granted the above simplifications (i) 
and (ii), we are thus left with the 
approximations
\begin{eqnarray}
v_\nu & = & y_\nu
\ ,
\label{s590}
\\
w_\nu & = & 
z_\nu - y_\nu\, \sum_k y_k^\ast z_k
\ ,
\label{s600}
\end{eqnarray}
where $y_\nu$ and $z_\nu$ can be considered as
independent Gaussian random variables with
mean values and variances as specified in 
(\ref{s400})--(\ref{s430}).

Introducing (\ref{s590}) and (\ref{s600}) into
(\ref{s580}) yields
\begin{subequations}
\begin{align}
F_{\mu\nu}^{mn} 
& = 
F_1-F_2-F_3+F_4 \ \ \mbox{for $m\not=n$\, ,}
\label{s601}
\displaybreak[0] \\[0.3cm]
F_1 & :=
\avx{y_{\mu} y^\ast_{\nu} 
z_{\nu} z^\ast_{\mu} }
\ ,
\label{s602}
\displaybreak[0] \\[0.2cm]
F_2 & := 
\sum_k\avx{y_{\mu} y^\ast_{\nu}
y_{\nu}y^\ast_k z_k z^\ast_{\mu} 
}
\ ,
\label{s603}
\displaybreak[0] \\
F_3 & := 
\sum_j \avx{y_{\mu} y^\ast_{\nu} 
z_{\nu}  y^\ast_{\mu} y_j z^\ast_j }
\ ,
\label{s604}
\displaybreak[0] \\
F_4 & := 
\sum_{j,k}\avx{y_{\mu} y^\ast_{\nu}
y^\ast_{\mu} y_j z^\ast_j y_{\nu} y^\ast_k z_k 
}
\ .
\label{s605}
\end{align}
\end{subequations}
Due to our assumption that the random variables
appearing in (\ref{s602})--(\ref{s605}) are
multivariate Gaussian variables with zero mean, 
the expectation values can be conveniently 
evaluated by means of the Isserlis theorem (\ref{c4}).

The term $F_1$ in (\ref{s602})
amounts to those contributions
which one would obtain in the absence of the last 
summand
in (\ref{s600}), corresponding to the simple
approximation adopted in Sec.~\ref{s44}.
Accordingly, $F_1$ must be identical to 
the previous finding (\ref{s310}) 
in the case $m\not=n$, i.e.,
\begin{eqnarray}
F_1 & = & \delta_{\mu\nu} \,
u(m-\mu) \, u(n-\mu)
\ .
\label{s606a}
\end{eqnarray}
Moreover, the remaining contributions $F_2$, $F_3$, $F_4$
must be the descendants of the last term in (\ref{s600}),
i.e., they are solely due to our improved 
approximation.

Focusing on the right hand side of (\ref{s603}), 
one readily sees that only partitions in the Isserlis 
theorem (\ref{c4}) may possibly result in non-zero 
contributions for which $z_k$ is paired with 
$z_{\mu}^\ast$, and for which $k$ agrees 
with $\mu$, i.e.,
\begin{eqnarray}
F_2 & = &
\avx{y_{\mu} y^\ast_{\mu} y_{\nu} y^\ast_{\nu}} 
\,
E[z_{\mu} z^\ast_{\mu}]
\ .
\label{s606}
\end{eqnarray}
The last factor is given by (\ref{s430}).
The remaining factor can be readily evaluated 
along similar lines, yielding
\begin{eqnarray}
F_2 & = &
u(m-\mu)\, u(m-\nu)\, u(n-\mu)
\nonumber
\\
& + &
\delta_{\mu\nu}\, u(m-\mu)\, u(m-\mu)\, u(n-\mu)
\ .
\label{s607}
\end{eqnarray}

Due to the extra Kronecker delta, the second
term on the right-hand side of (\ref{s607})
turns out to entail a correction
in the final result (\ref{s291})
which is of higher order 
in the small parameter
$s$ from (\ref{x10}), (\ref{eq:SmallParam})
compared to the contributions of the first term. 
In turn, the first term might seem 
to be a higher order correction compared to
the contributions of $F_1$ from (\ref{s606a}),
since the former is of third and the latter of
second order in $u$, see also 
(\ref{x10}) and (\ref{eq:SmallParam}).
However, it will turn out later that 
the ``missing Kronecker delta'' 
effectively compensates
for this extra factor $u$.
Heuristically, this can be seen
by summing over $m$, $n$, and $\nu$
both in (\ref{s606a}) and in the
first line in (\ref{s607}): In
both cases, the result is unity, suggesting
that both terms will also comparably contribute
to the more complicated sums in 
(\ref{s291}).
This ``effect'' pinpoints the above announced 
fundamental difference between 
approximating $a$ and $b$
in (\ref{s500})--(\ref{s530})
by unity, and approximating $\epsilon$ 
in (\ref{s530}) by zero:
$\epsilon$ entails new leading order
terms, while $a$ and $b$ only give 
rise to new subleading order terms and
to small corrections of the already
existing leading order terms.
Technically speaking, the basic 
mechanism is that the extra sum in 
(\ref{s600}) enables a pairing of the
four $y$ factors in (\ref{s606}) 
without an extra Kronecker delta,
while the same was not possible
in the $F_1$ term from (\ref{s602}) 
and (\ref{s606a}), to which the extra 
sum in (\ref{s606}) did not contribute.

After an analogous evaluation of $F_3$ and $F_4$,
one can rewrite (\ref{s601}) in the form
\begin{eqnarray}
\!\!\!\!\!\!\!\! & & F^{mn}_{\mu\nu}
=
\delta_{\mu\nu}\,
u(m-\mu)\, u(n-\mu)
\nonumber
\\
\!\!\!\!\!\!\!\! & & +
u(m-\mu)\, u(m-\nu)\,
[\ldosTilde(m-n)-u(n-\mu)- u(n-\nu)]
\nonumber
\\
\!\!\!\!\!\!\!\! & & 
+ \mbox{h.o.\ \ \ for $m\not=n$\,.}
\label{s610}
\end{eqnarray}
The first term on the right-hand side
of (\ref{s610}) derives from $F_1$
and thus agrees with the previous
finding (\ref{s310}) 
for $m\not=n$.
The second line in (\ref{s610})
represents the crucial modification due
to our present improved approximation.
The symbol ``h.o.'' in the last line
of (\ref{s610}) refers to a multitude of higher-order terms, which later turn out 
(by similar calculations as 
around (\ref{s380})) to entail 
subleading corrections in the 
final result (\ref{s291}).
One of them is the last summand 
in (\ref{s607}), and they all
have the property that 
they exhibit (at least) 
four factors of Kronecker deltas 
or $u$ functions,
while all leading order terms 
exhibit three such factors.

Turning to the case $m=n$ in (\ref{s250}), 
the question whether the two column 
vectors $U_{m\nu}$ and $U_{n\nu}$
($m$ and $n$ fixed, $\nu$ variable)
are strictly orthogonal or not is
obviously irrelevant.
Hence one recovers the same result
as previously in (\ref{s310}) if $m=n$.
The same applies to $G^{mn}_{\mu\nu}$
in (\ref{s260}) if $m=n$, while in
the case $m\not=n$ the previous result
in (\ref{s320}) will be modified
by additional terms, all of which 
however turn out to entail subleading 
corrections in the final result 
(\ref{s291}).

Introducing all these findings into 
(\ref{s291})
one obtains similarly as in
(\ref{s330})--(\ref{s390}) 
after a straightforward but somewhat
lengthy calculation the 
approximation
\begin{align}
\av{ \< A \>_{\!\rho(t)}}
\!\!\!  & = \!\!
\langle A \rangle_{\!\tilde\rho} + |g(t)|^2  \Big\{ \! \< A \>_{\!\rho_0\!(t)} 
- \langle A \rangle_{\!\tilde\rho} \Big\}
+ \tilde R(t) + \tilde S
\, ,\ \ 
\label{s620}
\displaybreak[0] \\
\tilde R(t) 
& :=
\sum_{m,n} \rho^0_{mm}(0)\, A^0_{nn}\, \tilde r(t,m-n)
\ ,
\label{s630}
\displaybreak[0] \\
\tilde r(t,n)
& :=
\!2 |g(t)|^2 \ldosTilde(n)
- g(t) [h(t,n) + h(t,-n)]
\,,
\label{s640}
\displaybreak[0] \\
h(t,n)
& :=
\sum_m \e^{-\I m\lvsp t}\, u(m-n)\, u(m)
\ ,
\label{s650}
\end{align}
with $\langle A \rangle_{\!\tilde\rho}$, $\ldosTilde(n)$, and $g(t)$ as in~\eqref{s330}.
The last summand $\tilde S$ in (\ref{s620}) 
originates from the higher-order terms in 
(\ref{s610}) as well as the corresponding
higher-order terms arising in 
$G^{mn}_{\mu\nu}$ (see above).
Similarly as in (\ref{s383}), one 
can show that
\begin{eqnarray}
|\tilde S| \leq 10\, s\,\norm{A}
\ ,
\label{s650a}
\end{eqnarray}
implying with (\ref{x10}) and (\ref{eq:SmallParam}) that
the last term in (\ref{s620}) 
is a negligible correction 
of higher order in $s$.

Until now, we approximated the coefficients
$a$ and $b$ in (\ref{s500})--(\ref{s530}) 
by unity
and the random variables $y_\nu$ and $z_\nu$ 
in (\ref{s590})--(\ref{s600}) as Gaussian 
distributed.
As pointed out below (\ref{s580}),
going beyond these approximations
gives rise to a flurry of additional
subleading-order corrections,
analogously to the correction $S$ 
encountered in (\ref{s370})--(\ref{s383}).
As a consequence, the factor $10$
on the right-hand side of (\ref{s650a})
will be replaced by some larger factor,
which, however, will usually still
remain on the order of 
$10$ to $100$.

Exploiting the definitions~\eqref{eq:g}, \eqref{eq:uTilde}, \eqref{eq:rhoTilde},
and (\ref{s650}),
one readily verifies that
$g(0)=1$, 
$h(0,n)=\ldosTilde(n)$, 
and $\sum_n h(t,n)=g^\ast\!(t)$.
It follows with (\ref{s630}), (\ref{s640})
that $\tilde R(0)=0$
and that if $A^0_{\nu\nu}$ is independent of
$\nu$ then $\tilde R(t)=0$ for all $t$.
Though there does not seem to be a reason
why this must always be so 
\cite{cas96}, in all
concrete examples known to us the function
$u(n)$ from (\ref{eq:AvgU2})
%always
exhibits
the symmetry
\begin{eqnarray}
u(-n)=u(n)
\ .
\label{s651}
\end{eqnarray}
It then readily follows from~\eqref{eq:g}
that $g^\ast\!(t)=g(t)$ and from~(\ref{s650})
that $h(t,-n)=h^\ast\!(t,n)$.

On the one hand, (\ref{s620}) still shares 
some similarities with our previous 
approximation in (\ref{s390}),
but it no longer exhibits the shortcomings
discussed below (\ref{s390}).
On the other hand, (\ref{s620}) generalizes~(\mref{16}), which we previously
derived in the main paper using the supersymmetry results from Sec.~\ref{sec:OverlapsFromSupersymmetry},
when $u(n)$ takes the form~\eqref{eq:u:BW}
and hence $g(t)$ in~\eqref{eq:g} is given 
(in very good approximation)
by $e^{-\Gamma |t|/2}$.
We note that $R(t)$ in~(\mref{17})
and $\tilde R(t)$ in (\ref{s630})
exhibit the same essential properties
(they vanish for $t=0$, for $t\to\infty$, and
if $A_{\nu\nu}^0$ is independent of $\nu$),
but quantitatively they are found to be 
not exactly identical in all details.
On the one hand, this corroborates \textit{a 
posteriori} the validity of our present 
\textit{ad hoc} approximation, on the other hand, 
it also indicates the limits which such an
approximation still may comprise
at least in principle.
However, in practice both
$R(t)$ and $\tilde R(t)$ can be
considered as negligible according 
to the line of reasoning below 
Eq.~(\mref{18})
in the main 
paper. In the same vein, we can
also adopt the approximation
$\langle A \rangle_{\!\tilde\rho} \simeq \Ath$ from the main 
paper, yielding
\begin{eqnarray}
\av{\< A \>_{\!\rho(t)}}
& = & 
\Ath + |g(t)|^2  \left\{ \< A \>_{\!\rho_0\!(t)} 
- \Ath\right\}
\, . \ \ \ \ \ \ \ 
\label{s660}
\end{eqnarray}
From this result together with (\ref{eq:g}) and 
the considerations below~(\mref{20}) in the main paper and Sec.~\ref{s5} below,
one finally recovers~(\mref{22}) with $g(t)$ from~(\mref{9}).

%%%%%%%%%%%%%%%%%%%%%%%%%%%%%%%%%%%%%%%%%%%%%%%%%%%%
%%%%%%%%%%%%%%%%%%%%%%%%%%%%%%%%%%%%%%%%%%%%%%%%%%%%
%
\section{Variance of the time evolution}
\label{s5}
%%%%%%%%%%%%%%%%%%%%%%%%%%%%%%%%%%%%%%%%%%%%%%%%%%%%
%
The objective of this section is to 
show that the deviation
\begin{eqnarray}
\xi_V(t)=\< A \>_{\!\rho(t)}- \av{ \< A \>_{\!\rho(t)}}
\label{s1010}
\end{eqnarray}
from the average $\av{ \< A \>_{\!\rho(t)}}$
is negligibly small for the vast 
majority of all members of the
$V$ ensemble.
In doing so, we will consider the
same ensembles for which the 
average expectation 
values $\av{ \< A \>_{\!\rho(t)}}$ have been
explored in the previous sections.
As outlined in the main paper,
this essentially amounts to establishing the upper bound
\begin{eqnarray}
\av{\xi_V^2(t)}\leq c \, s\, \norm{A}^2
\ ,
\label{s1021}
\end{eqnarray}
[cf.~(\mref{20})]
for 
the variance
\begin{eqnarray}
\av{\xi_V^2(t)}=\avv{\< A \>^2_{\!\rho(t)}}- \avv{ \< A \>_{\!\rho(t)}}^2
\ ,
\label{s1020}
\end{eqnarray}
where $s$ is our small parameter from (\ref{x10}) 
and (\ref{eq:SmallParam}),
$\norm{A}$ is the operator norm of $A$,
and $c$ is some positive real number
which does not depend on any further
details of the considered system
[in particular, it is independent of 
$s$, $A$, $\rho(0)$, and of
the perturbation strength 
$\lambda$ in (\ref{eq:H})].

%%%%%%%%%%%%%%%%%%%%%%%%%%%%%%%%%%%%%%%%%%%%%%%%%%%%
%
\subsection{Basic considerations}
\label{s52}
In this subsection, we further elaborate on the 
considerations in Sec.~\ref{sec:RandMatEns}, 
in particular on the 
heuristically expected (and numerically confirmed)
similarities between our present random matrix problem 
and the central limit theorem (CLT) for random 
variables.

The basic observation is that a large number 
of perturbation matrix elements $V_{\mu\nu}^0$
[see~(\ref{eq:V})] is
intuitively expected to be of comparable relevance
for the perturbed expectation value in (\mref{4}).
(For simplicity, we may imagine $t$ 
in (\ref{s1010})--(\ref{s1020}) as arbitrary 
but fixed.)
In other words, $\< A \>_{\!\rho(t)}$
on the left-hand side of (\ref{eq:V}) is a real-valued function of many, roughly speaking 
``equally important'' arguments $V_{\mu\nu}^0$.
If these arguments are furthermore randomly
sampled according to the specific $V$ ensemble
at hand, then it seems reasonable to expect --
similarly as in the CLT -- that the concomitant
probability distribution of the real-valued 
random variable $\< A \>_{\!\rho(t)}$ 
will be sharply peaked about the mean 
value $\av{ \< A \>_{\!\rho(t)}}$.
Or, in the words of Talagrand 
\cite{tal96}, 
``[a] random variable that depends (in a `smooth' way) 
on the influence of many independent variables (but not 
too much on any of them) is essentially constant''.
Basically (up to the ``quantitative  details''), 
this expectation is tantamount to the main result 
(\ref{s1021}) of the present section.

CLT-like phenomena of the above kind are often 
referred to as ``concentration of measure'', 
``typicality'', ``self-averaging'', 
or ``ergodicity'' properties.
Specifically in the context of random matrix 
theory, they are mostly taken for granted
without any further comment.
In other words, the main emphasis is put 
on evaluating ensemble-averaged properties, 
while the fluctuations about the average
are tacitly assumed to be negligible, but 
are hardly ever explicitly considered.

In our present work, we do not take 
such a property for granted, but 
rather we derive it.

%%%%%%%%%%%%%%%%%%%%%%%%%%%%%%%%%%%%%%%%%%%%%%%%%%%%
%
\subsection{Derivation of the upper bound (\ref{s1021})}
\label{s53}
To begin with, we rewrite $\av{ \< A \>_{\!\rho(t)}}$
by means of (\mref{4}) and (\mref{6})
in the compact form
\begin{eqnarray}
	\av{\< A \>_{\!\rho(t)} }
        & = & 
	\sum_{i...n} \rho_{ij} \, A_{kl}\, e^n_m\, K^{i...n}
	\ ,
\label{s1100}
\end{eqnarray}
where we introduced the abbreviations	
\begin{align}
\rho_{mn} &  :=  \braN{m} \rho(0) \ketN{n}
\ ,
\label{s1110}
\displaybreak[0] \\
A_{mn}  & :=  \braN{m} A \ketN{n}
\ ,
\label{s1120}
\displaybreak[0] \\
e^n_m  & :=  \e^{\I(n-m)\lvsp t}
\ ,
\label{s1130}
\displaybreak[0] \\
K^{i...n} & := 
\avv{U_{mi} U^*_{nj} U_{nk} U^*_{ml}}
\ ,
\label{s1140}
\displaybreak[0] \\
i...n & :=  i,j,k,l,m,n
\ ,
\label{s1150}
\end{align}
and where the previously employed Greek letters
have been replaced by Roman letters for notational 
convenience.
Note that $\rho_{mn}$ and $A_{mn}$ are 
identical to
the previous quantities $\rho^0_{mn}(0)$ and $A^0_{mn}$, 
and that 
\begin{eqnarray}
K^{i...n} = 
M_{mn}^{iklj}
\label{s1160}
\end{eqnarray}
according to (\ref{s210}).

As discussed in detail below (\ref{s230}),
the average in (\ref{s1140}) must vanish 
unless each of the $U$ matrix elements 
without a ``star'' has a ``partner'' among 
those with ``stars'', where being partners 
means that the second indices 
are equal.
Since there are two possibilities of
pairing the $U$'s along these lines, (\ref{s1140}) can be 
rewritten as the sum of two terms, namely
(see also (\ref{s240})--(\ref{s265}) and (\ref{s1160}))
\begin{subequations}
\label{s1190ff}
\begin{align}
K^{i...n}
%& =  \sum_{\nu=1}^2 \KK_{\nu}
& =  \KK_{1} + \KK_{2}
\ ,
\label{s1190}
\displaybreak[0] \\
\KK_1 & := 
\delta_{ij}\delta_{kl} 
\avv{U_{m i} \, U^*_{m k} \, U_{n k}\,  U^*_{ni} }
\ ,
\label{s1200}
\displaybreak[0] \\
\KK_2 & := 
\delta_{il}\delta_{jk}\overline{\delta}_{ij}
\avv{\,|U_{m i}|^2 \, |U_{n j}|^2}
\ ,
\label{s1210}
\end{align}
\end{subequations}
where the extra factor $\overline{\delta}_{ij}$
in (\ref{s1210}) is needed to avoid double
counting of the case $i=j=k=l$.
Finally, (\ref{s1100}) can be rewritten as
\begin{subequations}
\label{s1211ff}
\begin{eqnarray}
	\av{\< A \>_{\!\rho(t)} }
        & = & \sum_{\nu=1}^2 \tilde\KK_{\nu}
%        & = & \tilde\KK_{1} + \tilde\KK_{2}
\label{s1211}
\\
\tilde\KK_{\nu} & := & 
	\sum_{i...n} \rho_{ij} \, A_{kl}\, e^n_m\, \KK_\nu
	\ .
\label{s1212}
\end{eqnarray}
\end{subequations}
Similarly as in (\ref{s1110})--(\ref{s1150}) one finds that
\begin{eqnarray}
	\avv{\!\< A \>^2_{\!\rho(t)}\! }
        \!\! \!\! & = & \!\! \!\! \!\!  \!\! 
	\sum_{i...n,i'...n'} \!\! \!\! \rho_{ij} \, A_{kl}\, \rho_{i'j'} \, A_{k'l'}\, e^n_m\, e^{n'}_{m'}\, L^{i...n}_{i'...n'}
	\ ,
\label{s670}
\\
L^{i...n}_{i'...n'} \!\!  & := & \!\! 
\avv{U_{mi} U^*_{nj} U_{nk} U^*_{ml}U_{m'i'} U^*_{n'j'} U_{n'k'} U^*_{m'l'}}
. \ \ \ \ \ \ \ \ 
\label{s1180}
\end{eqnarray}
Analogously as above (\ref{s1190}), 
one can conclude that the average
in (\ref{s1180}) must vanish unless 
the $U$ matrix elements with and without
stars can be arranged into pairs with
identical second indices.
Since there are 24  possibilities of
pairing the $U$'s along these lines,
 (\ref{s1180}) must be of the form
\begin{eqnarray}
L^{i...n}_{i'...n'}
& = & \sum_{\nu=1}^{24} \LL_{\nu}
\ .
\label{s1220}
\end{eqnarray}
The explicit determination of the $\LL_\nu$
is a straightforward but somewhat tedious 
task, yielding
\begin{widetext}
\begin{subequations}
\begin{align}
\LL_1 & := 
\delta_{il}
\delta_{jk}
\delta_{i'l'}
\delta_{j'k'}
\,
\av{
\, |U_{mi}|^2
|U_{nj}|^2
|U_{m'i'}|^2
|U_{n'j'}|^2
}
\label{s1231}
\displaybreak[0]
\\
\LL_2 & := 
\delta_{il}
\delta_{jk}
\delta_{i'j'}
\delta_{k'l'}
\overline{\delta}_{i'k'}
\,
\av{
\, |U_{mi}|^2
|U_{nj}|^2
U_{m'i'}
U^*_{m'k'}
U_{n'k'}
U^*_{n'i'}
}
\label{s1232}
\displaybreak[0]
\\
\LL_3 & := 
\delta_{il}
\delta_{jk'}
\delta_{kj'}
\delta_{i'l'}
\overline{\delta}_{jk}
\,
\av{
\, |U_{mi}|^2
|U_{m'i'}|^2
U_{nk}
U^*_{nj}
U_{n'j}
U^*_{n'k}
}
\label{s1233}
\displaybreak[0]
\\
\LL_4 & := 
\delta_{il}
\delta_{ji'}
\delta_{kj'}
\delta_{k'l'}
\overline{\delta}_{jk}
\overline{\delta}_{jk'}
\,
\av{
\, |U_{mi}|^2
U_{nk}
U^*_{nj}
U_{m'j}
U^*_{m'k'}
U_{n'k'}
U^*_{n'k}
}
\label{s1234}
\displaybreak[0]
\\
\LL_5 & := 
\delta_{il}
\delta_{jk'}
\delta_{kl'}
\delta_{i'j'}
\overline{\delta}_{jk}
\overline{\delta}_{ki'}
\,
\av{
\, |U_{mi}|^2
U_{nk}
U^*_{nj}
U_{m'i'}
U^*_{m'k}
U_{n'j}
U^*_{n'i'}
}
\label{s1235}
\displaybreak[0]
\\
\LL_6 & := 
\delta_{il}
\delta_{ji'}
\delta_{kl'}
\delta_{j'k'}
\overline{\delta}_{jk}
\overline{\delta}_{jj'}
\overline{\delta}_{kj'}
\,
\av{
\, |U_{mi}|^2
|U_{n'j'}|^2
U_{nk}
U^*_{nj}
U_{m'j}
U^*_{m'k}
}
\label{s1236}
\displaybreak[0]
\\
\LL_7 & := 
\delta_{ij}
\delta_{kl}
\delta_{i'l'}
\delta_{j'k'}
\overline{\delta}_{ik}
\,
\av{
\, |U_{m'i'}|^2
|U_{n'j'}|^2
U_{mi}
U^*_{mk}
U_{nk}
U^*_{ni}
}
\label{s1237}
\displaybreak[0]
\\
\LL_8 & := 
\delta_{ij}
\delta_{kl}
\delta_{i'j'}
\delta_{k'l'}
\overline{\delta}_{ik}
\overline{\delta}_{i'k'}
\,
\av{
U_{mi}
U^*_{mk}
U_{nk}
U^*_{ni}
U_{m'i'}
U^*_{m'k'}
U_{n'k'}
U^*_{n'i'}
}
\label{s1238}
\displaybreak[0]
\\
\LL_9 & := 
\delta_{ij}
\delta_{kj'}
\delta_{lk'}
\delta_{i'l'}
\overline{\delta}_{il}
\overline{\delta}_{kl}
\,
\av{
\, |U_{m'i'}|^2
U_{mi}
U^*_{ml}
U_{nk}
U^*_{ni}
U_{n'l}
U^*_{n'k}
}
\label{s1239}
\displaybreak[0]
\\
\LL_{10} & := 
\delta_{ij}
\delta_{kj'}
\delta_{li'}
\delta_{k'l'}
\overline{\delta}_{il}
\overline{\delta}_{kl}
\overline{\delta}_{lk'}
\,
\av{
U_{mi}
U^*_{ml}
U_{nk}
U^*_{ni}
U_{m'l}
U^*_{m'k'}
U_{n'k'}
U^*_{n'k}
}
\label{s1240}
\displaybreak[0]
\\
\LL_{11} & := 
\delta_{ij}
\delta_{kl'}
\delta_{lk'}
\delta_{i'j'}
\overline{\delta}_{il}
\overline{\delta}_{kl}
\overline{\delta}_{ki'}
\,
\av{
U_{mi}
U^*_{ml}
U_{nk}
U^*_{ni}
U_{m'i'}
U^*_{m'k}
U_{n'l}
U^*_{n'i'}
}
\label{s1241}
\displaybreak[0]
\\
\LL_{12} & := 
\delta_{ij}
\delta_{kl'}
\delta_{li'}
\delta_{j'k'}
\overline{\delta}_{il}
\overline{\delta}_{kl}
\overline{\delta}_{kj'}
\overline{\delta}_{lj'}
\,
\av{
\, |U_{n'j'}|^2
U_{mi}
U^*_{ml}
U_{nk}
U^*_{ni}
U_{m'l}
U^*_{m'k}
}
\label{s1242}
\displaybreak[0]
\\
\LL_{13} & := 
\delta_{ij'}
\delta_{jk'}
\delta_{kl}
\delta_{i'l'}
\overline{\delta}_{ij}
\overline{\delta}_{ik}
\,
\av{
\, |U_{m'i'}|^2
U_{mi}
U^*_{mk}
U_{nk}
U^*_{nj}
U_{n'j}
U^*_{n'i}
}
\label{s1243}
\displaybreak[0]
\\
\LL_{14} & := 
\delta_{ij'}
\delta_{ji'}
\delta_{kl}
\delta_{k'l'}
\overline{\delta}_{ij}
\overline{\delta}_{ik}
\overline{\delta}_{jk'}
\,
\av{
U_{mi}
U^*_{mk}
U_{nk}
U^*_{nj}
U_{m'j}
U^*_{m'k'}
U_{n'k'}
U^*_{n'i}
}
\label{s1244}
\displaybreak[0]
\\
\LL_{15} & := 
\delta_{ij'}
\delta_{jk}
\delta_{lk'}
\delta_{i'l'}
\overline{\delta}_{ij}
\overline{\delta}_{il}
\overline{\delta}_{jl}
\,
\av{
\, |U_{nj}|^2
|U_{m'l}|^2
U_{mi}
U^*_{ml}
U_{n'l}
U^*_{n'i}
}
\label{s1245}
\displaybreak[0]
\\
\LL_{16} & := 
\delta_{ij'}
\delta_{jk}
\delta_{li'}
\delta_{k'l'}
\overline{\delta}_{ij}
\overline{\delta}_{il}
\overline{\delta}_{jl}
\overline{\delta}_{lk'}
\,
\av{
\, |U_{nj}|^2
U_{mi}
U^*_{ml}
U_{m'l}
U^*_{m'k'}
U_{n'k'}
U^*_{n'i}
}
\label{s1246}
\displaybreak[0]
\\
\LL_{17} & := 
\delta_{ij'}
\delta_{ji'}
\delta_{kl'}
\delta_{lk'}
\overline{\delta}_{ij}
\overline{\delta}_{il}
\overline{\delta}_{jk}
\overline{\delta}_{kl}
\,
\av{
U_{mi}
U^*_{ml}
U_{nk}
U^*_{nj}
U_{m'j}
U^*_{m'k}
U_{n'l}
U^*_{n'i}
}
\\
\LL_{18} & := 
\delta_{ij'}
\delta_{jk'}
\delta_{kl'}
\delta_{li'}
\overline{\delta}_{ij}
\overline{\delta}_{il}
\overline{\delta}_{jk}
\overline{\delta}_{jl}
\overline{\delta}_{kl}
\,
\av{
U_{mi}
U^*_{ml}
U_{nk}
U^*_{nj}
U_{m'l}
U^*_{m'k}
U_{n'j}
U^*_{n'i}
}
\label{s1248}
\displaybreak[0]
\\
\LL_{19} & := 
\delta_{il'}
\delta_{jk'}
\delta_{kl}
\delta_{i'j'}
\overline{\delta}_{ij}
\overline{\delta}_{ik}
\overline{\delta}_{ii'}
\,
\av{
U_{mi}
U^*_{mk}
U_{nk}
U^*_{nj}
U_{m'i'}
U^*_{m'i}
U_{n'j}
U^*_{n'i'}
}
\label{s1249}
\displaybreak[0]
\\
\LL_{20} & := 
\delta_{il'}
\delta_{ji'}
\delta_{kl}
\delta_{j'k'}
\overline{\delta}_{ij}
\overline{\delta}_{ik}
\overline{\delta}_{ij'}
\overline{\delta}_{jj'}
\,
\av{
\, |U_{n'j'}|^2
U_{mi}
U^*_{mk}
U_{nk}
U^*_{nj}
U_{m'j}
U^*_{m'i}
}
\label{s1250}
\displaybreak[0]
\\
\LL_{21} & := 
\delta_{il'}
\delta_{jk}
\delta_{lk'}
\delta_{i'j'}
\overline{\delta}_{ij}
\overline{\delta}_{il}
\overline{\delta}_{ii'}
\overline{\delta}_{jl}
\,
\av{
\, |U_{nj}|^2
U_{mi}
U^*_{ml}
U_{m'i'}
U^*_{m'i}
U_{n'l}
U^*_{n'i'}
}
\label{s1251}
\displaybreak[0]
\\
\LL_{22} & := 
\delta_{il'}
\delta_{jk}
\delta_{li'}
\delta_{j'k'}
\overline{\delta}_{ij}
\overline{\delta}_{il}
\overline{\delta}_{ij'}
\overline{\delta}_{jl}
\overline{\delta}_{lj'}
\,
\av{
\, |U_{nj}|^2
|U_{n'j'}|^2
U_{mi}
U^*_{ml}
U_{m'l}
U^*_{m'i}
}
\label{s1252}
\displaybreak[0]
\\
\LL_{23} & := 
\delta_{il'}
\delta_{ji'}
\delta_{kj'}
\delta_{lk'}
\overline{\delta}_{ij}
\overline{\delta}_{ik}
\overline{\delta}_{il}
\overline{\delta}_{jk}
\overline{\delta}_{kl}
\,
\av{
U_{mi}
U^*_{ml}
U_{nk}
U^*_{nj}
U_{m'j}
U^*_{m'i}
U_{n'l}
U^*_{n'k}
}
\label{s1253}
\displaybreak[0]
\\
\LL_{24} & := 
\delta_{il'}
\delta_{jk'}
\delta_{kj'}
\delta_{li'}
\overline{\delta}_{ij}
\overline{\delta}_{ik}
\overline{\delta}_{il}
\overline{\delta}_{jk}
\overline{\delta}_{jl}
\overline{\delta}_{kl}
\,
\av{
U_{mi}
U^*_{ml}
U_{nk}
U^*_{nj}
U_{m'l}
U^*_{m'i}
U_{n'j}
U^*_{n'k}
}
\label{s1254}
\end{align}
\end{subequations}
\end{widetext}
Note that the appearance of the 
$\overline{\delta}$-factors (to avoid 
double countings) depends on the 
order in which the 24 cases 
are executed.
In the above list, the first four Kronecker deltas
in every given $\LL_\nu$ uniquely determine with which 
among the 24 possible cases we are dealing, and the 
sequence of the indices
$\nu$ fixes the specific ordering of the 24 
cases which we have chosen.
Finally, (\ref{s1180}) can be rewritten as
\begin{subequations}
\begin{eqnarray}
	\avv{\< A \>^2_{\!\rho(t)}}
        & = & \sum_{\nu=1}^{24} \tilde\LL_{\nu} 
        \ ,
\label{s1300}
\\
\tilde\LL_{\nu} 
& := & \!\! \!\!
\sum_{i...n,i'...n'} \!\! \!\! \rho_{ij} \, A_{kl}\, \rho_{i'j'} \, A_{k'l'}\, e^n_m\, e^{n'}_{m'}\, \LL_\nu
\ . \ \ \ \ \ 	
\label{s1310}
\end{eqnarray}
\end{subequations}
Taking into account in (\ref{s1310}) the explicit
expressions for $\LL_\nu$ from 
(\ref{s1231})--(\ref{s1254}), one can infer that
\begin{subequations}
\begin{align}
\tilde\LL_2 & =  \tilde\LL_7
\label{s1320}
\displaybreak[0] \\
\tilde\LL_4 & =  \tilde\LL_{13}
\label{s1330}
\displaybreak[0] \\
\tilde\LL_5 & =  \tilde\LL_9
\label{s1340}
\displaybreak[0] \\
\tilde\LL_6 & =  \tilde\LL_{15}
\label{s1350}
\displaybreak[0] \\
\tilde\LL_{10} & =  \tilde\LL_{19}
\label{s1360}
\displaybreak[0] \\
\tilde\LL_{12} & =  \tilde\LL_{21}
\label{s1370}
\displaybreak[0] \\
\tilde\LL_{16} & =  \tilde\LL_{20}
\label{s1380}
\displaybreak[0] \\
\tilde\LL_{18} & =  \tilde\LL_{23}
\label{s1390}
\end{align}
\end{subequations}
Effectively, we are thus left with 16 summands in (\ref{s1300}).

The main remaining task is to evaluate the ensemble 
averages over eight $U$ matrix elements on the right-hand side of (\ref{s1231})--(\ref{s1254}).
In principle, this might be doable by a corresponding
generalization of the supersymmetric approach
from Sec.~\ref{sec:OverlapsFromSupersymmetry}.
In practice, already for the averages over four $U$ 
matrix elements in Sec.~\ref{sec:OverlapsFromSupersymmetry}, the explicit
evaluation of the pertinent saddle-point 
approximation turned out to be at the limit
of what seems analytically doable.
Therefore, such an extension of the supersymmetric
approach is beyond the scope of our present work. 
Instead, we will adopt the alternative
approach from Sec.~\ref{s4}.

We begin with recalling the evaluation of 
(\ref{s1190ff})--(\ref{s1211ff})
by means of the approach from Sec.~\ref{s45},
but employing a modified notation which is more 
suitable for the purpose of the subsequent 
generalizations.
Focusing first on the cases with $m\not=n$ in 
(\ref{s1200}) and (\ref{s1210}),
we introduce independent Gaussian random
variables $z_{1i}$ and $z_{2i}$
with (see also (\ref{s400})--(\ref{s420}))
\begin{subequations}
\begin{align}
\avvx{z_{\alpha i}} & = \avvx{\, (z_{\alpha i})^2 }= 0
\ ,
\label{s1400}
\displaybreak[0] \\
\avvx{\, |z_{\alpha i}|^2} & = u(\kappa_\alpha-i) \ ,
\label{s1410}
\\
%\alpha & \in \{1,2\}
%\label{s1420}
%\displaybreak[0] \\
\kappa_1 & :=  m \ ,
\label{s1430}
\displaybreak[0] \\
\kappa_2 & :=  n \ ,
\label{1440}
\end{align}
\end{subequations}
and $\alpha \in \{1, 2\}$.
In a next step, we define the random variables
(see also (\ref{s590}), (\ref{s600}))
\begin{eqnarray}
v_{1i} & := & z_{1i}
\ ,
\label{s1450}
\\
v_{2i}  & := & 
z_{2i} - z_{1i}\, \sum_k z_{1k}^* z_{2k}
\ ,
\label{s1460}
\end{eqnarray}
and we adopt the approximations (see also (\ref{s580}))
\begin{eqnarray}
\avv{U_{m i} \, U^*_{m j} \, U_{n k} \, U^*_{nl} }
& = &
\avvx{
v_{1i} v_{1j}^*  v_{2k}  v_{2l}^*
}
\label{s1470}
\end{eqnarray}
for the two averages appearing on the right-hand side of (\ref{s1200}) and (\ref{s1210}).
Note that in both cases the four indices $i,j,k,l$
actually amount to coinciding pairs.
While we assumed $m\not=n$ so far, the corresponding
approximations for $m=n$ are
\begin{eqnarray}
\avv{U_{m i} \, U^*_{m j} \, U_{m k} \, U^*_{ml} }
& = &
\avvx{
v_{1i} v_{1j}^*  v_{1k}  v_{1l}^*
}
\,.
\label{s1490}
\end{eqnarray}
After introducing (\ref{s1450}) and (\ref{s1460}) 
into (\ref{s1470}) and (\ref{s1490}), those averages 
can be explicitly evaluated by means of the
Isserlis  theorem (\ref{c4}) together with
(\ref{s1400}) and (\ref{s1410}).
These calculations and the subsequent
evaluation of  (\ref{s1190ff})--(\ref{s1211ff})
have been explained in detail in Sec.~\ref{s45}.

Next we turn to the averages over eight $U$ matrix 
elements in (\ref{s1231})--(\ref{s1254}).
As before, we first focus on the cases where
the four indices $m,n,m',n'$ are pairwise
distinct and we introduce four independent 
Gaussian random variables $z_{\alpha i}$
($\alpha \in \{1,2,3,4\}$)
with 
\begin{subequations}
\begin{align}
\avvx{z_{\alpha i}} & = \avvx{\, (z_{\alpha i})^2 }= 0
\ ,
\label{s1500}
\displaybreak[0] \\
\avvx{\, |z_{\alpha i}|^2} & = u(\kappa_\alpha-i)
\ ,
\label{s1510}
\displaybreak[0] \\
%\alpha & \in \{1,2,3,4\}
%\label{s1520}
%\displaybreak[0] \\
\kappa_1 & :=  m
\label{s1530}
\ ,
\displaybreak[0] \\
\kappa_2 & :=  n
\label{s1533}
\ ,
\displaybreak[0] \\
\kappa_3 & :=  m'
\label{s1536}
\ ,
\displaybreak[0] \\
\kappa_4 & :=  n'
\ .
\label{1540}
\end{align}
\end{subequations}
Next, the random variables $v_{\alpha i}$
are defined according to
\begin{eqnarray}
v_{\alpha i} & := & z_{\alpha i}
-\sum_{\beta=1}^{\alpha -1}\Big(
z_{\beta i}\, \sum_k z_{\beta k}^* z_{\alpha k}\Big)
\label{s1550}
\end{eqnarray}
and we adopt the approximations 
\begin{eqnarray}
\avv{U_{m i} \, U^*_{m j} \, U_{n k} \, U^*_{nl} U_{m' i'} \, U^*_{m' j'} \, U_{n' k'} \, U^*_{n'l'} }
& & 
\nonumber
\\
= 
\avvx{
v_{1i} v_{1j}^*  v_{2k}  v_{2l}^* v_{3i'} v_{3j'}^*  v_{4k'}  v_{4l'}^*
}
& & 
\label{s1570}
\end{eqnarray}
for the averages appearing on the right-hand side of (\ref{s1231})-(\ref{s1254}).
Again, in all those averages 
the 8 indices $i,j,k,l,i',j',k',l'$
actually amount to coinciding pairs.
In the remaining cases where the four indices 
$m,n,m',n'$ are not pairwise distinct, 
only a smaller set of indices 
$\alpha\in\{1,...,\beta\}$ with $\beta<4$ 
is actually needed.

After introducing (\ref{s1550}) into (\ref{s1570}) 
(or its counterparts if $m,n,m',n'$ 
are not pairwise distinct), the averages in
(\ref{s1231})--(\ref{s1254})
can be explicitly evaluated by means of the
Isserlis  theorem (\ref{c4}) together with
(\ref{s1500}) and (\ref{s1510}).
(Due to (\ref{s1320})--(\ref{s1390}) not all of 
them are actually needed.)
Finally, one is left with evaluating
(\ref{s1300}) and (\ref{s1310}).
These calculations are in principle
not very difficult, but in practice
they amount to a daunting task due to
the huge number of different terms.
Omitting the details of those very lengthy 
calculations, the main results are as follows:

The second moment in (\ref{s1300}) is dominated by
the contributions of the summands $\tilde\LL_\nu$
with indices $\nu\in\{1,2,7,8\}$.
To leading order, they exactly cancel the 
corresponding dominating contributions to 
the last term in (\ref{s1020}) which one 
obtains via (\ref{s1211}).
All the remaining contributions by the two
terms on the right hand side of (\ref{s1020})
are found to be of subleading order in 
the small parameter $s$ from (\ref{x10}) 
and (\ref{eq:SmallParam}).
Yet another flurry of subleading order 
terms arises for the same reasons as 
discussed below (\ref{s650a}).
Explicitly, one finally arrives at
the upper bound (\ref{s1021}).
As a first, rather conservative 
estimate for the constant $c$ 
in (\ref{s1021}) we furthermore 
obtained $c\leq \ord(10^3)$.
It seems likely that this rough upper bound could 
still be substantially reduced by painstakingly
searching among the very numerous contributions 
to (\ref{s1211}) and (\ref{s1300}) for
terms which cancel each other partially
or even completely. 
Here, we have confined ourselves to separately
bounding every single term relatively 
generously and without taking into 
account possible cancellations.
Even in this case, our detailed 
calculations extended over a very 
large number of pages.

\end{document}